\documentclass[final,oneside,12pt]{book}       
\usepackage{uf_thesis}	
\usepackage{epsf}	
\Title{{\rightline{\bf UF-IHEPA 00-01}}
\vskip 1cm
Exclusive Measurements in $B \rightarrow D^* N \bar{N} X$}
\Author{Antonio I. Rubiera}
\Degree{Doctor of Philosophy}
\WorkName{Dissertation}
\DegreeYear{2000}
\DegreeMonth{August}
\Department{Physics}
\College{Liberal Arts and Sciences}
\AddChairman{Professor of Physics}{J. Yelton}

\newcommand{\bDnnx}{$B \rightarrow [D]  N {\bar{N}} X $}
\newcommand{\bdnnx}{$B \rightarrow D^{*\pm} N {\bar{N}} X $}

\newcommand{\bbar}{$B$$\bar{B}$}

\newcommand{\bdeca}{$B^0$$\rightarrow$ $D^{*-}$ $p$ ${\bar{n}}$ }
\newcommand{\bdecb}{$B^0$$\rightarrow$ $D^{*-}$ $p$ ${\bar{p}}$ ${\pi}^+$ }

\newcommand{\bdecc}{$B^0$$\rightarrow$ $D^{**}$ $p$ ${\bar{n}}$ }

\newcommand{\dzz}{$D^0$}

\newcommand{\dst}{$D^*$}  
\newcommand{\ds}{$D^{*-}$}
\newcommand{\dsdec}{$D^* \rightarrow D^0 {\pi}_{soft}$}

\newcommand{\dsspn}{${D^{+}_{s}}$ $\rightarrow$ $p$ ${\bar{n}}$}

\newcommand{\bdsspn}{$B^0$ $\rightarrow$ ${D^{+}_{s}}$ $D^{*-}$}
\newcommand{\bdsstpn}{$B^0$ $\rightarrow$ ${D^{*+}_{s}}$ $D^{*-}$}

\newcommand{\dzdeca}{$\bar{D}^0$$\rightarrow$$K^+$${\pi}^-$${\pi}^+$${\pi}^-$}
\newcommand{\dzdecb}{$\bar{D}^0$$\rightarrow$$K^+$${\pi}^-$${\pi}^0$}

\newcommand{\dzdecc}{$\bar{D}^0$$\rightarrow$$K^+$${\pi}^-$}

\newcommand{\mdif}{$m_{D^*}$-$m_{D^0}$}
\newcommand{\mdz}{$m_{D^0}$}

\newcommand{\bbaryons}{$B \rightarrow$ Baryons}

\newcommand{\blcppi}{$B^+$ $\rightarrow {\bar{\Lambda}}_{c} p {\pi}^+$ }
\newcommand{\blcppipi}{$B^0$ $\rightarrow$ ${\bar{\Lambda}}_{c} p {\pi}^+$ ${\pi}^-$ }

\newcommand{\pizero}{${\pi}^{0}$}

\newcommand{\pisoft}{${\pi}_{s}$}

\newcommand{\bdelpp}{$B^0 \rightarrow D^{*-} {\bar{p}} \Delta^{++}$}

\newcommand{\bdelz}{$B^0 \rightarrow D^{*-} p \bar{\Delta}^{0}$}

\newcommand{\bpnpzds}{$B^0 \rightarrow D^{*-} p \bar{n} {\pi}^0$}

\begin{document}
   \frontmatter		
      \maketitle
      
   \newpage
   \addcontentsline{toc}{chapter}{\acknowledgmentsname}
   \begin{center} \acknowledgmentsname \end{center}
   \bigskip
   \doublespacing   

	If the pursuit of an undergraduate degree is comparable 
to a 500-meter race, the pursuit of a doctorate is more like a 
marathon. Many people have been instrumental to me finishing 
this marathon. 

	The idea for this analysis came from my advisor, 
John Yelton, who played a principal role in its success. 
His patience and 
wisdom have been instrumental in my development as a 
scientist. Paul Avery offered helpful criticisms along the 
way which helped me improve my delivery of the results. 
I would also like to thank 
the University of Florida faculty members who have been 
most helpful to me, for the courses they 
taught, and the professional guidance they willingly 
volunteered: Pierre Ramond, Charles Thorn, Zongan Qiu, 
and Bernard Whiting. 
While at Cornell I was guided and helped 
by David Besson, Brian Helstley, David Jaffe and Andy Foland. 
Many of the suggestions that have improved the quality of 
this analysis came from these colleagues. 
My fellow graduate students in the CLEO Florida 
group, Jiu Zheng and Craig Prescott, were patient 
in their guidance. Andy Foland and Craig Prescott 
proved to me that brilliance can be achieved without 
arrogance. 

	The long and tortuous road to the finish line 
would not be possible without the unflinching 
support of my family: my grandfather and grandmother, 
my father and mother. I fail to find words that
accurately describe how deeply I feel my debt to them. 
Neither my grandfather nor my mother lived to 
see their seeds bear fruit. Their positive influence 
is sorely missed. 

	The lunch CLEO software elite endured 
my opinions: Andreas Warburton, David Urner, 
Peter Gaidarev, Martin Lohner, Chris Jones and 
Adam Lyon. The Chapter House gang, Rahida, Samina, 
Basit and Mike Marsh, made my Friday nights during 
the nine months of Ithaca summer considerably 
more enjoyable than they would have been otherwise. 
I thank the deplorable upstate New York weather 
for forcing me to work harder. 
The Chapter House gang also endured my opinions, 
but with the added advantage of a few beers. 
Herbert, Pia, and baby Gabriel offered 
me company off-CLEO while I lived in Ithaca. 
Lauren Hsu and Antonella Cipollone allowed me 
to pass on some of my analysis experience. I thank 
Jean Duboscq and Bonnie Valant-Spaight, and 
Stefan Anderson. 

	I have been fortunate to be graced with 
friends who have offered me 
their company and their understanding during the 
bad times and loads of fun during the good times: 
From Cornell EE, Wolfgang Hofman and 
Jason Reed; From UF, Steve Thomas (who shared with me 
his deep insights into French culture), Dawn Shuler, 
Mike (DR) Jones, Richard Pietri, Richard Haas 
and Ilsa Webeck; From Miami High/Miami/Cornell, 
Christine Sobilo, Luis (Kike) Ramos, 
George and Oscar Hernandez, Armando Garcia de la Torre, 
Elizabeth San Martin, Elizabeth Padron, 
Mario and Blanca Berrios, Jimmy Windsor, 
Jimmy Windsor Jr, Tiburon, and others who 
I may have unwittingly forgotten. Barbara 
Tuchman and Henry Kissinger provided invaluable 
reading material. Madonna, Depeche Mode, 
and the Orb provided great music. 

I hope a new generation of graduate students 
is able to profit from this analysis, and thank 
the CLEO collaboration for all its support.

   \singlespacing

      \tableofcontents
      
   \newpage
   \singlespacing
   \addcontentsline{toc}{chapter}{\abstractname}
   \begin{center} 
      Abstract of \WorkNameValue\ Presented to the Graduate School \\
      of the University of Florida in Partial Fulfillment of the \\
      Requirements for the Degree of \DegreeValue
      \par\bigskip
      \TITLEVALUE
      \par\bigskip
      By
      \par\bigskip
      \AuthorValue
      \par\bigskip
      \DegreeMonthValue\ \DegreeYearValue
      \par\bigskip
   \end{center}
   \begin{flushleft}
   Chairman: \Chairman \\
   Major Department: \DepartmentValue
   \end{flushleft}

   \doublespacing   
We report the first observation of exclusive decays of the 
type $B \rightarrow D^*N\bar{N}X$, where $N$ 
is a nucleon. Using a sample of 9.7 $\times$ $10^{6}$ 
$B\bar{B}$ pairs collected with the CLEO detector 
operating at the Cornell Electron Storage Ring, 
we measure the branching fractions ${\cal B}$
({$B^0 \rightarrow D^{*-} p {\bar{p}} {\pi}^+$}) = 
(${6.5}^{+1.3}_{-1.2} \pm 1.0) \times 10^{-4}$, and 
${\cal B}$({$B^0$$\rightarrow D^{*-} p {\bar{n}}$
}) = (${14.5}^{+3.4}_{-3.0} \pm 2.7) 
\times 10^{-4}$. 
The charge conjugate process is implied in 
the reconstruction of 
$B^0$$\rightarrow$ $D^{*-}$ $p$ ${\bar{p}}$ ${\pi}^+$. 
However, in the reconstruction of 
$B^0$$\rightarrow$ $D^{*-}$ $p$ ${\bar{n}}$, 
only the mode with the antineutron is used in our 
measurement because neutrons do not have a 
distinctive annihilation signature.

Antineutrons are identified by their 
annihilation in the CsI electromagnetic calorimeter. 
Since we are unable to isolate a sample of 
antineutrons in data, we use antiproton annihilation 
showers in a $\bar{\Lambda} \to \bar{p} \pi^+$ sample 
to define the antineutron selection criteria. 
We find a discrepancy 
for antiproton annihilation showers 
between the Monte Carlo and data, which we assume 
affects antineutrons as well. 
We increase the raw yield for {\bdeca} by 21$\%$ 
to correct for this discrepancy. 

The possible contributions from 
{$B^0$ $\rightarrow$ $D^{*-}$ ${D^{+}_{s}}$} 
with {${D^{+}_{s}}$ $\rightarrow$ $p$ ${\bar{n}}$} 
and {$B^0$ $\rightarrow$ $D^{*-}$ ${D^{*+}_{s}}$} 
with {$D^{*+}_{s} \rightarrow D^{+}_{s} \gamma$} and 
{${D^{+}_{s}}$ $\rightarrow$ $p$ ${\bar{n}}$} are 
eliminated from the analysis by rejecting events with 
1.91 GeV $<$ $M_{p + \bar{n}}$ $<$ 2.04 GeV for a 
loss of 9$\%$ in the reconstruction efficiency. 
We fail to find evidence for the decay 
${D^{+}_{s}}$ $\rightarrow$ $p$ ${\bar{n}}$.

We search for possible contributions 
to the resonant substructure of {\bdeca} 
and {\bdecb} due to 
a heavy charmed baryon decaying 
strongly to $\bar{p}$ $D^{*-}$ for 
$B^0$$\rightarrow$ $D^{*-}$ $p$ ${\bar{p}}$ ${\pi}^+$ 
and $\bar{n}$ $D^{*-}$ for 
$B^0$$\rightarrow$ $D^{*-}$ $p$ ${\bar{n}}$, as well as 
a resonance of the virtual W 
decaying to $p \bar{p} \pi^+$. We also 
study the possible effect of feed-down 
$\Delta$ baryon contributions to the 
background for both modes, as well as 
the $B^0$$\rightarrow$ $D^{*-}$ $p$ ${\bar{p}}$ ${\pi}^+$ 
signal. No conclusive evidence is found for a measurable 
contribution from the aforementioned contributions to the 
resonant substructure.

Antineutrons are used for the first time in the 
exclusive reconstruction of a $B$ meson. 
By finding conclusive evidence for the existence 
of decay modes of the type $B \to DN\bar{N}X$, 
we challenge the assumption that the 
$B \rightarrow Baryons$ rate is dominated by 
decays to charmed baryons.


   \singlespacing

      \listoffigures
      \listoftables
   \mainmatter		
\chapter{Introduction}

The study of the elementary particles at a wide 
range of interaction energies has occupied scientists 
since the discovery of the electron in 1896 \cite{pais}. 
Particle physics has evolved from a field involved in 
the discovery of new particles to one devoted to 
their systematic study. A logical structure, currently 
explained by the Standard Model of elementary particles 
\cite{basic,Peskin,DGH}, has been unveiled. 

The Standard Model, however, offers an incomplete picture 
of some experimental high energy physics results. The 
results we describe are amongst these. Despite some 
weaknesses, the Standard Model has withstood intense 
experimental scrutiny, and while some of the equations 
are currently not calculable, evidence has not been 
found for physics beyond the Standard Model. Particles 
acquire their masses in the Standard Model via 
the Higgs mechanism, that requires the existence 
of a massive gauge boson, the Higgs boson, that 
has yet to be found. 

We first introduce particle physics phenomenology
-particles and their decays, which is the information 
most often used by the practitioner of experimental 
high energy physics. A discussion of the current 
status of B physics phenomenology follows. 
Finally, we review {\bbaryons} previous 
to our results, concluding with an overview of our 
work and its significance. 

\section{Matter}  

Matter is composed of 
three types of indivisible constituents: 
leptons, gauge bosons, and quarks. Leptons and 
gauge bosons are found individually in Nature.  
Quarks combine in two types of arrangements to form 
hadrons. The first arrangement is of the form 
quark-antiquark, and is called a meson. The 
second arrangement, three quarks, is called a baryon. 
Mesons and baryons, collectively known as hadrons, 
comprise all the known forms of matter consisting of quarks.

Hadronic matter is said to be colorless. A color 
is assigned to each quark in a meson or a baryon. 
The three quarks in a baryon each 
have a different color$-$red, green, blue$-$
and the combination of all 
three colors yields a colorless hadron. 
The quark and the antiquark in a meson form 
a color-anticolor pair (e.g. $q_{blue} \bar{q}_{\overline{blue}}$). 

\subsection{Hadrons}

	There are three families of quarks, each consisting of 
two types of quark: an up-type quark, with 
electromagnetic charges +2/3 that of the electron, or 
(+2/3) $q_{e}$, and a down-type quark, with (-1/3) $q_{e}$. 
Every type of quark is called a flavor of quark.

	The first and lightest family consists of the up ($u$) and 
down ($d$) quarks. The next family, with heavier quarks, 
is the charm ($c$) and strange ($s$) family. 
Even heavier still is the third family: the top ($t$) and 
bottom ($b$) quarks. All are shown in Table \ref{quarks}. 

\begin{table}[htb]
\caption{Quark families
\label{quarks} }
\begin{center}
\begin{tabular}{|c|c|c|c|}
\hline 
up-type (+2/3) $q_{e}$  & up: $u$ & charm: $c$ & top: $t$ \\
down-type (-1/3) $q_{e}$  & down: $d$ & strange: $s$ & bottom: $b$ \\ 
\hline 
\end{tabular}
\end{center}
\end{table}

	For every matter constituent there is an 
anti-matter constituent with opposite electromagnetic 
charge and equal mass, as shown in 
Table \ref{quarks2}.

\begin{table}[htb]
\caption{Anti-quark families
\label{quarks2} }
\begin{center}
\begin{tabular}{|c|c|c|c|}
\hline  
up-type (-2/3) $q_{e}$  & $\bar{u}$ & $\bar{c}$ & $\bar{t}$ \\ 
down-type (+1/3) $q_{e}$  & $\bar{d}$ & $\bar{s}$ & $\bar{b}$ \\ 
\hline 
\end{tabular}
\end{center}
\end{table}

	Quarks are not found individually in Nature. 
Their masses can be estimated by the mass spectrum of 
mesons and baryons that has been measured to date. 
In Table \ref{qmasses} we show current best 
estimates of the lower and upper limits for 
quark masses. The estimated masses for quarks 
are useful in current Standard Model calculations. 
However, the quark mass estimates have large 
uncertainties, especially in the case of the 
up and down quarks. The values quoted in this Table and 
the next values quoted in this section 
are from the 1998 Review of particle physics \cite{pdg99}. 
The mass of the top quark $m_{t} = 173.8 \pm 5.2$ GeV.

\begin{table}[htb]
\caption{Estimated quark masses
\label{qmasses} }
\begin{center}
\begin{tabular}{|c|c|c|}
\hline 
Quark & Low mass & High mass \\ \hline
u & 1.5 MeV & 5 MeV  \\ 
d & 3 MeV & 9 MeV \\ 
s & 60 MeV & 170 MeV \\ 
c & 1.1 GeV &  1.4 GeV \\ 
b & 4.1 GeV &  4.4 GeV \\ \hline 
\end{tabular}
\end{center}
\end{table}

        The proton, for example, has the quark content ($uud$), 
and one of the pions, the $\pi^+$, has the quark content ($u\bar{d}$). 
Each type of quark is considered a distinct flavor. 
The group of mesons containing a charm quark and a light 
antiquark (one of $\bar{s}$, $\bar{u}$, $\bar{d}$) 
is called the charm mesons. 

	Meson and baryon masses are known 
to varying degrees of accuracy, as shown in 
Table \ref{mesons} for selected mesons, and 
Table \ref{baryons} for selected baryons. 

\begin{table}[htb]
\caption{Selected meson masses 
\label{mesons} }
\begin{center}
\begin{tabular}{|c|c|c|}
\hline
Meson & Quark content & Mass \\ \hline
$\pi^\pm$ & $u\bar{d}$ & $139.56995 \pm 0.00035$ MeV  \\ 
$K^\pm$ & $s\bar{u}$ & $493.677 \pm 0.016$ MeV  \\ 
$D^0$ & $c\bar{u}$ & $1864.6 \pm 0.5$ MeV \\ 
$B^0$ & $b\bar{d}$ & $5279.2 \pm 1.8$ MeV \\ \hline 
\end{tabular}
\end{center}
\end{table}

\begin{table}[htb]
\caption{Selected baryon masses 
\label{baryons} }
\begin{center}
\begin{tabular}{|c|c|c|}
\hline 
Meson & Quark content & Mass \\ \hline
proton & $uud$ & $938.27231 \pm 0.00028$ MeV  \\ 
$\Lambda$ & $sud$ & $1115.683 \pm 0.006$ MeV  \\ 
$\Lambda_{c}$ & $cud$ & $2284.9 \pm 0.6$ MeV \\ 
$\Lambda_{b}$ & $bud$ & $5624 \pm 9$ MeV \\ \hline 
\end{tabular}
\end{center}
\end{table}

\subsection{Leptons.}

	The leptons, the $e$, the $\mu$, and the $\tau$, 
are fundamental particles. Each can be produced with an 
accompanying neutrino, $\nu_{l}$, with $l=e,\mu$, or $\tau$. 
Neutrino physics in the near future will yield a 
better understanding than is currently available. 
Neutrinos have been thought to be massless and not 
to mix (that is, each neutrino flavor was thought 
only to interact with 
its lepton partner). The recent observation of 
neutrino mixing, however, implies that neutrinos 
have mass \cite{nmixing}. 
Unlike quarks, leptons are observed as single 
particles with well-measured masses. In Table \ref{leptons} 
we show the current mass measurements. 
Large differences in masses are found between the 
electron, the muon, and the tau.

\begin{table}[htb]
\caption{Lepton masses 
\label{leptons} }
\begin{center}
\begin{tabular}{|c|c|}
\hline 
Lepton & Mass \\ \hline
electron: $e^\pm$ & $0.51099907 \pm 0.00000015$ MeV   \\ 
muon: $\mu^\pm$ & $105.658389 \pm 0.000034$ MeV  \\ 
tau: $\tau^\pm$ & ${1777.05}^{+0.29}_{-0.26}$ GeV  \\ \hline 
\end{tabular}
\end{center}
\end{table}

Helicity is the orientation of a particle's 
momentum vector in respect to its spin. Helicity 
can be right-handed or left-handed for all 
fermions except neutrinos, which are 
only left-handed. Antineutrinos, likewise, 
are only right-handed. 

\subsection{Gauge Bosons}

	The mediators of the physical forces, the particles 
that allow decays to take place, are called gauge bosons. 
In Table \ref{bosons} we outline their masses and 
the types of interaction that each mediates. 
$l=e,\mu,\tau$, $u$=up-type quark($u,c,t$), 
and  $d$=down-type quark($d,s,b$).

\begin{table}[htb]
\caption{Some characteristics of bosons 
\label{bosons} }
\begin{center}
\begin{tabular}{|c|c|c|c|}
\hline 
Boson & Force & Mass & Decays \\ \hline 
photon & electromagnetic & massless & 
$\gamma \rightarrow l^+l^-$ \\ 
gluons & strong & massless & 
$g \rightarrow q \bar{q}$ \\ 
$W^\pm$ & weak & $80.41 \pm 0.10$ GeV & 
$W^+ \rightarrow l^+\nu_{l}, u\bar{d}$  \\ 
$Z^0$ & weak & $91.187 \pm 0.007$ GeV & 
$W^+ \rightarrow l^+l^-, \nu\bar{\nu}, q^+q^-$ \\ \hline
\end{tabular}
\end{center}
\end{table}

\subsection{Spin and Statistics}

Matter is also characterized by the statistics 
obeyed. Leptons and quarks are fermions, 
obeying Fermi statistics, 
an example of which is the Pauli exclusion 
principle for electrons occupying the same 
shell in an atom. The spin of leptons 
and quarks is ($\pm\frac{1}{2}$). 
Bosons obey Bose statistics, 
which allow an infinite number of particles 
to occupy the same energy state. 
Bosons have integral spin (0, $\pm$1). 
The different relative alignments of the 
spins of the individual quarks, together 
with the addition of angular momentum, 
results in a large number of possible 
states. 

\subsection{The CKM Matrix} 

	The interactions of quarks from different 
flavor families are suppressed with respect to 
those within the same family. 
In order to correct for this discrepancy, the 
characteristics of a given decay as prescribed 
by weak theory are adjusted using the 
flavor mixing 3 $\times$ 3 unitary and complex matrix $V$, 
the Cabibbo-Kobayashi-Maskawa matrix \cite{pdg99}, 
which transforms the weak $d, s, b$ quark 
states to the physically measured $d^{'}, s^{'}, b^{'}$ 
quark states, to arrive at a correct theoretical 
understanding of a weak decay for all quark 
flavors:

\vskip 20pt 

\centerline{$V_{CKM} =$
$\left( \begin{array}{ccc} 
V_{ud} (0.9745-0.9760) & 
V_{us} (0.217-0.224) & 
V_{ub} (0.004-0.013) \\ 
V_{cd} (0.217-0.224) & 
V_{cs} (0.9737-0.9753)& 
V_{cb} (0.036-0.042)\\ 
V_{td} (0.004-0.013)& 
V_{ts} (0.035-0.042)& 
V_{tb} (0.9991-0.9994)
\end{array} \right)$
}

\vskip 20pt 

with 
\vskip 20pt 

\centerline{$\left( \begin{array}{c} 
d^{'} \\ s^{'} \\ b^{'} \\ \end{array} \right) = 
V \left( \begin{array}{c} 
d \\ s \\ b \\ \end{array} \right)$ 
}

\vskip 20pt 

	The diagonal elements of $V$ are $\approx$ 1, 
implying that decays which involve these CKM matrix 
elements are Cabibbo-allowed, whereas all other 
decays, which involve off diagonal elements, are 
Cabibbo-suppressed. For example, a 
$b \rightarrow cW^{-}, W^{-} \rightarrow \bar{c}s$ decay 
has a much larger decay rate than 
a $b \rightarrow cW^{-}, W^{-} \rightarrow \bar{c}d$ 
decay. 

\subsection{Symmetries}

	A particle and its antiparticle are said to 
be symmetric under CPT transformations, where C is 
charge conjugation, P is parity, and T is 
time reversal. The inversion of parity acts like 
a mirror in the inversion of space coordinates. 
CPT symmetry is valid for all forces. 

	A subset of this symmetry is CP. The product 
of a charge conjugation and a parity inversion affect 
a particle the same as its antiparticle. When 
CP is violated, a particle will prefer a different 
subset of decays than its antiparticle. Such is 
the case for neutral kaons.The 
$K^0$ and $\bar{K}^0$ states in weak theory are 
different from the physically observed 
strong states, the $K_{S}$, or $K$-short and 
the $K_{L}$, or $K$-long, each with its respective 
antiparticle. Short and long refer to the respective 
decay lifetimes. Charge symmetry is obeyed 
implicitly in neutral kaon decay, but parity is 
violated. CP violation is evident in the difference 
in decay rates of $K_{S}$'s and $K_{L}$'s to final 
states of two and three pions, the former being 
even under parity, and the latter odd under parity 
\cite{cpkaon}. CP violation 
in $B$ decays and will be the focus of many studies 
in the near future. 

\section{Decays} 

	All hadrons, except the proton and electron, 
are unstable and decay to lighter hadrons, leptons, 
and bosons. The top quark decays well before the time 
required to form hadrons (baryons or mesons). Therefore, 
the bottom group of baryons (e.g. $\Lambda_{b}$, with 
quark content $bud$) and mesons (e.g. $B_{c}$, 
with quark content $b\bar{c}$) is the 
group of hadrons with the largest masses 
currently found in Nature. 

\subsection{Weak Decays}

\begin{figure}[ht]
   \centering \leavevmode
        \epsfysize=6cm
   \epsfbox[170 525 575 699]{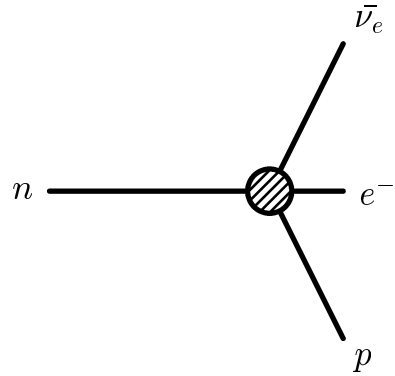}
   \caption{A Feynman diagram for neutron beta decay in Fermi 
Weak Theory} 
\label{fermi}
\end{figure}

\begin{figure}[ht]
   \centering \leavevmode
        \epsfysize=3cm
   \epsfbox[70 525 575 699]{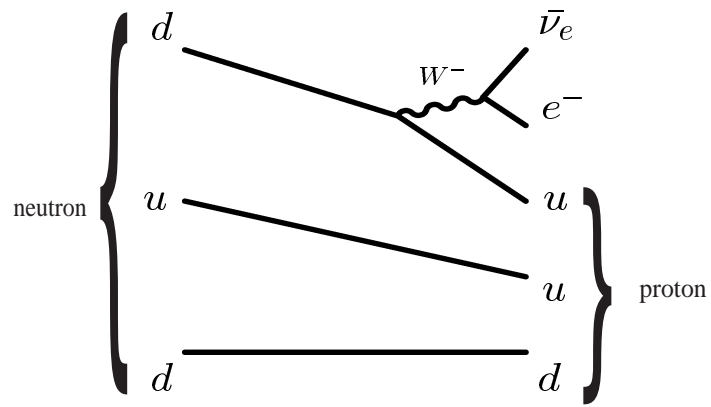}
\vskip 90pt
   \caption{A Feynman diagram for neutron beta decay in 
the Standard Model} 
\label{fermi2}
\end{figure}

	The b quark in the B meson decays via the weak 
interaction. The b quark decay is often accompanied by 
the strong decays of soft gluons which allow for the 
hadronization of a large number of final states. 
The Fermi Electroweak theory, 
which aimed to explain neutron $\beta$ decay, 
$n \rightarrow pe^-\bar{\nu_{e}}$, as shown by 
the Feynman diagram Figure \ref{fermi}, introduced the 
neutrino to particle physics, serving as the precursor 
of the Standard Model. A full theoretical understanding 
of this decay was accomplished by the Standard Model, 
in which this decay is mediated by the W vector boson, 
not present in the Fermi Electroweak theory. The 
decay $n \rightarrow pe^-\bar{\nu_{e}}$ is correctly 
described as the quark level process 
$d \rightarrow W^-u$, followed by 
$W^- \rightarrow e^-bar{\nu_{e}}$, as shown by the 
Feynman diagram in Figure \ref{fermi2}. 

\clearpage

	Examples of weak semi-leptonic decays of mesons, 
analogous to neutron $\beta$ decay, are 
$K^+ \rightarrow \pi^0e^+\nu_{e}$, 
$D^0 \rightarrow K^-e^+\nu_{e}$, and 
$B^0 \rightarrow D^-e^+\nu_{e}$. 
The weak semi-leptonic decays 
of baryons follow a similar pattern. 
These decays are referred to as semi-leptonic decays 
since the final products are a combination of a 
weak leptonic decay and quark hadronization. 
Leptonic weak decays, in which the unstable 
particle annihilates into an $l^+\nu_{l}$ pair, examples 
of which are $\pi^+ \rightarrow \mu^+\nu_{\mu}$ and 
$K^+ \rightarrow \mu^+\nu_{\mu}$ are the most 
theoretically well understood type of decays, 
as it lacks any final state hadronization. 
	
\subsection{Strong Decays}
 
The weak decays of the heavy (charm and bottom) 
quarks in mesons and baryons often involve the secondary 
emission of strongly-interacting soft gluons. While 
we understand the weak component of these decays 
using the Standard Model, the strong component is 
not calculable. 

The strong interaction hadronization process 
is not well understood theoretically when soft, or 
low momentum, gluons mediate the decay. The Standard 
Model is based on perturbation expansions which rely 
on convergence. A process involving soft gluons 
yields equations that are no longer perturbatively 
convergent. This stumbling block has prevented us 
from understanding many details of unstable particle 
decay, particularly for the case of heavy hadrons, 
in which the large mass available in the decay 
implies a very large number of possible final 
decay products from an equally large number 
of soft gluons. 

\begin{figure}[ht]
   \centering \leavevmode
        \epsfysize=4cm
   \epsfbox[70 525 575 699]{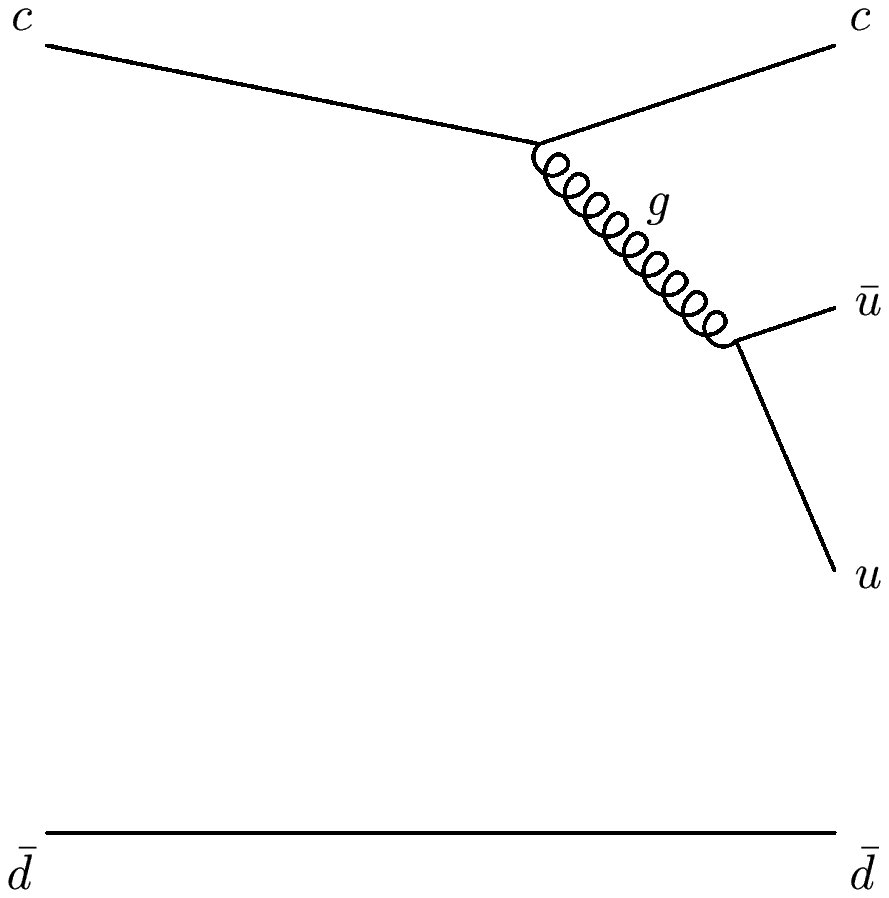}
\vskip 140pt
   \caption{A Feynman diagram for {\dsdec} } 
\label{dsstrong}
\end{figure}

\begin{figure}[ht]
   \centering \leavevmode
        \epsfysize=4cm
   \epsfbox[70 525 575 699]{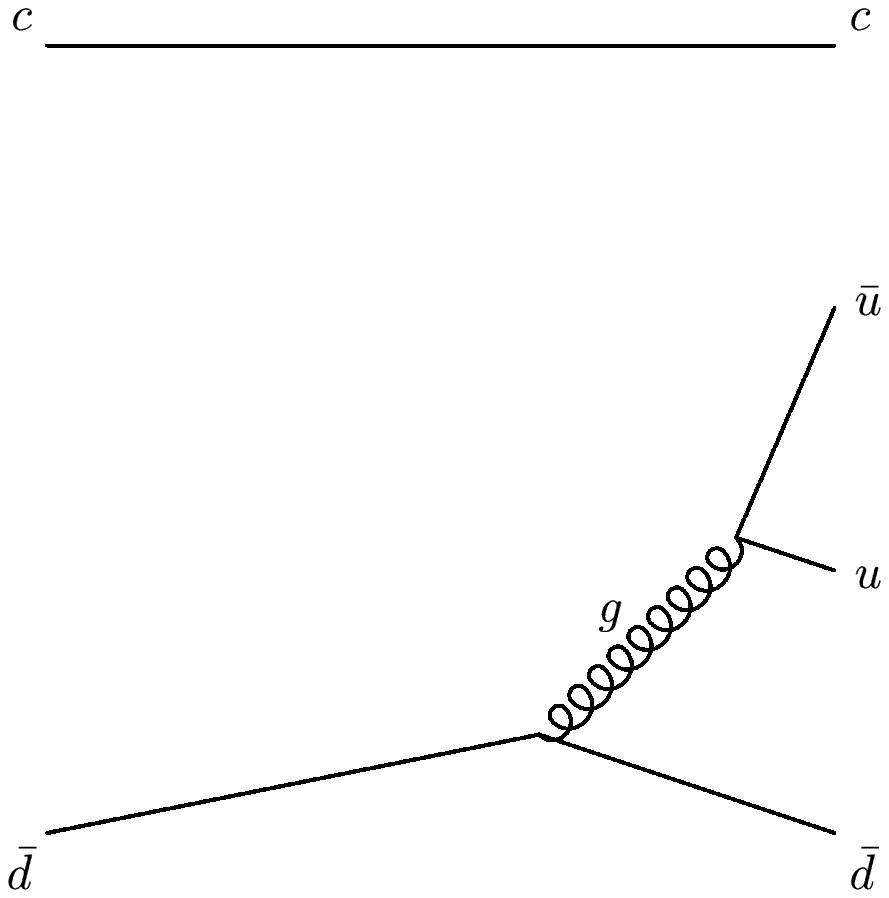}
\vskip 200pt
   \caption{A second Feynman diagram for {\dsdec} } 
\label{dsstrong2}
\end{figure}

	The {\dst} meson, for example, 
which we reconstruct in this analysis, is a 
spin 1 meson that decays via the strong interaction. 
Two possible Feynman diagrams for this 
decay are shown in Figures \ref{dsstrong} 
and \ref{dsstrong2}. 
Our ignorance about soft gluon strong 
interactions forbids us from knowing 
the proportion of the total decay rate 
due to any one diagram. 

\vskip 100pt

\clearpage

\clearpage

\section{B Meson Decays \label{bdecays} }

	The B meson, which is the topic of our study, 
can decay to a large number of lighter particles by 
various mechanisms, some of which are detailed 
in Table \ref{mech}. The $b \rightarrow cW^{-}$ 
type decays account for more than 95$\%$ of 
the $B$ decays that are possible in 
the Standard Model. The combined semileptonic decay rate 
for the three lepton families is $\approx$ 25$\%$, 
with hadronic decays accounting for almost all of 
the remaining decay rate. 

\begin{table}[htb]
\caption{Some B meson decay mechanisms 
\label{mech} }
\begin{center}
\begin{tabular}{|c|c|}
\hline
Quark-level mechanism & Sample decay mode \\ \hline 
Semileptonic decay: & \\ 
$b \rightarrow cW^{-}, W^{-} \rightarrow l^- \nu_{l}$ 
& $B \rightarrow D l^- \nu_{l}$ \\ 
Hadronic decays: & \\ 
$b \rightarrow cW^{-}, W^{-} \rightarrow \bar{c}s$ 
& $B \rightarrow D D_{s}$ \\ 
$b \rightarrow cW^{-}, W^{-} \rightarrow \bar{u}d$ 
& $B \rightarrow D \pi$ \\ \hline 
\end{tabular}
\end{center}
\end{table}	

\begin{figure}[ht]
   \centering \leavevmode
        \epsfysize=4cm
   \epsfbox[70 525 575 699]{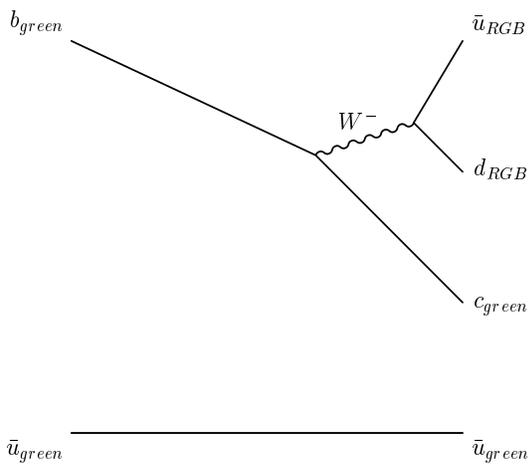}
\vskip 180pt
   \caption{A color-allowed Feynman diagram for 
$B^- \to D^0\pi^-$ for one quark color } 
\label{color}
\end{figure}

\begin{figure}[ht]
   \centering \leavevmode
        \epsfysize=4cm
   \epsfbox[70 525 575 699]{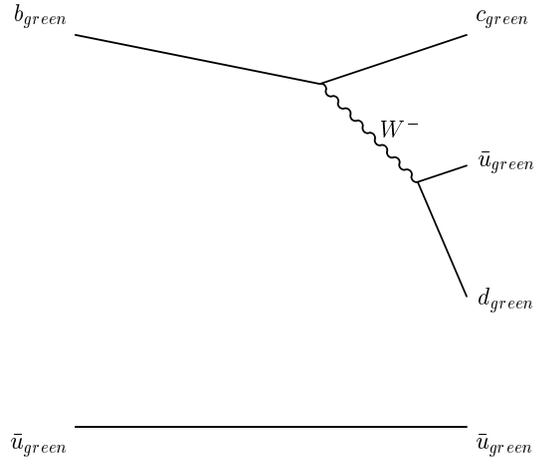}
\vskip 180pt
   \caption{A color-suppressed Feynman diagram for 
$B^- \to D^0\pi^-$ for one quark color } 
\label{color2}
\end{figure}

	The $W$ boson is colorless. 
Its decay to quarks constrains 
the color of both quarks to cancel. 
The number of Feynman diagrams for a 
given $B$ decay by the color of the quarks. 
Whereas decays mediated by an external $W$ boson 
allow for any of the three possible colors 
(for example, the $\pi^-$ in $B^- \to D^0\pi^-$, 
as shown in the Feynman diagram in Figure 
\ref{color}). Those in which the $W$ boson 
decays internally limit the color of all 
the quarks to be the same as that of the 
parent meson, as shown in the Feynman diagram in Figure 
\ref{color2}. The latter quality is referred to 
as color suppression. Decays that only  
have an internal $W$ boson-mediated diagram, 
such as $B^0 \to D^0 \pi^0$, are color-suppressed decays. 

\clearpage
	
\clearpage

	The decays $b \rightarrow (c,u)W^{-}$ 
are mediated by the $W^{-}$ vector boson, while 
decays of type $b \rightarrow (s,d)\gamma$ and 
$b \rightarrow (s,d)g$ are mediated by neutral bosons. 
All decays except $b \rightarrow cW^{-}$ 
contribute a small fraction of the total decay rate. 
However, it is for many of these rare decays that 
our current phenomenological understanding allows 
for the closest agreement between theory and 
experiment. 

	It is theoretically allowed, and has been 
experimentally measured, that a $B^0$ meson 
can oscillate to a $\bar{B}^0$ before decay, allowing 
for neutral $\Upsilon(4S) \rightarrow B\bar{B}$ 
events with either 2 $B^0$'s, or 2 $\bar{B}^0$'s 
\cite{andy}. 

\subsection{Quantum Chromodynamics}

	The Standard Model unifies the electromagnetic 
and the weak interactions. It also encompasses the 
strong interactions in the form of Quantum 
Chromodynamics (QCD) \cite{qcd,qcd2}. 
The strong coupling constant 
$\alpha_{s}$ is smallest for short 
range interactions, or for large momentum 
transfer, a quality of QCD called asymptotic 
freedom. 

	The QCD Lagrangian is given by \cite{quigg}: 
\vskip 20pt
\centerline{${\cal L}_{QCD} 
= (\bar{q}_{red},\bar{q}_{blue},\bar{q}_{green})
(i\gamma^{\mu}D_{\mu}- m)\left(\begin{array}{c} 
q_{red} \\ q_{blue} \\ q_{green} \\ \end{array}\right) 
+ \frac{1}{2}tr(G_{\mu\nu}G^{\mu\nu})$+ h.c.,}
\vskip 20pt 
where the covariant derivative is:
\vskip 20pt 
\centerline{$D_{\mu} = \partial_{\mu} + 
\frac{1}{2}ig\lambda^l{A}_{\mu}$}
\vskip 20pt 

	$\frac{1}{2}\lambda^l$ are the SU(3) flavor 
matrices, and ${A}_{\mu}$ are the eight color 
gauge fields. $G_{\mu\nu}$ is the gluon 
field-strength tensor.  

\subsection{Heavy Quark Effective Theory}

	The properties of the $D$ and $B$ mesons, 
which are composed of a heavy quark and a light 
anti-quark, have been described using 
heavy-quark symmetry \cite{voloshin,shifman} by 
Heavy Quark Effective Theory (HQET). 
In a meson, a quark and an 
antiquark are confined to a bound state in a cloud 
of virtual quarks and gluons which need 
to be incorporated into any calculation. 
In the case of a heavy meson, such as 
a $B$ meson, the heavy quark has a 
substantially larger mass than the light 
antiquark. In HQET, the heavy quark is 
independent of the light anti-quark. 
HQET assumptions simplify the 
Standard Model equations, allowing, 
for instance, the comparison with theory 
of experimental values 
of $V_{ub}$ and $V_{cb}$.

The effective Lagrangian that 
is used to characterize $B$ decays is given 
by \cite{mathias,wise}: 

\vskip 20pt

\centerline{${\cal L}_{eff} 
= -2\sqrt(2) G_F {J}_{CC}^{\mu} {J}_{CC,\mu}^{\dagger}$ 
+ h.c.,}
\vskip 20pt

where $G_F$, the Fermi constant, is 1.17 GeV$^{-2}$, and 
${J}_{CC}^{\mu}$ is the charged weak current given by: 
\vskip 20pt
\centerline{${J}_{CC}^{\mu} 
= (\bar{\nu}_e,\bar{\nu}_{\mu},\bar{\nu}_{\tau})
\gamma^{\mu}\left(\begin{array}{c} 
e_L \\ \mu_L \\ \tau_L \\ \end{array}\right) + 
(\bar{u}_L,\bar{c}_L,\bar{t}_L)
\gamma^{\mu}V_{CKM}\left(\begin{array}{c} 
d_L \\ s_L \\ b_L \\ \end{array}\right)$}

\vskip 20pt

 	The assumption that the 
heavy quark mass $m_{Q}$ is effectively 
infinite is used to simplify the 
QCD Lagrangian. The heavy quark and the 
light quark are decoupled, and the effect 
of the cloud of virtual light quarks, 
light antiquarks, and gluons is assumed 
to be small enough to be ignored. 
The QCD Lagrangian, of which ${\cal L}_{eff}$ 
is a simplified version, is simplified to: 
\vskip 20pt

\centerline{${\cal L}_{Q} 
= \bar{Q} (i\gamma^{\mu}D_{\mu}- m_Q)Q$}

\vskip 20pt

where $D^{\mu}$ is the QCD covariant derivative. 

	In the limit $m_Q \to \infty$ the strong 
interactions of the heavy quark become independent 
of its mass and spin and the effective Lagrangian 
is further simplified to:
\vskip 20pt
\centerline{${\cal L}_{HQET} 
= {\bar{h}}_{\it{v}} i\it{v} \cdot D h_{\it{v}}$}
\vskip 20pt

where $h_{\it{v}}$ is the effective heavy quark field 
and $\it{v}$ is the hadron's velocity, which is 
close to that of the heavy quark.  

	In the calculation of HQET quantities, the 
strong interaction effects that are non-perturbative 
are grouped into a form factor that includes a 
dimensionless probability function, the Isgur-Wise function 
$\xi(v \cdot v^{'})$ \cite{ISG}, where $\it{v}$ and 
$\it{v^{'}}$ are respectively the initial and 
final velocities in a scattering or decay process. 
An example of the role this function plays is the 
elastic scattering of a $B$ meson. 
The hadronic matrix element for this process is: 
\vskip 20pt
\centerline{$\frac{1}{m_B} 
\langle \bar{B}
(v^{'})|{\bar{b}}_{v^{'}}\gamma^{\mu}b_v|\bar{B}(v)\rangle
= \xi(v \cdot v^{'})(v + v^{'})^{\mu}$}
\vskip 20pt

where $b_v$ and $b_{v^{'}}$ are the heavy quark fields. 
The heavy quark symmetry used in $B$ physics phenomenology 
represents substantial progress in the theoretical 
description of $B$ decays. In the next section we discuss 
semileptonic $B$ decays, for which HQET has been successfully 
used to derive decay rates. 

\subsection{Semileptonic Decays to Mesons}

	The combination of large branching fractions and 
high reconstruction efficiencies have allowed experiments 
such as CLEO to measure several semileptonic $B$ decays 
with high accuracy \cite{karl}. This wealth of experimental 
results has allowed phenomenologists to compare theory 
and experiment. The decay kinematics of a specific $B$ 
semileptonic decay dictate the type of form factor contributions to 
the decay rate. In the case of $\bar{B} \to D^* l \bar{\nu}$, 
for example, there are no $\frac{1}{m_Q}$ corrections 
to the decay rate. This behavior, which is explained by 
Luke's theorem \cite{luke}, implies that the HQET-derived 
decay rate for $\bar{B} \to D^* l \bar{\nu}$ has low 
sensitivity to higher order perturbative corrections 
as well as non-preturbative effects.

\subsection{Hadronic Decays to Mesons}

	Whereas HQET has been useful in describing 
semileptonic $B$ decays, an equally accurate description 
of hadronic $B$ decays using HQET has only recently 
begun to be pursued for two-body decays to mesons 
\cite{bsw,mathias2,buras}. Whereas in 
semileptonic $B$ decays one of the two currents 
in a matrix element is weak, and therefore 
calculable to all orders in perturbation theory, 
in the hadronic case we have matrix elements 
of four-quark operators with hadronic uncertainties 
due to the exchange of gluons and quarks. 

	In energetic hadronic two-body $B$ decays 
hadronization is assumed to take place after the 
quarks that form each of the two hadrons have traveled 
sufficiently long distances for there to have been 
no significant exchange of gluons or quarks between them. 
This decay characterisitic is referred to as factorization, 
in which the matrix elements of four-quark operators 
factorized to independent current elements for 
each hadron. By using the operator product 
expansion (OPE) \cite{wilson,buras2}, the weak 
interaction effects are treated as separate from 
the long range strong interaction effects. The 
HQET effective weak hamiltonian for $b \to c,u$ 
transitions after this procedure is given by: 

\vskip 20pt
\centerline{${\cal H}_{eff} 
= \frac{G_F}{2} ( V_{cb} [c_1(\mu){Q_1}^{cb} + 
c_2(\mu){Q_2}^{cb}]$ + h.c.) 
+ $Q_{b \to u}$ + $Q_{penguin}$}
\vskip 20pt
where $c_1,2(\mu)$ are scale dependent Wilson 
coefficients. ${Q_{1,2}}^{cb}$, $Q_{b \to u}$, 
and $Q_{penguin}$ are, respectively, four-quark 
operators for $b \to c$, $b \to u$, and penguin 
decays \cite{mathias2}. The relative strength of 
each type of operator as well as the Wilson 
coefficients are decay-dependent. 

	Applying the factorization hypothesis 
to ${\cal H}_{eff}$ to, for example, the 
decay amplitude of the decay $\bar{B}^0 \to D^+\pi^-$, 
results in: 
  
\vskip 20pt
\centerline{$A_{factorized} 
= \frac{G_F}{2} V_{cb} {V_{ud}}^* a_1 
\langle \pi^-|{(\bar{d}u)}_{axial}|0 \rangle
\langle D^+|{(\bar{c}b)}_{vector}|\bar{B}^0 \rangle$}
\vskip 20pt
where $a_1$ can be verified with experiment. As the 
number of hadronic two-body $B$ decays and the accuracy 
with which their decay rates are measured increases 
in the near future, it will be possible to test 
the decay rates derived using HQET similarly to 
how it has been done for semipletonic $B$ decays. 

\section{{\bbaryons}}
 
A distinctive feature of the $B$ meson system is 
that the large mass of the b-quark allows for many 
of the weak decays of the $B$ meson to include 
the creation of a baryon-antibaryon pair. The 
as yet unmeasured decay 
{\dsspn} bars this feature from being unique. 
In Table \ref{bbdec} we outline the {\bbaryons} 
decays allowed in the Standard Model. 

\begin{table}[htb]
\caption{Representative {\bbaryons} decay mechanisms 
\label{bbdec} }
\begin{center}
\begin{tabular}{|c|c|}
\hline
Quark-level mechanism & Sample decay mode \\ \hline 
Semileptonic decays: & \\ 
$b \rightarrow cW^{-}, W^{-} \rightarrow l^- \nu_{l}$ 
& $B \rightarrow \Lambda_c \bar{p} l^- \nu_{l}$ \\ 
$b \rightarrow uW^{-}, W^{-} \rightarrow l^- \nu_{l}$ 
& $B \rightarrow p \bar{p} l^- \nu_{l}$ \\ 
Hadronic decays: & \\
$b \rightarrow cW^{-}, W^{-} \rightarrow \bar{c}s$ 
& $B \rightarrow J/\psi K$,  $J/\psi \to Baryons$ \\ 
$b \rightarrow cW^{-}, W^{-} \rightarrow \bar{c}s$ 
& $B \rightarrow \Xi_{c} \bar{\Lambda_{c}}$\\ 
$b \rightarrow cW^{-}, W^{-} \rightarrow \bar{u}d$ 
& $B \rightarrow \Lambda_c \bar{p} \pi$ \\ 
$b \rightarrow cW^{-}, W^{-} \rightarrow \bar{u}d$ 
& {\bdeca} \\ 
$b \rightarrow sg$, 
& $B \rightarrow \Lambda \bar{p}$,$B \rightarrow p\bar{p}K$ \\ 
$b \rightarrow uW^{-}, W^{-} \rightarrow \bar{u}d$ 
& $B \rightarrow p \bar{n} \pi$ \\ \hline 
\end{tabular}
\end{center}
\end{table}	

\subsection{Results to Date}
 
The $B \rightarrow pX$ was measured by 
CLEO to be $8.0\pm0.5\pm0.3 \%$\cite{cleo92}, 
assuming $B \rightarrow pX = B \rightarrow nX$. 
Based on this number we expect roughly 8$\%$ of all 
$B$ mesons in our data to be a $B\to Baryons$ 
event. The $\Lambda_c$, $\Sigma_c$ and $\Xi_c$ 
charmed baryons have been inclusively measured 
in $B$ decays \cite{sigc,casc}. Upper limit 
exclusive measurements have been reported to date 
for $B \rightarrow \Lambda_c \bar{p} l^- \nu_{l}$ 
with $l=e,\mu$ \cite{bbsemil}, and selected 
two-body rare decays \cite{brare}.

\begin{figure}[ht]
   \centering \leavevmode
        \epsfysize=4cm
   \epsfbox[70 525 575 699]{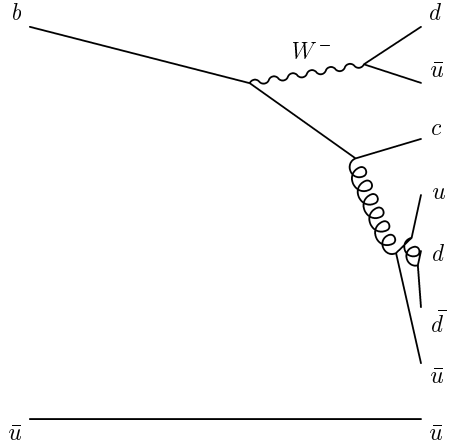}
\vskip 170pt
   \caption{A Feynman diagram for 
$B^- \rightarrow {\Lambda_{c}}^+ \bar{p} \pi^-$. } 
\vskip 20pt
\label{blamc}
\end{figure}

Based on the inclusive measurements reported to date, 
$B \rightarrow Baryons$ decays to date have 
been expected to be produced predominantly 
in decays of the type $B\to\Lambda_c\bar{p}X$, 
and this is the type of decay that has been 
exclusively reconstructed to date \cite{jjo}. 
A typical Feynman diagram for a 
$B\to\bar{\Lambda_c}pX$ decay is shown 
in Figure \ref{blamc}. 

\subsection{The Argument for {\bDnnx} modes}

Combining the recently measured value 
${\cal B}(\Lambda_c^+\to pK^-\pi^+) =
5.0\pm0.5\pm1.2 \%$ \cite{russ-paper} with estimates
of the product branching fraction
${\cal B}(B\to\Lambda_c X)\times
{\cal B}(\Lambda_c\to pK^-\pi^+) = 0.18 \%$ \cite{zoeller}
we can determine that $B\to{\overline\Lambda_c}pX$ modes
account for only $3.6\%$ of {\bbaryons}, which is 
approximately half of the total $B\to Baryons$
rate, as measured in the inclusive 
$B \to pX$ measurement \cite{cleo92}.

Based on our current knowledge of $B\to Baryons$, 
there must be processes other than $\Lambda_c$ 
production that contribute to the $B\to Baryons$ 
rate. It is our goal to find evidence for 
$B\to Baryons$ decays which do not involve 
a $\Lambda_c$. 
Dunietz \cite{dunietz} suggested that modes 
of the type {$B \rightarrow DN{\bar{N}} X $},
in which $D$ represents any 
charmed meson, and $N$ a proton or neutron, are 
likely to be sizeable. $B \rightarrow DN{\bar{N}} X $ 
final states can arise either from the hadronization 
of the W boson into a baryon-antibaryon pair, or 
from the production of a highly excited 
charmed baryon that decays strongly into 
a baryon plus a charmed meson. 
CLEO previously reported an inclusive 
upper limit for ${\cal B}$($B \rightarrow DN{\bar{N}} X $) at 
the 90$\%$ confidence level of $<$ 4.8 $\%$\cite{cleo92}. 

\subsection{Thesis Overview}

\begin{figure}[ht] 
   \centering \leavevmode
        \epsfysize=4cm
   \epsfbox[70 525 575 699]{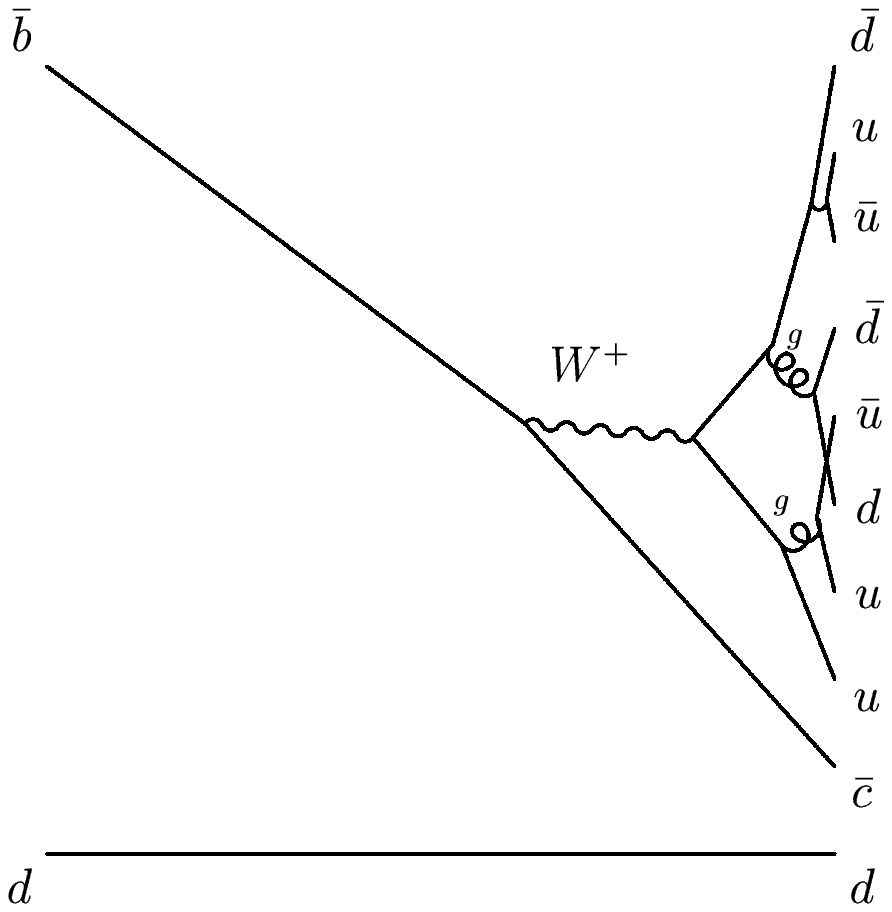}
\vskip 100pt
        \epsfysize=4cm
   \epsfbox[70 525 575 699]{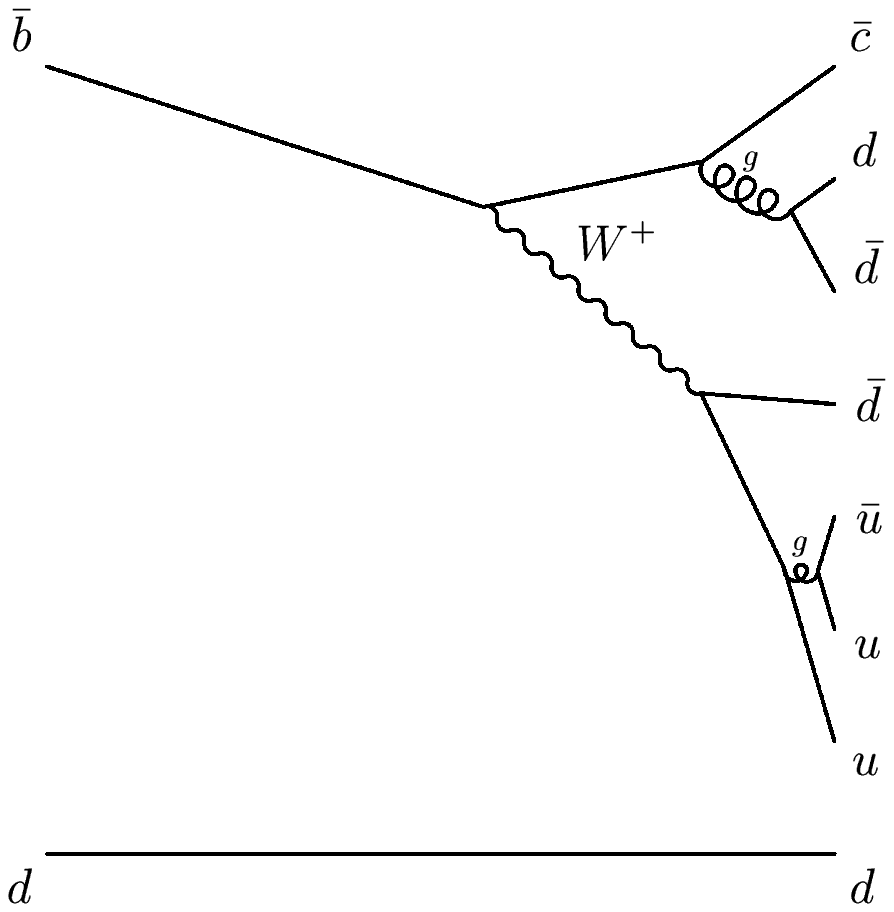}
\vskip 190pt
   \caption{Two Feynman diagrams for {\bdeca} } 
\vskip 20pt
\label{feynbdeca}
\end{figure}

\begin{figure}[ht]
   \centering \leavevmode
        \epsfysize=4cm
   \epsfbox[70 525 575 699]{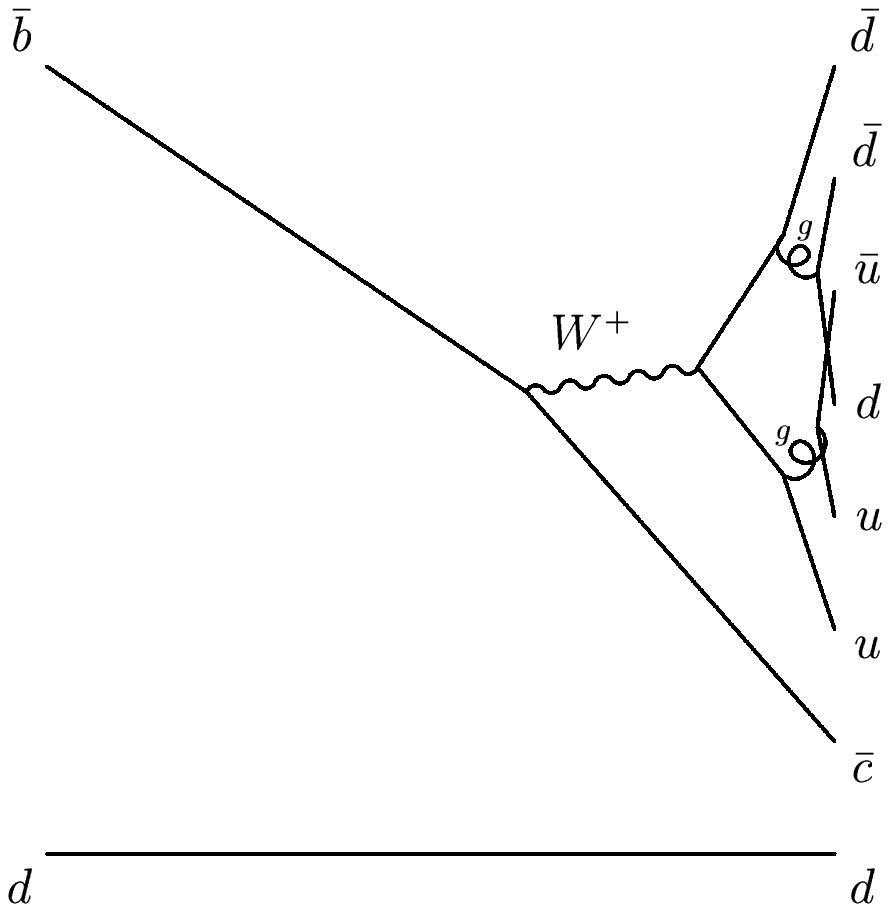}
\vskip 190pt
   \caption{A Feynman diagram for {\bdecb} } 
\vskip 20pt
\label{feynbdecb}
\end{figure}

In this thesis we will attempt the exclusive reconstruction 
of two specific {\bDnnx} decay modes, 
$B^0$$\rightarrow$ $D^{*-}$ $p$ ${\bar{p}}$ ${\pi}^+$ 
and $B^0$$\rightarrow$ $D^{*-}$ $p$ ${\bar{n}}$. 
Typical Feynman diagrams for {\bdeca} and {\bdecb} 
are shown in Figures \ref{feynbdeca} and 
\ref{feynbdecb}, respectively. 

The choice of these two modes is guided by the following 
criteria: 

\renewcommand{\theenumi}{\arabic{enumi}}
\begin{enumerate}
\item $D^*$ mesons have the lowest signal contamination 
among the $D$ mesons.
\item Both decays can occur via external $W$ decay. 
Although this characteristic is not a principle, to date 
only the $b \to c$ decay modes that have been measured 
share this characteristic. 
The reasons for the predominance of these decays 
are not known.
\item These are the two modes with the lowest decay 
daughter multiplicity, which translates into the 
highest reconstruction efficiency.
\item {\bdeca} and {\bdecb} have low combinatoric 
backgrounds. 
\end{enumerate}

We report here, for the first 
time, evidence for decays of the type {\bDnnx}, 
and present measurements of the branching fractions 
${\cal B}$({$B^0$$\rightarrow$ $D^{*-}$ $p$ ${\bar{p}}$ ${\pi}^+$}) 
and ${\cal B}$({$B^0$$\rightarrow$ $D^{*-}$ $p$ ${\bar{n}}$}). 
The charge conjugate process is implied in 
the reconstruction of 
$B^0$$\rightarrow$ $D^{*-}$ $p$ ${\bar{p}}$ ${\pi}^+$. 
However, in the reconstruction of 
$B^0$$\rightarrow$ $D^{*-}$ $p$ ${\bar{n}}$, 
only the mode with the antineutron is used in our 
measurement because neutrons do not have the 
distinctive annihilation signature. These measurements 
invalidate the previous assumption that {\bbaryons} 
is dominated by $\Lambda_c$ decays, while establishing 
evidence for the existence of a new type of 
$B$ decay mechanism with a sizeable decay rate. 

The thesis is divided as follows: 

In Chapter 2 we describe the CLEO detector. 
We place special emphasis on the electormagnetic 
calorimeter, which we use to select antineutron 
candidates. 

In Chapter 3 we outline the selection criteria 
for the {\bdecb} and {\bdeca} decay daughters. 
We place special emphasis on our selection of 
antineutron showers due to the novelty of their use.

In Chapter 4 we describe our measurement of {\bdecb}, 
in which we use a recontruction technique used 
to reconstruct decays in which the energy of all 
the decay daughters is well determined.   

In Chapter 5 we describe our measurement of {\bdeca}, 
in which we use a recontruction technique which 
is similar to that used to reconstruct other $B$ decays 
with missing energy. In our case the missing 
energy is due to the antineutron candidate. 

In Chapter 6 we conclude by summarizing 
our results, stressing on their significance, 
and outlining possible decay modes that we believe 
are important and measurable with the expectedly 
larger datasets available to future studies.

      \chapter{CLEO II Detector}

\begin{figure}[htbp]
  \begin{center}
   \centering \leavevmode
        \epsfxsize=6in
   \epsfbox{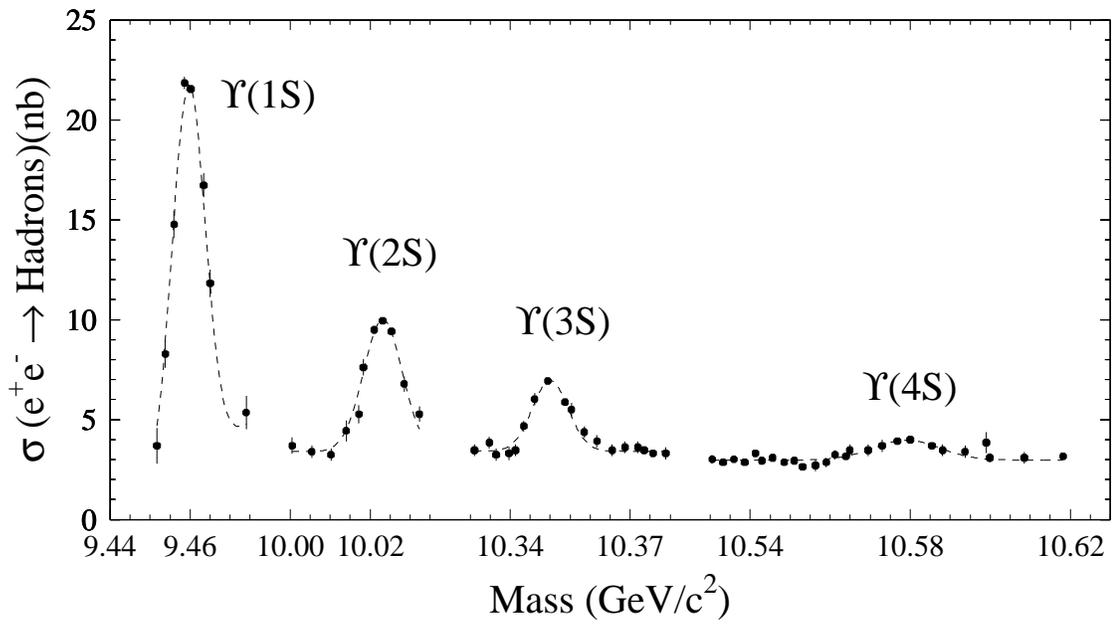}
  \end{center}
   \caption{Cross section into hadrons 
from the collision of $e^+e^-$ beams at CESR 
as measured by the CLEO II detector in 
the energy range 9.44 GeV to 10.62 GeV
\label{resonance} }
\end{figure}

	The CLEO II detector \cite{nim} is 
centered at the interaction point resulting from the 
collision of an electron beam and a positron beam, each 
at $\approx$ 5.3 GeV beam energy, located 
at the Cornell Electron Storage Ring (CESR). The beams 
collide almost head-on, resulting in a center-of-mass total 
energy of $\approx$ 10.6 GeV. It is in the energy range 
near this value that the $\Upsilon$ resonances, 
with the quark content $b\bar{b}$, are produced. 
As shown in Figure \ref{resonance}, the 
$\Upsilon(1S)$ to $\Upsilon(4S)$ resonances are 
found in the energy range 9.44 GeV to 10.62 GeV. 
The $\Upsilon(4S)$ is above the threshold for 
strong decay to $B\bar{B}$ pairs to take place. 
Close to 100$\%$ of the $\Upsilon(4S)$ decay 
rate is to nearly equal numbers of charged and 
neutral $B\bar{B}$ pairs.

	Data is taken at the $\Upsilon(4S)$ resonance (to be 
referred to as ON resonance) to study B decays. 
Because approximately 2/3 of the ON resonance data 
are composed of events in which the initial 
$q\bar{q}$ pair is not an $\Upsilon(4S)$, a sample 
of data is also taken 60 MeV below the resonance 
(to be referred to as OFF resonance) in order to subtract 
the non-$\Upsilon(4S)$, or continuum 
component of the ON resonance data. 
The $e^+e^-$ annihilation at or near the $\Upsilon(4S)$ 
resonance yields a wide variety 
of possible final states, some which are 
shown in Table \ref{annidec}.

\begin{table}[htb]
\caption{Some $e^+e^-$ annihilation final states 
\label{annidec} }
\begin{center}
\begin{tabular}{|c|}
\hline 
$e^+e^- \to \gamma\gamma$ \\ 
$e^+e^- \to l^+l^-$ (with l=$e,\mu,\tau$)  \\ 
$e^+e^- \to \Upsilon(4S) \to B\bar{B}$ (requires an 
energy threshold) \\
$e^+e^- \to q\bar{q}$ (followed by hadronization)
\\ \hline 
\end{tabular}
\end{center}
\end{table}

	The decay product of an $e^+e^-$ annihilation 
is called an event. Because final states differ greatly 
in cross section (the frequency with which events of a 
given topology are produced) 
some event types are produced with 
high frequency, while other event types are 
produced with low frequency. Electronic triggering on 
an event-by-event basis is used to select some or all 
of a given type of event. Triggering allows us to 
record only the events that we are interested 
in studying, most of which are of types produced 
with low cross sections. 

\section{Sub-detector Components}

\begin{figure}[htbp]
  \begin{center}
   \centering \leavevmode
        \epsfxsize=6in
   \epsfbox[20 143 575 699]{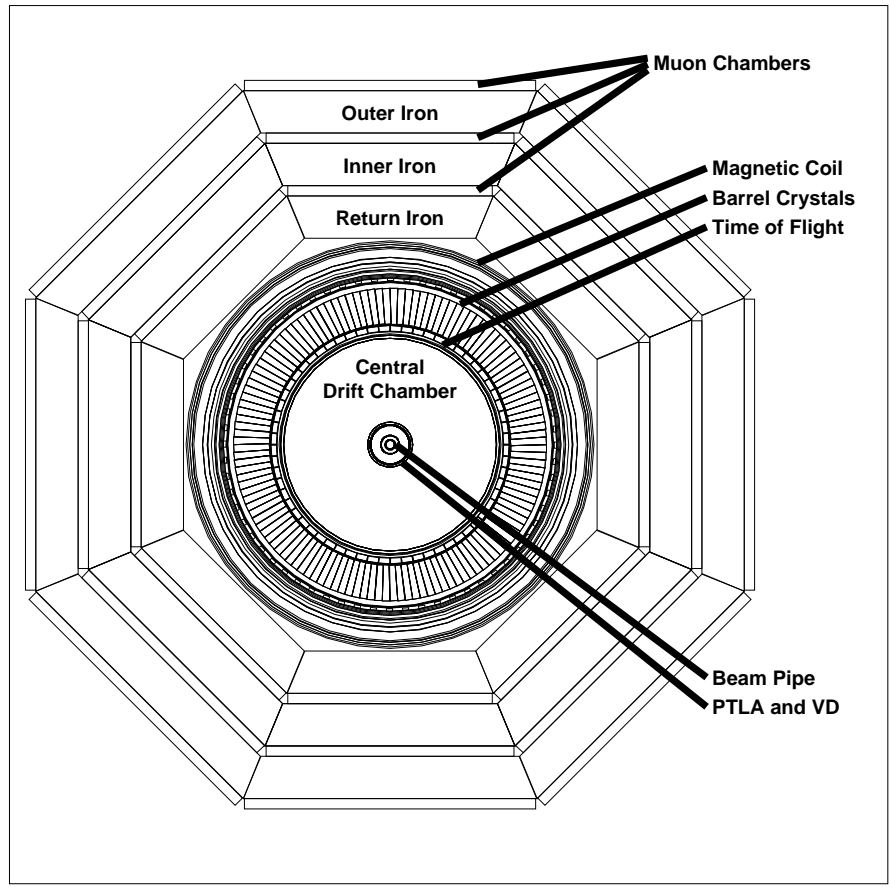}
  \end{center}
   \caption{Front view of the CLEO II detector
\label{detector_front} }
\end{figure}

\begin{figure}[htbp]
  \begin{center}
   \centering \leavevmode
        \epsfxsize=6in
   \epsfbox[20 143 575 699]{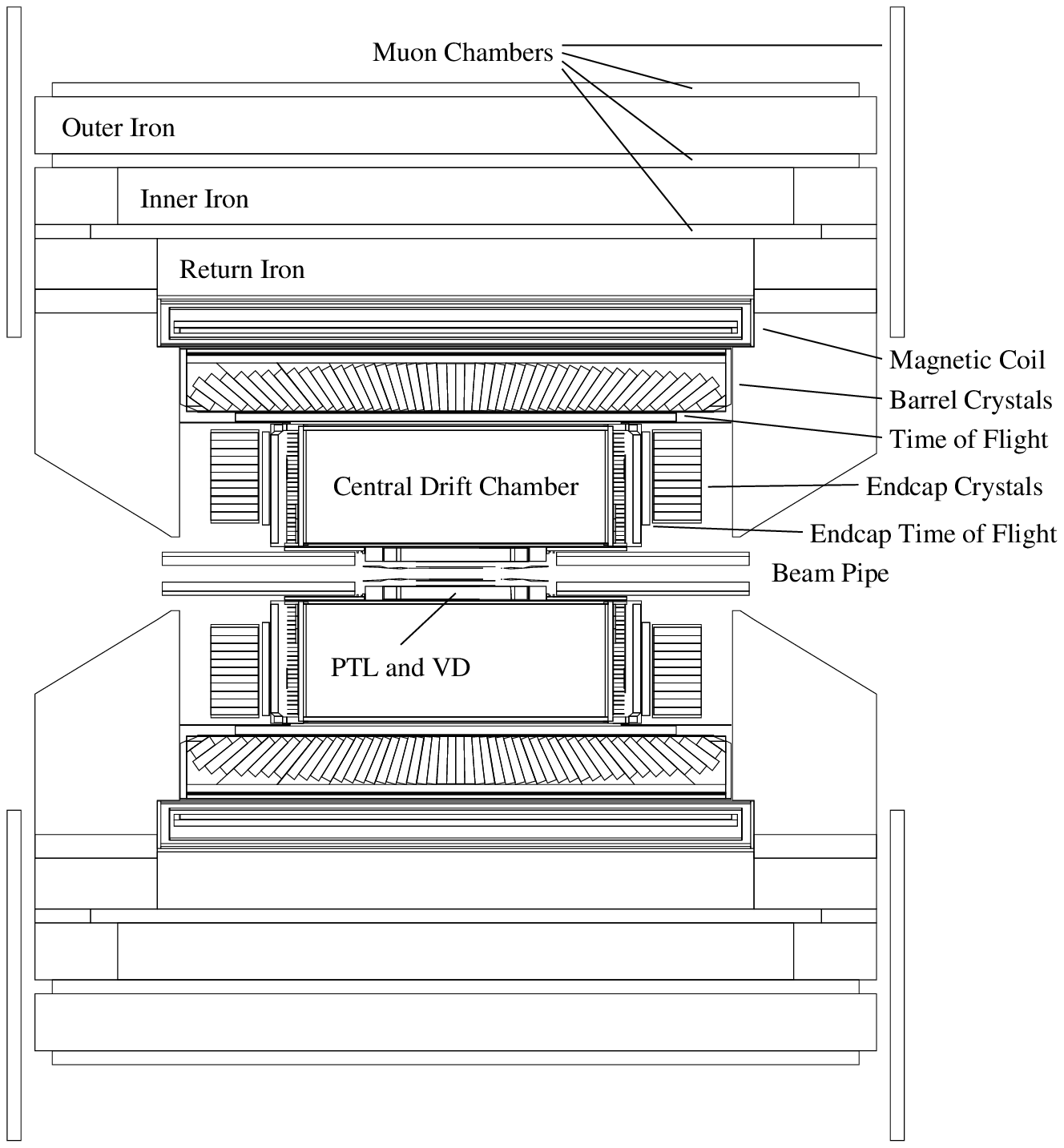}
  \end{center}
   \caption{Side view of the CLEO II detector
\label{detector_side} }
\end{figure}

In order to get enough information out of an 
event, we use an ensemble of sub-detectors, each yielding 
incomplete information about the event. 
When the information from all sub-detectors is combined, 
we have sufficient information to measure useful physics 
properties. A front view and a side view of the 
CLEO II detector are shown in Figure \ref{detector_front} and 
Figure \ref{detector_side}.

	From the innermost to the outermost (with 
respect to the beam pipe, which is located at the center of 
the detector), the sub-detectors are:

\renewcommand{\theenumi}{\arabic{enumi}}
\begin{enumerate}
\item Vertex detector
\renewcommand{\theenumi}{\arabic{enumi}}
\begin{enumerate}
\item PTL (precision tracking layers) detector, 
used during the earlier part of data recording. 
These data will be referred to as CLEO II data.
\item 3-layer SVX (silicon vertex detector), 
used during the later part of data recording. 
These data will be referred to as CLEO II.5 data.
\end{enumerate}
\item Drift chamber.
\item TOF (Time-of-Flight) detector.
\item Electromagnetic calorimeter.
\item Muon detector.
\end{enumerate}
	
The volume including all but the muon detector 
is enclosed in a 1.5 Tesla superconducting magnet. An 
important feature of this magnet is 
the uniformity of its magnetic field, which ensures 
that charged particles bend uniformly regardless of 
where in the detector the particle travels. 
A clear introduction to detectors as well 
as experimental methods in high energy physics 
is found in Perkins \cite{perkins}.

\section{Tracking System}	

	A charged particle traversing a magnetic field 
in the presence of charged wires in a chamber containing 
gas will ionize this gas as it loses energy. 
We measure the time at which this process takes place 
as well as the energy collected by each wire. 
These measurements allow us to know the position of the 
particle in time and the energy released at a number of 
points, or hits, along its trajectory. 
For a given momentum, the rate at which a particle loses 
energy along this trajectory, measured as dE/dx, is 
dependent on its mass, thus allowing us to separate 
protons, kaons, and pions. 
The TOF (Time-of-Flight) of a particle in a scintillating 
medium is also dependent on its mass and momentum. 
TOF measurements yield a second way to separate 
protons, kaons, and pions. 

\subsection{PTL Detector}	

	The PTL detector is an inner drift chamber composed 
of six layers of straw tubes. There are 64 axial wires 
for each layer, and there is a half cell stagger between 
sequential layers. The PTL detector 
does not measure the longitudinal, 
or z-axis, position of the particle. 
The PTL transverse position measurements are more 
precise than those from the drift chamber. 

\subsection{SVX Detector}	

	In later running of CLEO II, the PTL drift 
chamber was replaced by a 3-layer SVX detector capable 
of longitudinal as well as axial measurements 
\cite{silicon}, each measurement taking place on 
the two sides of each of the silicon wafers. 
The radii of the SVX layers are 2.35 cm, 
3.25 cm, and 4.75 cm, for layers 1, 2, and 3, 
respectively. It is composed of 96 wafers arranged into 
8 octants of 12 wafers each, with 26,208 data 
readout channels. The intrinsic resolution from 
$e^+e^- \rightarrow \mu^+\mu^-$ events at normal 
incidence is 29 $\mu$m.

	The improved measurement resolution of the 
SVX detector in comparison with the PTL detector 
allows for more accurate determination of the 
event vertex. This advantage is most useful 
to lifetime studies, yet it does not affect greatly 
the results presented here. 

\subsection{Drift Chamber}
	  
The drift chamber system (the main drift chamber 
and the vertex detector), together with the 
SVX or PTL, are used to measure the momentum 
of charged particles. 
Some vertex detector and drift chamber parameters 
are shown in Table \ref{drift}. 
The beam pipe is located at radius 3.5 cm in the 
CLEO II data and at radius 2.0 cm in the 
CLEO II.5 data. 

\begin{table}[htb]
\caption{Vertex detector and drift chamber parameters 
\label{drift} }
\begin{center}
\begin{tabular}{|c|c|c|c|}
\hline 
Detector & Layers & Radius (cm) & Wires per layer \\ \hline 
PTL (CLEO II only) & 6 & 4.7 to 7.2 & 64 \\ 
Vertex Detector (VD) & 10 & 8.4 to 16 
& 64 (first 5), 96 (second 5) \\ 
Outer Drift chamber & 51  & 17.5 to 95
& 96 to 384 \\ \hline
\end{tabular}
\end{center}
\end{table}
 
 	The $r-\phi$ and z measurement resolutions 
for each of the sections are shown on Table \ref{res}.

\begin{table}[htb]
\caption{Drift chamber resolution 
\label{res} }
\begin{center}
\begin{tabular}{|c|c|c|}
\hline
Detector & $r-\phi$ resolution & z  resolution  \\ \hline 
PTL (CLEO II only) & 90 $\mu$m & N/A \\ 
Vertex Detector (VD) & 150 $\mu$m & 0.75 mm \\ 
outer drift chamber & 110 $\mu$m & 3 cm \\ \hline
\end{tabular}
\end{center}
\end{table}
 	
	The Vertex Detector (VD) is bounded by concentric 
inner and outer cathode strips which provide z measurements. 
The segmentation of the VD cathode strips is 5.85(6.85) mm 
along z, which is the beam direction, on the inner(outer) 
cathode. Segmented cathodes also bound layers 1 and 51 of 
the outer drift chamber. Segmentation is about 1 cm along z. 

\subsection{Momentum and Angular Resolution}

	There are two factors that affect the 
track momentum resolution: 
	
	1. 	The error in the measurement of the 
track curvature due to the hit-level measurement error 
in drift distance. This resolution component is 
parametrized by a term linear in $p_{t}$, the 
transverse momentum. 

	2. 	Multiple scattering at material 
boundaries which cause the track trajectory to deviate 
from a helix. This resolution component is 
parametrized by a constant term.

	The parametrization for CLEO II data is, in GeV:

{\large{ {\centerline{ $(\frac{\delta{p_{t}}}{p_{t}})^2 = 
(0.0011 p_{t})^2 + (0.0067)^2$
}} }}

	In Table \ref{restrack} we show this resolution 
in MeV for selected values of $p_{t}$.

\begin{table}[htb]
\caption{Momentum resolution ($\delta{p_{t}}$) 
for the CLEO II data at selected values of $p_{t}$ 
\label{restrack} }
\begin{center}
\begin{tabular}{|c|c|}
\hline 
$p_{t}$ & $\delta{p_{t}}$  \\ \hline 
0.5 GeV & 3.3 MeV \\ 
1.0 GeV & 6.8 MeV \\ 
1.5 GeV & 10.4 MeV \\
2.0 GeV & 14.1 MeV \\ \hline
\end{tabular}
\end{center}
\end{table}

	The angular resolution is measured using 
$e^+e^- \rightarrow \mu^+\mu^-$ events, in which the 
typical $p_{t}$ = 5.0 GeV. $\delta{\phi}$ = 1 mrad 
and $\delta{\theta}$ = 4 mrad. We expect 
$\delta{\phi}$ and $\delta{\theta}$ to be higher 
for the tracks we use in our analysis since muons 
have a much lower probability of multiple scattering than 
do other charged particles. 

\begin{figure}[htbp]
  \begin{center}
   \centering \leavevmode
        \epsfxsize=5in
   \epsfbox[20 143 575 699]{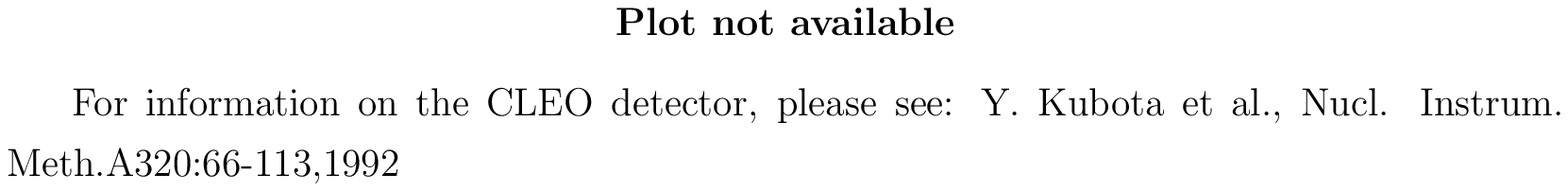}
  \end{center}
   \caption{dE/dx vs. track momentum
\label{dedx} }
\end{figure}

\subsection{dE/dx Measurements}

	 dE/dx is a function of particle mass and 
momentum, since $p = m\beta\gamma$, where 
$\beta = v/c$. The degree of separation we are 
able to achieve is shown in Figure \ref{dedx} 
for the CLEO II data. The 51 layers of the outer, 
or main, drift chamber$-$to be referred to as DR$-$are 
used to measure the specific ionization 
energy loss (dE/dx) of particles.

	A mixture of argon-ethane gas was used 
in the main drift chamber for the CLEO II data. 
This mixture was changed to helium-propane during 
data taking for CLEO II.5, which allowed for 
an improvement in the dE/dx resolution, resulting 
in better charged particle separation. 
In Figure \ref{dedx} each of the particle 
bands has been plotted after a subtraction of 
some of the higher dE/dx data points in a given 
particle band. 

\begin{figure}[htbp]
  \begin{center}
   \centering \leavevmode
        \epsfxsize=4.5in
   \epsfbox{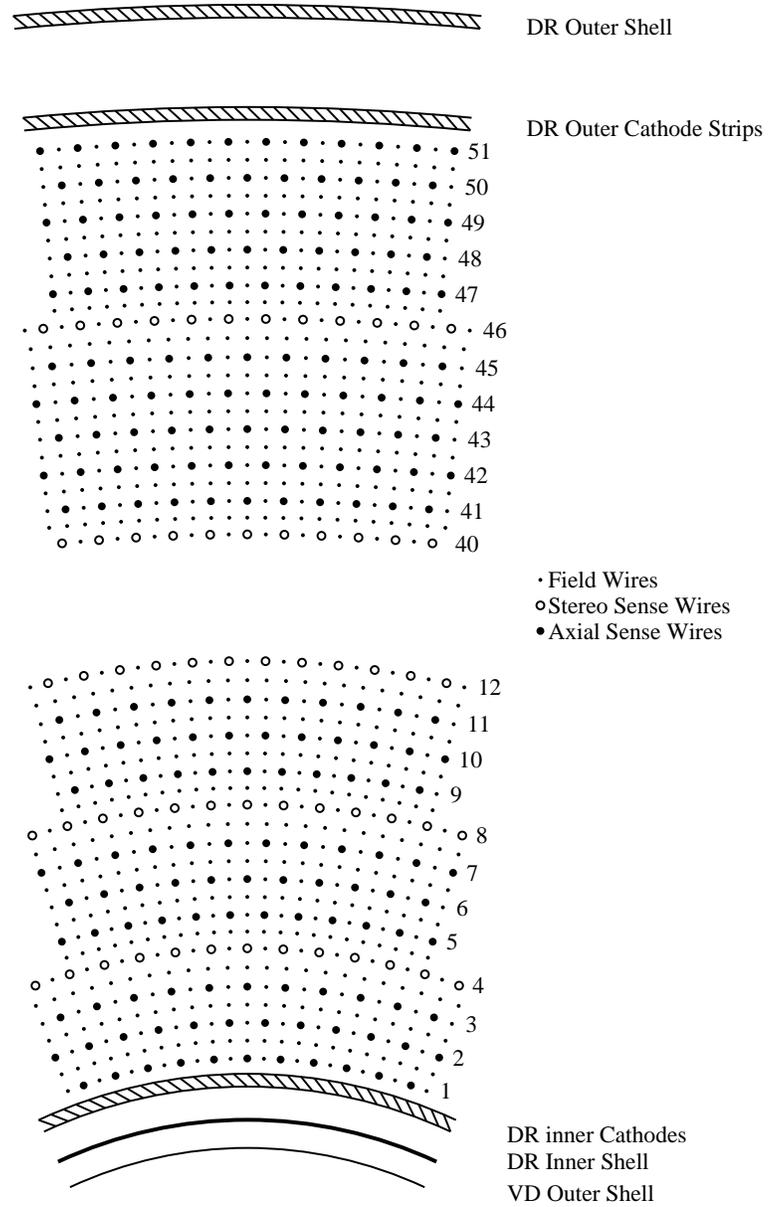}
  \end{center}
   \caption{Main drift chamber (DR) wire arrangement
\label{wires} }
\end{figure}

 	The main drift chamber has contiguous cells 
each with a sense wire surrounded by field wires, 
as shown in Figure \ref{wires}. Overall, there are 
three field wires for every sense wire in the 
main drift chamber. A number of corrections are 
applied to optimize the resolution: 

\renewcommand{\theenumi}{\arabic{enumi}}
\begin{enumerate}
\item Dip angle saturation: tracks perpendicular 
to the sense wires have the highest density 
of ionization along the z direction. The amount 
of collected charge is reduced by electric 
shielding for these tracks.
\item Drift distance: varies depending on 
the field configuration of each cell.
\item $(r,\phi)$ entrance angle: 
its magnitude as well as its sign is 
field dependent.
\item Axial-stereo layer: cells for axial 
and stereo layers have a different field 
dependence. 
\end{enumerate}

\subsection{Time-of-Flight Measurements}

\begin{figure}[htbp]
  \begin{center}
   \centering \leavevmode
        \epsfxsize=5in
   \epsfbox[20 143 575 699]{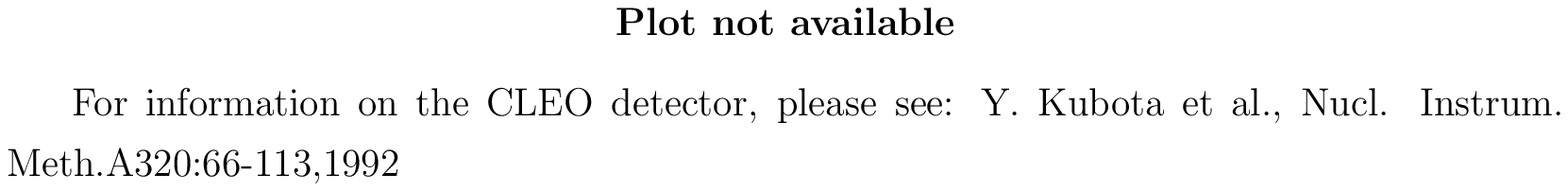}
  \end{center}
   \caption{Time-of-Flight vs. track momentum
\label{tof} }
\end{figure}

	Time-of-flight detectors surround the DR. 
A bar of organic scintillating material 
5 cm thick has photomultiplier tubes at each end 
for the barrel detectors, and at one end for those 
in the endcaps. A time measurement is made, 
with a 154 ps resolution for the barrel, and 
272 ps resolution for the endcaps. From the time and 
distance travelled, a $1/\beta$ quantity is defined. 
$1/\beta$ varies by particle type, as shown in 
Figure \ref{tof}. 

\section{Electromagnetic Calorimeter}

	The electromagnetic calorimeter is 
used to measure the electromagnetic energy deposition of 
charged and neutral particles. It is composed 
of 7800 CsI crystals. A clustering algorithm 
is used to combine the energy deposition in a 
crystal region, which is called a shower.

	The information from detector components 
is used in our analysis in a way that is consistent 
with previous CLEO II analyses with the exception 
of measurements from the electromagnetic calorimeter. 
The electromagnetic calorimeter has been used previously 
to measure electron and photon energy deposition. 
In our measurement of {\bdeca} we are required to 
select showers that are consistent with being 
due to antineutrons annihilating with the CsI. 
The antineutron selection procedure 
is successful for the first time at CLEO. 
	
\subsection{Dimensions}

\begin{figure}[htbp]
  \begin{center}
   \centering \leavevmode
        \epsfxsize=6in
   \epsfbox{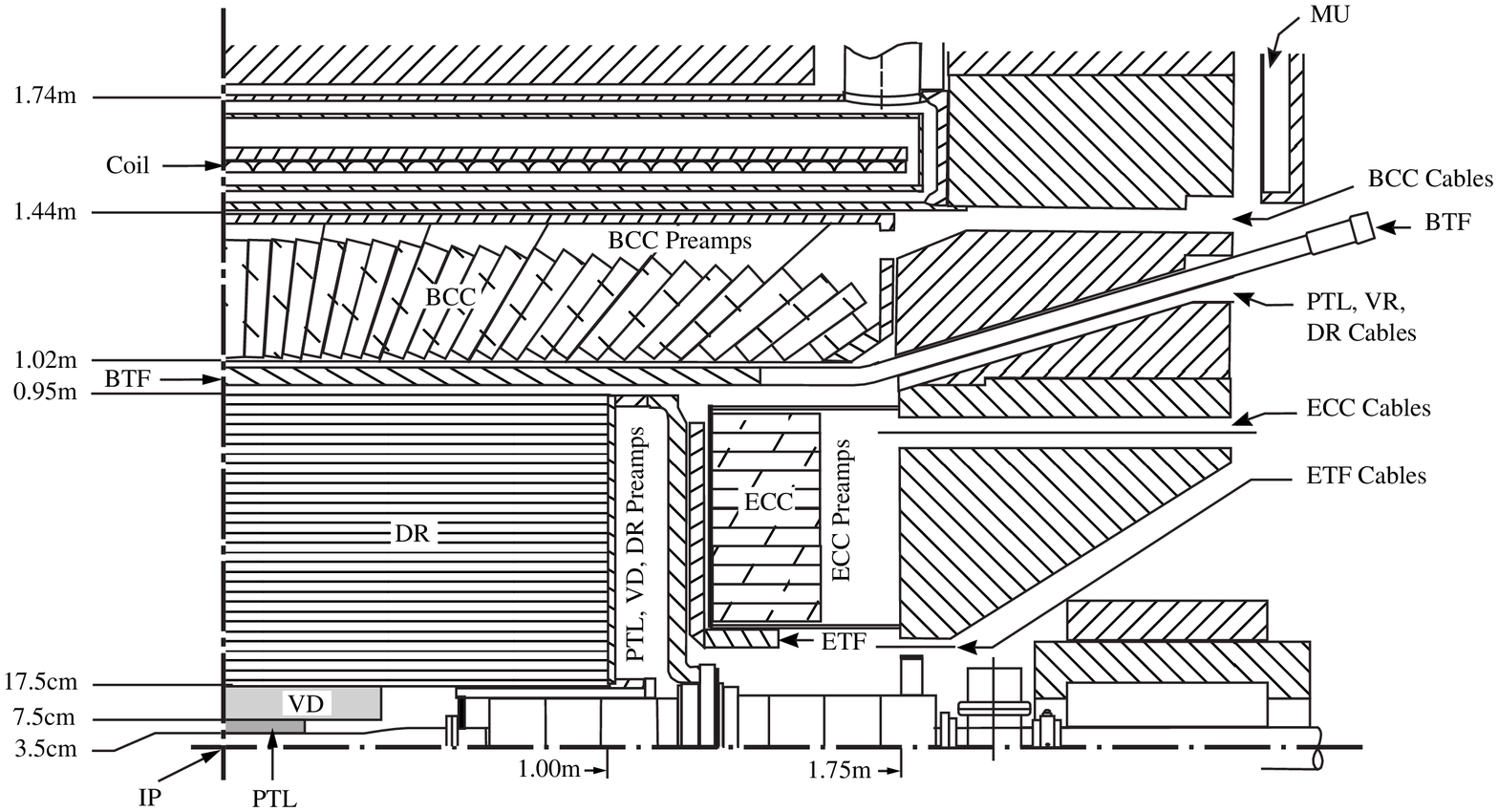}
  \end{center}
   \caption{Layout of CLEO II detector showing 
barrel and endcap calorimeter sections
\label{cal} }
\end{figure}

	The calorimeter is within the 1.5 Tesla 
magnetic field. All crystal faces are at 1 m 
from the interaction point, facing it in such a 
way that showers reach all crystals at normal 
incidence. A partial diagram showing some of the 
barrel and one of the endcaps is shown in 
Figure \ref{cal}. Each calorimeter crystal is 5-cm 
$\times$ 5-cm $\times$ 30-cm, where the 
latter is the length of the crystal. The choice of 
thalium doped CsI for the calorimeter crystals took 
into consideration factors such as cost, resistance 
to cracking, high density, and short radiation length 
(1.83 cm). Because the calorimeter is 16.4 radiation 
lengths deep, $\approx$ 1$\%$ of the energy of a 
5 GeV electron leaks out of it. There are 6,144 barrel 
crystals and 828 crystals for each endcap. 

	The crystals are arranged in a rectilinear grid 
with care taken to have a minimum of material 
between crystals. The material in front of the 
barrel section is described by radiation lengths in 
Table \ref{rad}. 

\begin{table}[htb]
\caption{ Radiation lengths of material before 
barrel section of the calorimeter
\label{rad} }
\begin{center}
\begin{tabular}{|c|c|}
\hline 
Material  &  Radiation length ($\%$)  \\ \hline 
beam line to layer 1 of DR & 2.5 \\ 
argon-ethane gas and rest of DR & 0.4 \\
DR outer cathode layer  & 1.0 \\ 
outer DR wall & 2.0 \\ 
TOF counters & 12.0 \\ \hline
\end{tabular}
\end{center}
\end{table}

	There is more material in front of 
the endcap calorimeter sections, which degrades 
measurement quality. We use endcap information 
in the $\pi^0$ reconstruction, but not for 
the antineutron selection criteria. 
The central barrel, which covers 71$\%$ of the 
polar angle ($45^\circ$ $<$ $\theta$ $<$ $135^\circ$), 
has less material in front of it, thus providing 
the measurements with highest resolution. 

\subsection{Clustering}

	The algorithm involved in associating a group 
of nearby cells with energy above a threshold 
in the calorimeter is called clustering. 
An average of 430 crystals have energy recorded 
in a hadronic event (either $B\bar{B}$ or continuum). The 
raw ADC count measurements need to be calibrated using 
a crystal-to-crystal calibration. The sample used is 
Bhabba events ($e^+e^- \rightarrow e^+e^-$). This sample 
has high statistics and a beam energy constraint. The 
electrons and positrons in Bhabba events deposit almost 
100$\%$ of their energy in the calorimeter. Crystal 
noise, which is in the range of a few MeV, 
does not affect this sample appreciably. The set of constants 
obtained from the crystal-to-crystal calibration needs 
to be updated only every few months. The main change 
is the preponderance of a few crystals to become noisy. 	

	A clustering algorithm for CLEO should accomplish 
the following:

\renewcommand{\theenumi}{\arabic{enumi}}
\begin{enumerate}
\item Match tracks to showers.
\item Allow $\pi^0$ reconstruction.
\item Allow separation of photons and 
$K_{L}$'s from antineutrons.
\end{enumerate}

	There are two clustering packages in the CLEO II 
data: CCFC, and XBAL. 
Both accomplish these three requirements. 
Even though we use CCFC, using XBAL would have 
yielded results similar at the 
few $\%$ level. Our discussion of clustering is limited to 
CCFC. A discussion of clustering using XBAL would 
be similar to ours. 			
	 	 
	In order to separate shower clusters from noise, 
only cells with energy $>$ 10 MeV are considered as 
candidates for the center of a shower. Only cells that 
are at most two cells away from a cell in the cluster 
are added to the cluster, which allows cells without 
an energy measurement to be located inside a cluster. 
The number of cells N that is considered a cluster is 
a function of the total energy of the cluster. 
For example, 4 cells correspond to a 25 MeV shower, 
17 cells to a 4 GeV shower. 	

	The position vector, which contains the 
directional cosines, is determined using a weighted 
sum of the energy measured with each cell, using 
the geometric center of each to define a centroid. 
Two corrections, one lateral, and another longitudinal, 
are applied to this position vector. These corrections 
account for calorimeter segmentation as well 
as depth into the CsI crystal of the shower center. 

	The cluster energy $E$ and angular resolutions, using 
Monte Carlo, are: 
\vskip 10pt
	Barrel: {\large{ 
	$\frac{\sigma_{E}}{E}[\%] = 
	\frac{0.35}{E^{0.75}} + 1.9 - 1.0 E$
	}} 

	Endcap: {\large{ 
	$\frac{\sigma_{E}}{E}[\%] = 
	\frac{0.26}{E^{0.75}} + 2.5$
	}} 

	Barrel: {\large{ 
	$\sigma_{\phi}[mrad] =  
	\frac{2.8}{\sqrt{E}} + 1.9$
	}} 

	Endcap: {\large{ 
	$\sigma_{\phi}[mrad] =  
	\frac{3.7}{\sqrt{E}} + 7.3$
	}} 

	Barrel: {\large{ 
	$\sigma_{\theta}[mrad] =  
	0.8 \sigma_{\phi} sin(\theta)$
	}} 

	Endcap: {\large{ 
	$\sigma_{\theta}[mrad] =  
	\frac{1.4}{\sqrt{E}} + 5.6$
	}} 
\vskip 20pt

	Using the example of a photon in the barrel, 
we list in Table \ref{photres} energy and azimuthal 
angular resolutions at two cluster energies. Their 
low values are a testament of our ability to properly 
reconstruct electromagnetic showers. A by-product 
of this ability is our success in separating 
baryon-antibaryon annihilation showers from 
electromagnetic showers. 

\begin{table}[htb]
\caption{Energy and angle resolutions for a photon 
in the barrel at two values of cluster energy 
\label{photres} }
\begin{center}
\begin{tabular}{|c|c|c|}
\hline
Cluster energy  & $\frac{\sigma_{E}}{E}[\%]$ 
&  $\sigma_{\phi}[mrad]$ \\ \hline 
100 MeV & 3.8 & 11 \\
5 GeV & 1.5 & 3 \\ \hline 
\end{tabular}
\end{center}
\end{table}

\section{Muon Detector}

	By the time a particle has traversed all the 
sub-detector components enclosed within the muon 
detector, all particles except muons have deposited 
most of their energy in the calorimeter and/or have 
been deflected by the magnetic field within the 
drift chamber. Since the muon detector is, above 
everything else, a large volume of iron, only 
muons are expected to pass through a significant 
section of the muon detector.

	 The nuclear absorption length $\lambda_{i}$ 
for a muon in iron is 16.8 cm. It is considerably 
less for all other charged tracks we study. The 
maximum depth of iron to be traversed by a particle 
is between 7.2$\lambda_{i}$ and 10$\lambda_{i}$. 
We do not use the muon detector in our analysis.

\chapter{Particle Selection}

In order to reconstruct {\bdecb} and 
{\bdeca} we need to have low background 
samples of all the decay daughters. 
Protons, antiprotons and pions are 
selected as single tracks, the 
{\mdif} distribution is used to 
select $D^*$'s, and shower selection 
criteria is used to select 
antineutrons. We select the $B$ decay 
daughters from ON resonance hadronic events.  
The ON resonance candidates can be fully 
or partially reconstructed $B$ mesons, or background 
candidates from ON resonance continuum events. 
The OFF resonance candidates can only be 
continuum events as the total event energy 
is below the energy 
threshold for $B\bar{B}$ production . 
We use the Mn$\_$fit histogram package \cite{mnfit}, 
and its function fitting utility, MINUIT \cite{minuit}, 
both of which are widely used by experimental 
high energy physicists.		

	To suppress the already small number of 
continuum background candidates in our reconstructions, 
we select events using the parameter 
R2GL \cite{wolfrum}. R2GL is a measure of how the 
momentum is distributed for the event. A high 
value of R2GL corresponds to continuum events, 
in which the initial quarks hadronize to form a two-body 
decay. The $B\bar{B}$ momentum distributions are the result 
of two separate two-body or higher decays that 
can be most approximately described by a 
spherically symmetric momentum distribution, 
which tends to yield a low value of R2GL. 

\section{Data Sample}

	The CLEO data is collected at the $\Upsilon(4S)$ 
resonance (ON resonance) and 60 MeV below the resonance 
(OFF resonance). Roughly 2/3 of the ON resonance data 
is continuum and the remaining 1/3 is $B\bar{B}$ pairs. 
We show in Table \ref{data} the integrated luminosity 
by dataset.
	
\begin{table}[htb]
\caption{The CLEO data \label{data} }
\begin{center}
\begin{tabular}{|c|c|c|c|}
\hline
Dataset& ON luminosity & OFF luminosity & {\bbar}'s\\
\hline
CLEO II & 3.14 ${fb}^{-1}$ & 1.61 ${fb}^{-1}$ 
& 3.30 $\times$ $10^{6}$\\ 
CLEO II.5 &  6.03 ${fb}^{-1}$  & 2.94 ${fb}^{-1}$ 
& 6.35 $\times$ $10^{6}$\\ 
\hline
\end{tabular}
\end{center}
\end{table}

\section{Monte Carlo Sample \label{genmc}}

	We use Monte Carlo generated events to model 
the data in order to study the efficiency of our 
selection criteria as well as the effect 
of backgrounds. Two types of Monte Carlo samples are used: 

\renewcommand{\theenumi}{\arabic{enumi}}
\begin{enumerate}
\item Generic Monte Carlo, in which many decays 
are generated using our current knowledge of B decays. 
\item Signal Monte Carlo, in which one 
of the $B$ mesons in the event decays via a 
predetermined mode, such as {\bdeca}. The 
remaining $\bar{B}$ meson in the event decays according 
to the prescription used to generate generic 
Monte Carlo events. 
\end{enumerate}

	The generic Monte Carlo sample consists of 
$B$ decays of type $b \to cW^-$. 
A detailed explanation of the criteria 
used for the generic Monte Carlo sample is found 
in the Appendix of \cite{cleo92}. This sample does not 
include any {\bDnnx} decays because until now 
we have had no evidence for their existence. 
The {\bbaryons} decays in the generic Monte Carlo 
sample are of the type $B \rightarrow \Pi W^+$, 
in which $\Pi \rightarrow {\Lambda_{c}}^- \bar{N}$, 
where $\bar{N}$ is an antiproton or an antineutron. The 
conjugate modes are also generated.  

\section{Track Selection}

	A combination of measurements is used to select 
quality tracks that are most likely to be the particle 
type we need for a given reconstruction-pion, kaon, or 
proton. The tracks we use in our reconstruction are 
particles that are stable (in the case of protons), 
or decay outside our detector (in the case of kaons and pions). 
Therefore, a track is assumed to 
begin near the interaction point, which is within 
a solid volume defined by the uncertainty with which 
we can measure the beam spot. This volume is a few microns 
wide in x and y and a few hundred microns wide in z.

\subsection{Fitting Algorithm} 

	Tracks are fitted using DR hits by a Kalman 
algorithm that is based on the assumption 
that the track is a helix inside a vacuum. 
The change of trajectory due to material interaction 
is taken into consideration in the track fit. 
The fit is performed from both ends of the track. The 
inward fit begins a hit-to-hit swim, adding 
hits to form a track, analogous to the way beads 
are strung to form a necklace. Where there is more 
than one hit to a layer of the drift chamber, the hit 
with the better resolution is chosen. The outward fit 
begins at the outermost layer of the drift chamber, 
performing a hit swim into the drift chamber.

\subsection{Drift Chamber Track Variables 
\label{driftqual}} 

	Variables we use which are defined with main 
drift chamber measurements (DR) are: 

\renewcommand{\theenumi}{\arabic{enumi}}
\begin{enumerate}
\item PQCD: signed 3-momentum of track. This vector 
does not have a correction assuming the mass of a 
particle type.
\item Z0CD: distance of closest approach along the 
z-axis. The z measurement resolution is considerably 
worse than the $r-\phi$ measurement resolution. However,
Z0CD is effective in separating tracks from the 
interaction point from material interaction tracks 
and long decay particle daughter tracks. 
\item DBCD: radial distance of point of closest 
approach to the interaction point. DBCD is usually 
more accurately determined than Z0CD.
\end{enumerate}
	
        All tracks are required to pass the following track 
quality cuts:

\renewcommand{\theenumi}{\arabic{enumi}}
\begin{enumerate}
\item PQCD $<$ 1 GeV: DBCD (in meters) $<$ 
0.005-(0.0038) $\times$ $|PQCD|$.
\item Low momentum tracks are more likely to have a poorly 
measured DBCD. PQCD $>$ 1 GeV: DBCD (in meters) $<$ 0.001. 
\item Z0CD $<$ 0.05 meter.
\end{enumerate}

\subsection{The TRKMNG Package} 

	A software package is used to reject duplicated 
tracks from curlers \cite{trkmng}. A curler is a low momentum track 
which is bent more than one revolution inside the drift 
chamber. A curler spirals inside the drift chamber as 
the momentum of the particle decreases.	
The tracking algorithm will often assign a track number 
in the list of tracks for each half-revolution. The 
TRKMNG package selects the half-revolution 
that is most likely the closest to where the track 
began to curl, rejecting all other half-revolutions. 
The variable used in 
the TRKMNG Package is tng(track) $\ge$ 0. All 
negative numbers of tng(track) are for the redundant 
half-revolutions. 

\section{Particle Separation}	

	By population size alone, all tracks are pions. 
The purpose of particle separation, 
or particle ID, is to separate tracks that are likely 
to be something other than a pion from the overwhelmingly 
more numerous  pions. 
Protons are produced with a significantly 
reduced rate, therefore constituting a small 
background to all other charged particles. 
The predominant backgrounds to protons are 
kaons, pions, electrons and muons. For the case 
of kaons, pions represent the largest background. 
The kaon in {\dzdecc} has more phase space 
available than the kaon in {\dzdecb} and {\dzdeca}, 
which allows its kinematic separation from backgrounds. 
Even for the low $x_{p}$ in {\dst}'s from B decays, 
the kaon in {\dzdecc} can be easily separated from the pion 
by the decay kinematics. 

	Charged particle identification is accomplished 
by combining the specific ionization (dE/dx) measurements 
from the drift chamber with time-of-flight 
(TOF) measurements. A normalized 
probability ratio $L_{i}$ is used for charged particle 
separation. 

	$L_{i}$ is definded as 
$L_{i} = P_{i}/(P_{pion}+P_{kaon}+P_{proton})$, and 
$P_{i}$ is the particle hypothesis $\chi_{i}^2$ probability 
combining dE/dx and TOF measurements, with $\chi_{i}^2$ 
defined as: 

\vskip 10pt
 
{\large{ {\centerline{ $\chi_{i}^2 = 
[\frac{ (dE/dx)_{measured} - (dE/dx)_{expected}}
{ \sigma_{dE/dx} }]^2 + 
[\frac{ (TOF)_{measured} - (TOF)_{expected}}
{ \sigma_{TOF} }]^2
$
}} }}
\vskip 10pt 

	$\sigma_{dE/dx,TOF}$ are the deviations from the 
mean and vary by particle type. We define the respective 
$\chi_{i}^2$ for each particle type using the 
$\sigma_{dE/dx,TOF}$ for the identical particle type. 
$L_{kaon}$, for example, is defined using the $\sigma_{dE/dx,TOF}$'s 
for kaons. In Table \ref{pid} we show how our criteria varies, 
depending on how much information 
is available. By defaulting to a pion we are setting 
$P_{i}$ = 1.0. Useful dE/dx information is available for the 
vast majority of tracks passing the track quality 
requirements outlined in section \ref{driftqual}. 
Low momentum particles ($<$ 200 MeV) curl 
before reaching the TOF detectors. 

\begin{table}[htb]
\caption{Particle identification criteria
\label{pid} }
\begin{center}
\begin{tabular}{|c|c||c|}
\hline 
$\sigma_{dE/dx}$ & $\sigma_{TOF}$ & Action \\ \hline 
$<$ 3 & $<$ 3 & use both \\
$<$ 3 & $>$ 3 & use $\sigma_{dE/dx}$ only \\ 
$>$ 3 & $<$ 3 & use $\sigma_{TOF}$ only \\
$>$ 3 & $>$ 3 & default to pion \\ \hline 
\end{tabular}
\end{center}
\end{table}	
 
	We place loose requirements of $L_{pion}$ 
$>$ 0.001 on all pion candidates, and the kaon candidates in 
{\dzdecc}. Kaon candidates in {\dzdecb}, and {\dzdeca} 
are required to satisfy  $L_{kaon}$ $>$ 0.4. Proton 
candidates in {\bdecb} and {\bdeca} are required to 
satisfy  $L_{proton}$ $>$ 0.9. 
In addition, electron backgrounds to protons 
and antiprotons are rejected using the variable cut 
R2ELEC $<$ 0. R2ELEC is a logarithmic probability that 
a particle is an electron. Negative values of R2ELEC 
are least likely to be for an electron. R2ELEC is 
defined using $\frac{E}{p}$ from the electromagnetic 
calorimeter, which is very close to 1 for electrons, 
which leave all of their energy in the 
calorimeter, as well as a non-negligible number of protons. 

\section{$\pi^0$ Reconstruction}

	The $\pi^0$'s used to reconstruct {\dzdecb} 
decay via $\pi^0 \rightarrow \gamma\gamma$. 
The $\pi^0$'s in this decay are soft, 
with momenta $<$ 500 MeV. The decay angle for 
$\pi^0 \rightarrow \gamma\gamma$ is large enough 
at this momentum range for the individual $\gamma$'s 
not to overlap. On average there are 7 $\pi^{\pm}$'s 
in a $B\bar{B}$ event. The average number of $\pi^0$'s, 
as of yet unmeasured, should be near 
this number. Since shower multiplicity increases 
with decreasing shower energy, as $\pi^0$ momentum 
decreases, the combinatoric shower background increases. 
A minimum momentum of 100 MeV is required of our 
$\pi^0$ candidates to suppress this background.

	A shower is considered a $\gamma$ if the ratio 
between the summed energy of a 3 by 3 grid of shower cells 
over the same are extended to a 5 by 5 grid of shower cells 
is almost 1. This ratio is referred to as E9OE25. 
Most photons have most of their energy concentrated within 
a 3 by 3 grid of shower cells. 

	We require at least one of the $\gamma$'s to 
be in the good barrel section of the calorimeter, which 
is defined as $|cos(\theta_{\gamma})| <$ 0.71. When 
one of the $\gamma$'s is not in the barrel, we 
require that it not be from the overlap region 
between the barrel and endcap sections of the calorimeter 
or near the beam pipe, both of which yield degraded 
shower energy measurements. We also require the candidate 
$\pi^0$ mass to be within 2.5 $\sigma$ 
($\pi^0$ $\sigma$ $\sim$ 5 MeV) from 135 MeV central value. 

\section{$D^*$ Reconstruction}

	Finding {\dst} candidates with high efficiency 
and low backgrounds is an important part of this 
analysis. In the decay {\dsdec} {\mdif} is only 
slightly larger than the mass of the {\pisoft}. 
Since the {\mdif} resolution is smaller than the $m_{D^*}$ 
resolution, we use the former to select {\dst} candidates. 
We verify our {\dst} yield with a previous analysis. 

In order to reconstruct $B \rightarrow D^* X$ 
with {\dsdec} correctly, we check our yield for the 
scaled momentum ${X}_{p}$ range 
0.25 $<$ ${X}_{p}$ $<$ 0.35. It is in this range 
that we find $\approx$ 50 $\%$ of $D^*$'s to be 
generated in signal {\bdecb} and {\bdeca} Monte Carlo. 
We check our yield with a previous CLEO 
result \cite{bdstarx}.

\subsection{The KNLIB Fitting Package}

The KNLIB fitting package is a collection of 
kinematic routines used for track vertexing \cite{avery}. 
We use it to vertex {\dzz}'s and {\dst}'s. 
{\dzz}'s are vertexed using the KNLIB routine KNVTX3. 
A new invariant mass is calculated with this vertex, 
and only the {\dzz} mass range 1.70 to 2.03 GeV 
is allowed to pass on to {\dst} reconstruction. 
To form a {\dst} vertex, we use KNLIB routine KNBVXK 
which combines the {\dzz} decay products and the 
{\pisoft} track to make a vertex-constrained mass from 
which {\mdif} is calculated. We use KNBVXK with the 
flag IVOPT set to 1 for the {\pisoft}-i.e. the
{\pisoft} is constrained to pass through the vertex 
formed by the other particles, but not used to determine 
that vertex. The latter method offers a noticeable 
improvement over other options tested, including use of 
{\pisoft} track when vertexing, as well as using the {\dzz} 
vertex instead of the individual {\dzz} tracks when 
vertexing the {\dst}. We place a loose cut of 
$\chi^2 < 100$ on the {\dst} vertex fit for all three modes. 

\subsection{Fit Optimization}

	We reconstruct {\dzz}'s, selecting candidates 
near the mean mass. After the addition of a {\pisoft}, 
we select candiates within a range of {\mdif} 
values. We use the 1998 Review of Particle Properties 
values for {\mdz}, 1.8646 GeV, 
and {\mdif}, 0.1454 GeV, as our central values \cite{pdg99}. 
The double Gaussian fits model {\mdif} better than the 
single Gaussian fits. 

        The optimization procedure is as follows:

\renewcommand{\theenumi}{\arabic{enumi}}
\begin{enumerate}
\item The background is fitted with a second order Chebyshev 
polynomial. 
\item The signal is fitted with a double Gaussian distribution.
\item The continuum subtracted data {\mdz} distribution 
is fitted using abs({\mdif}-0.1454) $<$ 0.002 GeV, 
which is a wide cut.
\item The widths, as well as the means of each of the 
double Gaussians are allowed to float. The best fits are 
obtained with non-zero ${\Delta}_{MEAN}$'s. 
${\Delta}_{MEAN}$ is the difference 
between the mean value of each of the two Gaussians. We 
integrate the double Gaussian, keeping 95$\%$ of the signal 
symmetrically from the PDG mean.  
\item The final set of double Gaussian {\mdz} and {\mdif} 
cuts from data are used in the B reconstructions, and the 
same cuts are applied to the Monte Carlo to find 
${\epsilon}_{MC}$, the signal Monte Carlo reconstruction 
efficiency.
\end{enumerate}

\begin{figure}[ht]
   \centering \leavevmode
        \epsfysize=10cm
   \epsfbox[20 143 575 699]{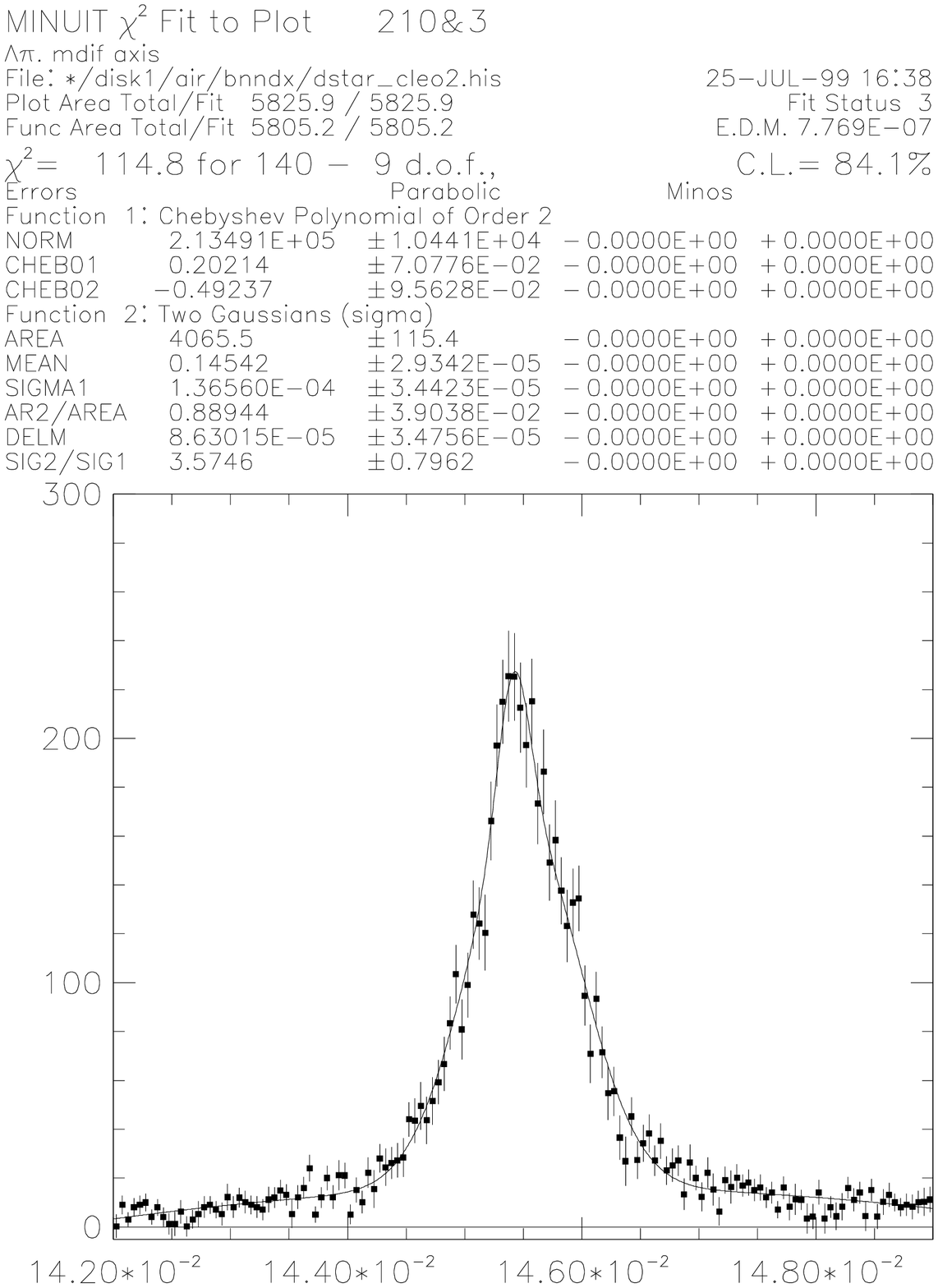}
\vskip 40pt 
   \caption{ \label{dzkp20} 
{\mdif} in GeV for $B \rightarrow D^* X$ with {\dzdecc} in CLEO II }
\vskip 60pt
\end{figure}

\begin{figure}[ht]
   \centering \leavevmode
        \epsfysize=10cm
   \epsfbox[20 143 575 699]{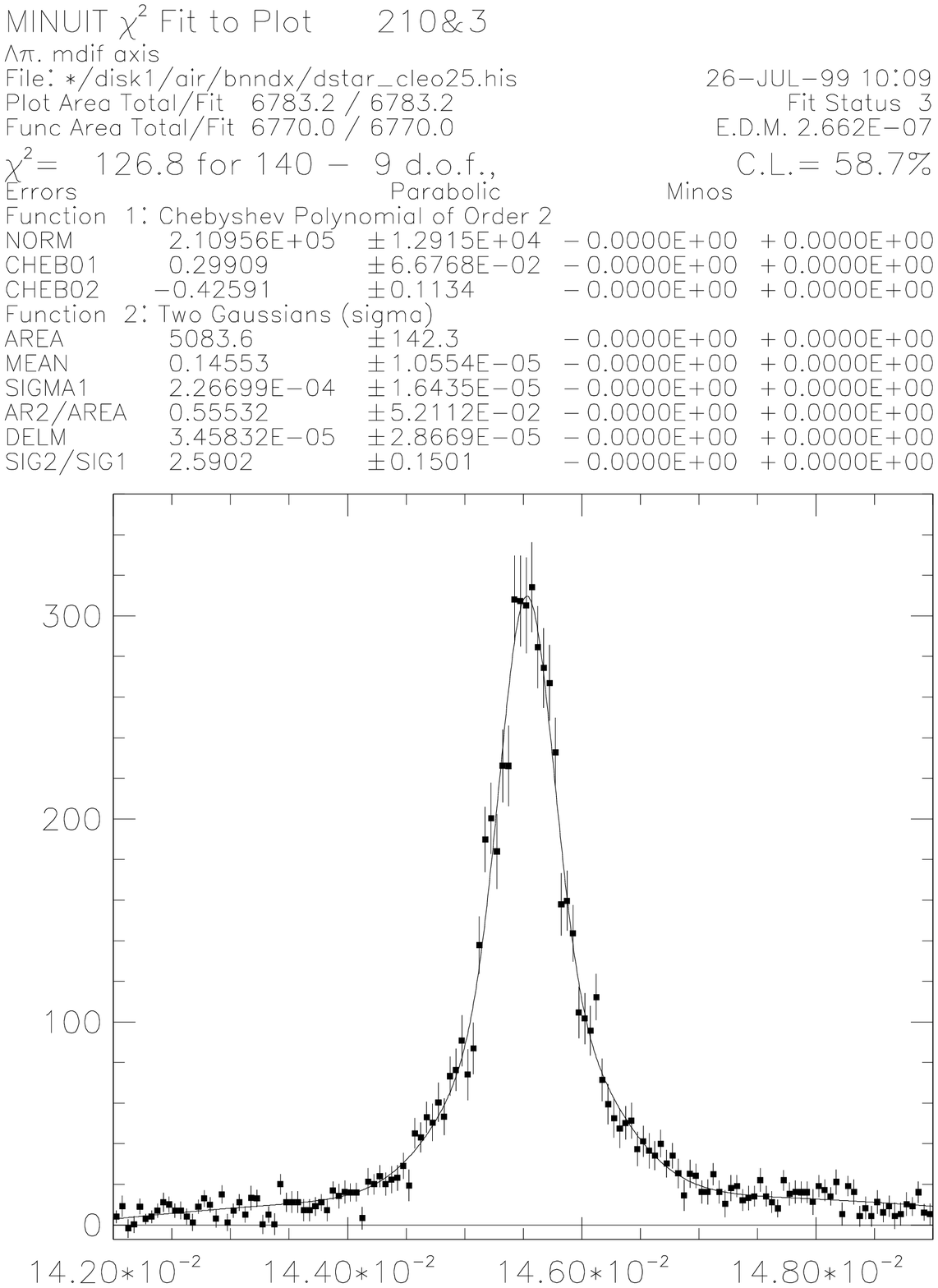}
\vskip 40pt 
   \caption{ \label{dzkp25} 
{\mdif} in GeV for $B \rightarrow D^* X$ with {\dzdecc} in CLEO II.5 }
\vskip 60pt
\end{figure}

\begin{figure}[ht]
   \centering \leavevmode
        \epsfysize=10cm
   \epsfbox[20 143 575 699]{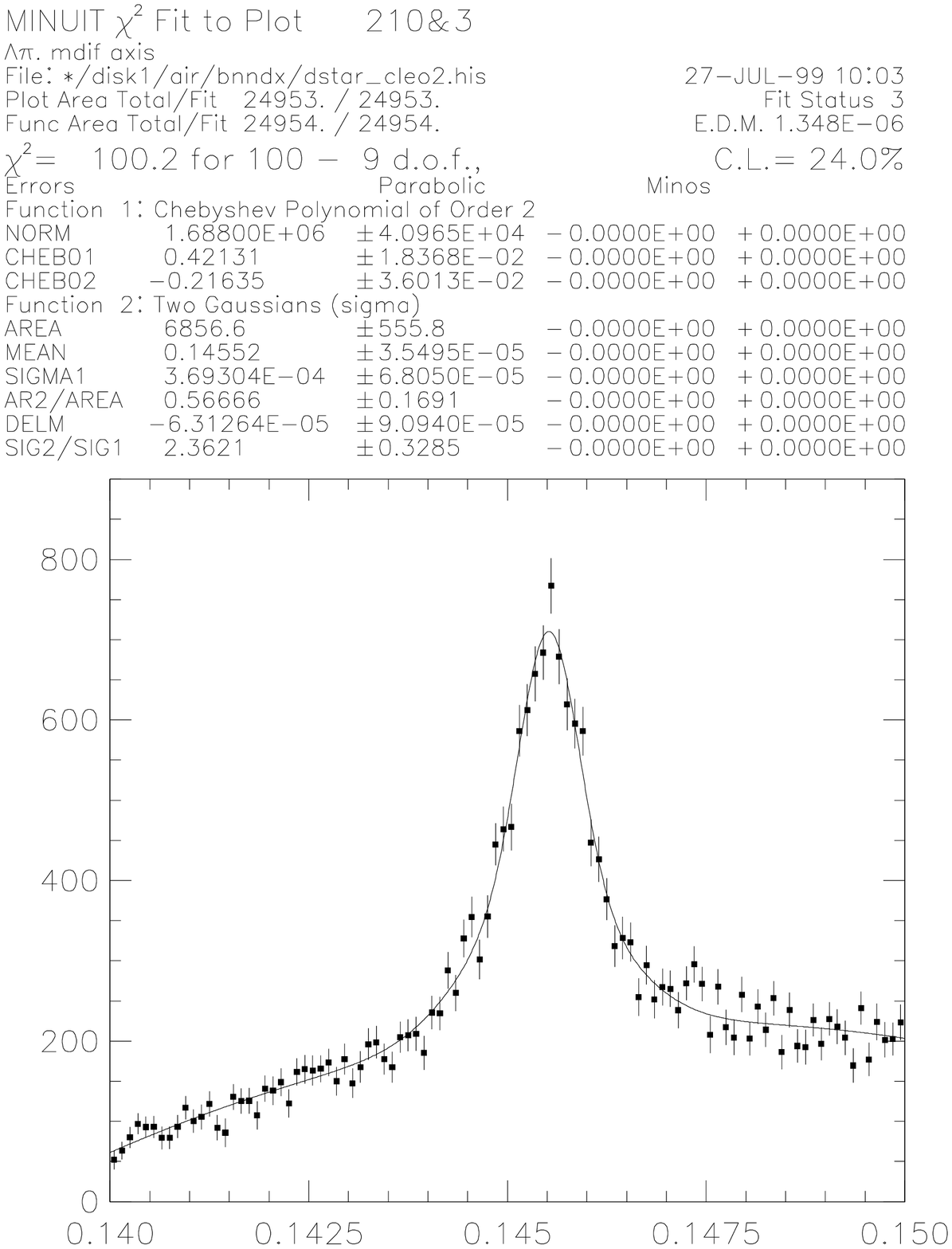}
\vskip 40pt 
   \caption{ \label{dzkppz20} 
{\mdif} in GeV for $B \rightarrow D^* X$ with {\dzdecb} in CLEO II }
\vskip 60pt
\end{figure}

\begin{figure}[ht]
   \centering \leavevmode
        \epsfysize=10cm
   \epsfbox[20 143 575 699]{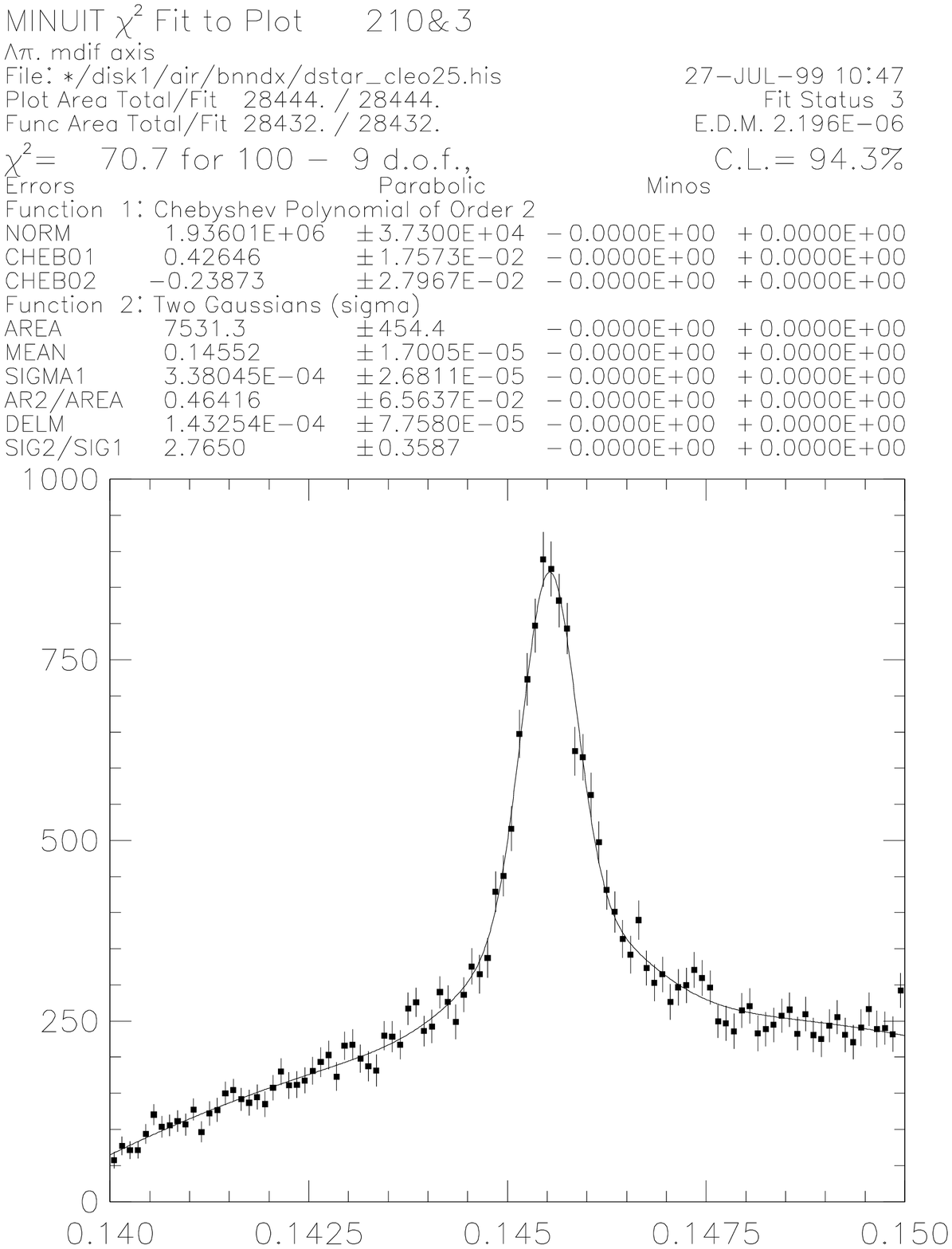}
\vskip 40pt 
   \caption{ \label{dzkppz25} 
{\mdif} in GeV for $B \rightarrow D^* X$ with {\dzdecb} in CLEO II.5 }
\vskip 60pt
\end{figure}

\begin{figure}[ht]
   \centering \leavevmode
        \epsfysize=10cm
   \epsfbox[20 143 575 699]{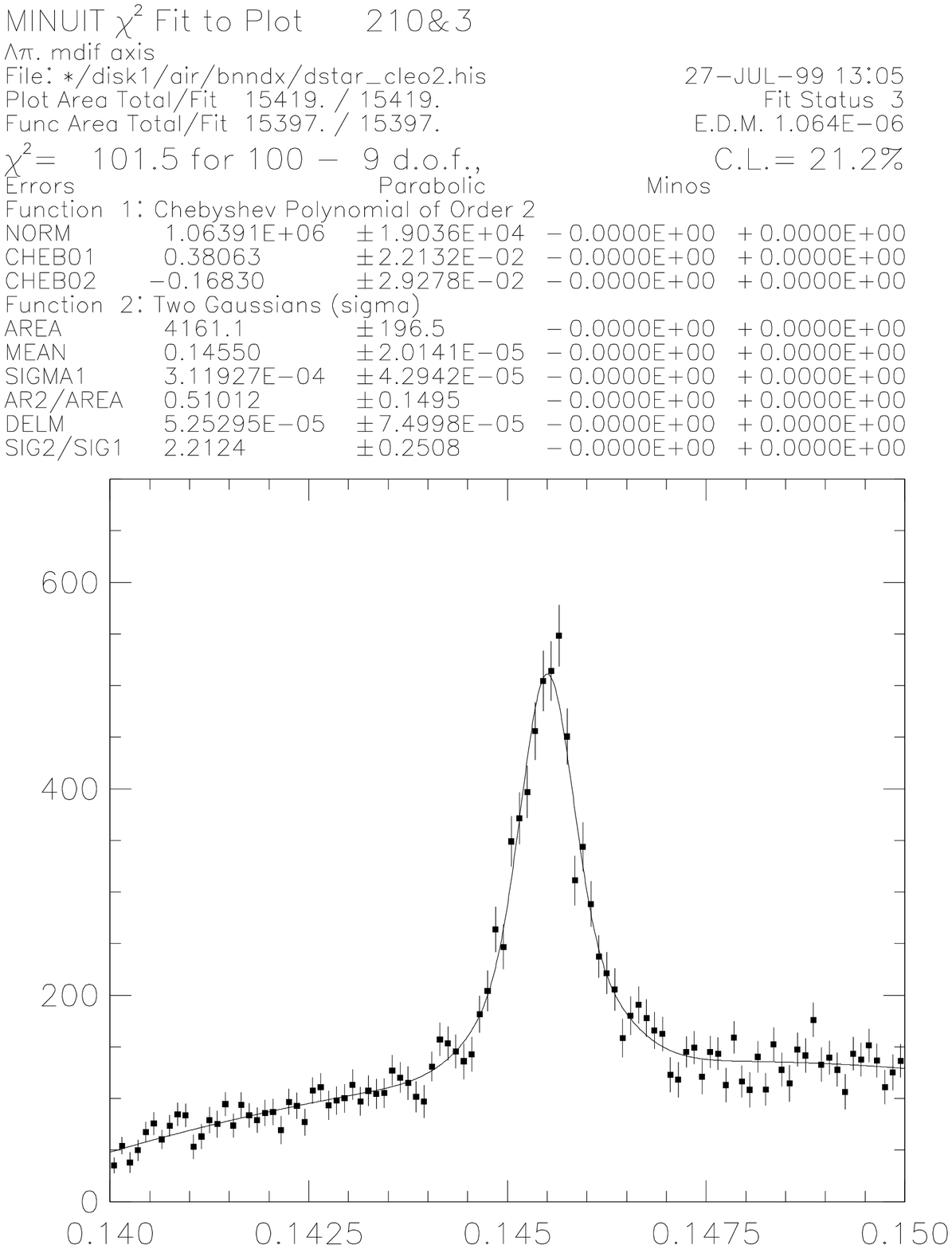}
\vskip 40pt 
   \caption{ \label{dzk3p20} 
{\mdif} in GeV for $B \rightarrow D^* X$ with {\dzdeca} in CLEO II  }
\vskip 60pt
\end{figure}

\begin{figure}[ht]
   \centering \leavevmode
        \epsfysize=10cm
   \epsfbox[20 143 575 699]{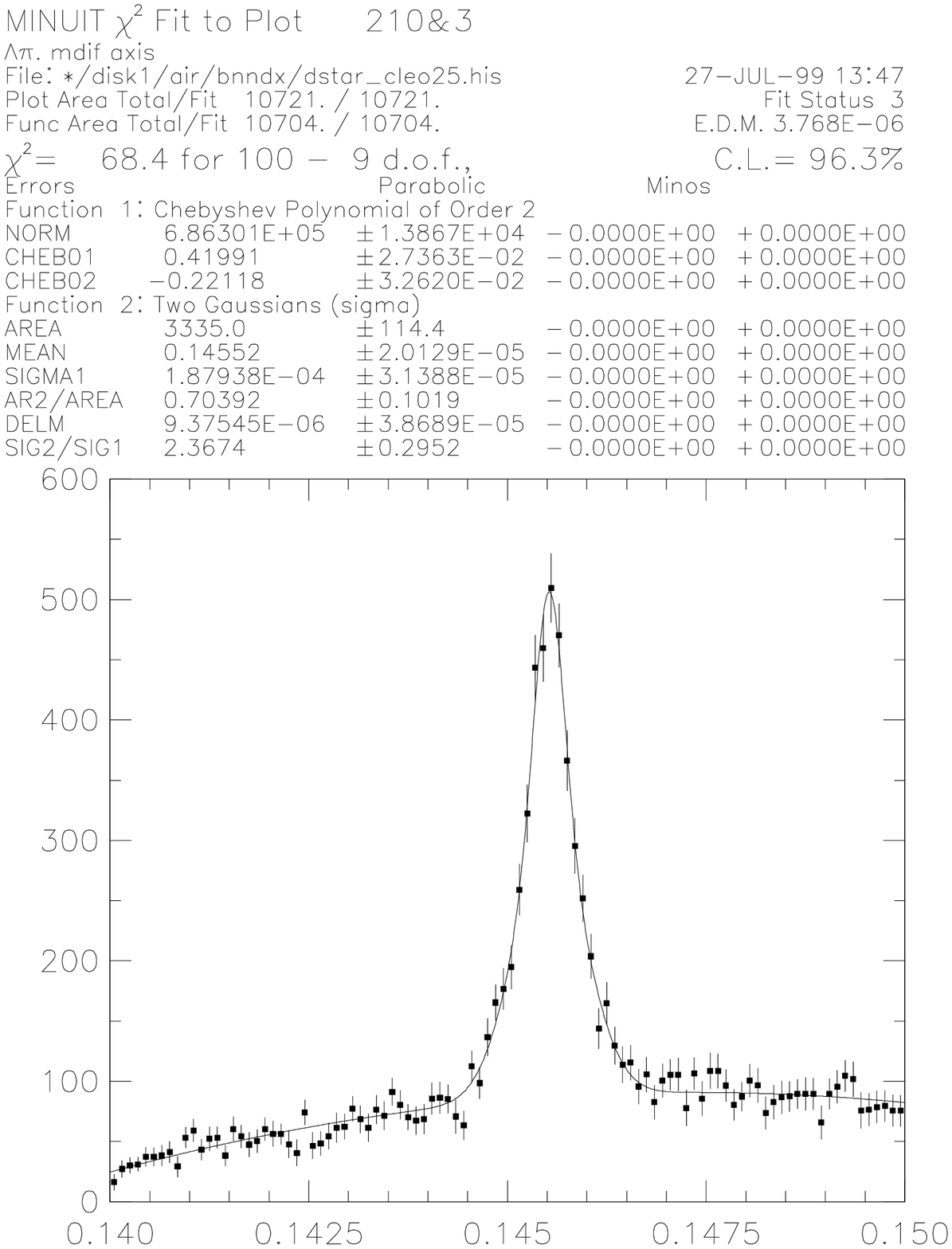}
\vskip 40pt 
   \caption{ \label{dzk3p25} 
{\mdif} in GeV for $B \rightarrow D^* X$ with {\dzdeca} in CLEO II.5 }
\vskip 60pt
\end{figure}

	Shown in Figures \ref{dzkp20}, \ref{dzkp25}, 
\ref{dzkppz20}, \ref{dzkppz25}, \ref{dzk3p20} 
and \ref{dzk3p25} are the double Gaussian fits 
for each of the three $D^0$ modes for either dataset, 
all in the ${X}_{p}$ range 0.25 $<$ ${X}_{p}$ $<$ 0.35. 

\vskip 40pt 

\vskip 60pt

\clearpage

	Shown in Table \ref{dstarcuts} are the 
{\mdz} and  {\mdif} double Gaussian data cuts for 
all three $D^0$ modes and for each dataset.

\begin{table}[htb]
\caption{ {\mdz} and {\mdif} double Gaussian data cuts 
\label{dstarcuts} }
\begin{center}
\begin{tabular}{|c|c|c|c|}
\hline 
 && {\mdz} & {\mdif} \\
Mode  & Dataset & in $\pm$ MeV  & in $\pm$ MeV \\ \hline 
{\dzdecc} & CLEO II & 17.5 & 1.15 \\
{\dzdecc} & CLEO II.5 & 15.0 & 1.10  \\ 
{\dzdecb} & CLEO II & 26.0 & 1.50 \\
{\dzdecb} & CLEO II.5 & 27.0 & 1.50  \\ 
{\dzdeca} & CLEO II & 14.0 & 1.30 \\ 
{\dzdeca} & CLEO II.5 & 12.0 & 0.90  \\ \hline 
\end{tabular}
\end{center}
\end{table}

\subsection{Comparison with $B \to D^* X$}

        To verify the accuracy of our {\dst} reconstruction, 
we compare our results with the PDG value Br($B \to D^* X$) 
= 22.7 $\%$ \cite{pdg99}. 
Since ${\epsilon}_{MC}$ varies with momentum, we 
need to compare our fully reconstructed sample with the 
expected number in the same momentum range. 
        In {\bdeca} and {\bdecb} signal Monte Carlo is 
$\approx$ 50 $\%$ of the generated {\dst}'s 
are in the scaled momentum (${X}_{p}$) range from 
0.25 to 0.35. Approximately 32 $\%$ of Br($B \to D^* X$) 
lies in this momentum range \cite{bdstarx}. 
The expected {\dst} yield is given by: 
\vskip 10pt
(Number of charged/neutral B's) $\times$ 

(Br(B $\rightarrow$ {\dst} X) in ${X}_{p}$ range 0.25 to 0.35) $\times$ 

(Br({\dzz} mode)) $\times$ 

(Br({\dsdec}) = 68.3 $\%$) $\times$ 

({\dst} ${\epsilon}_{MC}$) $\times$

(0.32/0.50: efficiency correction).
\vskip 10pt

This number is efficiency corrected for 
single and double Gaussian signal fits using {\bdeca} Monte Carlo. 
The corrected product is the expected yield that we compare to 
our respective fitted yield for each mode. 

        The close agreement between the expected yield from 
Monte Carlo and the data result for each mode, as shown on 
Tables \ref{dz1}, \ref{dz2}, and \ref{dz3}, is a function 
of how well our Monte Carlo models {\dst} decays from B mesons 
and of the accuracy of our reconstruction code. The 
double Gaussian fits yield better agreement than the single 
Gaussian fits. There is a significant drop in reconstruction 
efficiency for all three $D^0$ modes in the CLEO II.5 
data. This drop is due to the significantly reduced 
reconstruction efficiency of the soft pion in {\dsdec} 
in CLEO II.5. Whereas in CLEO II the PTL drift chamber 
is used in the tracking algorithm, the SVX silicon 
detector which replaces it in the CLEO II.5 data, 
in addition to having more material, is not used 
in the tracking algorithm.

\begin{table}[htb]
\caption{{\dst} yield: {\dzdecc} \label{dz1} }
\begin{center}
\begin{tabular}{|c|c|c|c|c|c|c|c|}
\hline
&Single&&&Double&&\\
Dataset&Gaussian&Expected&Found&Gaussian&Expected&Found\\
& ${\epsilon}_{MC}$   &yield&& ${\epsilon}_{MC}$   & yield&\\
\hline
CLEO II & 30.8 $\%$ &3,862.3&4,117.8& 33.0  $\%$ &4,138,1&4,065.5 \\ 
CLEO II.5 & 16.3 $\%$ &3,933.1&4,455.2 & 20.7 $\%$ &4,994,9&5,083.6 \\ 
\hline 
\end{tabular}
\end{center}
\end{table}

\begin{table}[htb]
\caption{ {\dst} yield: {\dzdecb} \label{dz2} }
\begin{center}
\begin{tabular}{|c|c|c|c|c|c|c|c|c|}
\hline
&Single&&&Double&&&\\
Dataset&Gaussian&Expected&Found&Gaussian&Expected&Found
& S / \\
& ${\epsilon}_{MC}$   &yield&& ${\epsilon}_{MC}$   &yield&& $\sqrt{S + B}$\\
\hline
CLEO II & 14.0 $\%$ &6,371,3&5,955.8& 15.3 $\%$ &6,963.0&6,856.6& 59.7 \\ 
CLEO II.5 & 8.4 $\%$ &7,356.0&6,138.5 & 8.9 $\%$ &7,793.9&7,531.3&61.4 \\ 
\hline 
\end{tabular}
\end{center}
\end{table}

\begin{table}[htb]
\caption{{\dst} yield: {\dzdeca}\label{dz3} }
\begin{center}
\begin{tabular}{|c|c|c|c|c|c|c|c|c|}
\hline
&Single&&&Double&&&\\
Dataset&Gaussian&Expected&Found&Gaussian&Expected&Found
& S / \\
& ${\epsilon}_{MC}$   &yield&& ${\epsilon}_{MC}$   &yield&& $\sqrt{S + B}$\\
\hline
CLEO II & 14.5 $\%$ &3,560.6&3,637.6& 16.0 $\%$ &3,928.9&4,161.6 & 48.3 \\ 
CLEO II.5 & 6.5 $\%$ & 3,071.3&3,100.5 & 6.9 $\%$ &3,260.3&3,335.0 & 47.7\\ 
\hline 
\end{tabular}
\end{center}
\end{table}

\section{Antineutron Showers}

We need to define a set of criteria which allows 
us to select antineutrons with high accuracy without 
incurring a large loss in efficiency. The 
following characteristics, some limiting, 
some exploitable, apply to 
antineutron showers in the CLEO II 
electromagnetic calorimeter: 
\renewcommand{\theenumi}{\arabic{enumi}}
\begin{enumerate}
\item The antineutron shower yields an 
incomplete measurement of its energy and 
an accurate measurement of its direction. 
We use the antineutron shower energy 
to select candidates, but cannot use 
this energy when reconstructing the 
$B$ candidate. We can use the well-measured 
shower direction.  
\item Antineutrons frequently annihilate with 
matter in the calorimeter. Since antineutron 
annihilation showers have 
distinctive characteristics that enable us 
to separate them from other showers, we 
use these characteristics to select them.
\item By baryon number conservation, if the 
event has an antineutron, it must have a 
corresponding baryon. In an exclusive 
reconstruction, as is ours, the selection of 
an proton increases the probability of there 
being an antineutron in the event substantially 
due to baryon number conservation. 
In addition, once we have selected a {\dst}, 
only shower candidates within a narrow 
momentum cone can be selected in the event. 
\end{enumerate}

We refer the reader to other studies of 
baryon-antibaryon annihilation \cite{barrel,barrel2}. 
In Table \ref{showers} we outline the types 
of showers encountered in a hadronic event.

\begin{table}[htb]
\caption{Shower types and energy measured in 
calorimeter \label{showers} }
\begin{center}
\begin{tabular}{|c|c|c|}
\hline 
Particle & Shower type & Energy measured \\ \hline 
$\mu^{\pm}$, $\pi^{\pm}$, $K^{\pm}$, $p^+$ &  minimum ionizing & 
small fraction \\ 
$e^\pm$,$\gamma$ & electromagnetic &  full measurement \\
$K_{L}$ & soft annihilation & 
small fraction  \\ 
neutron & small electromagnetic & 
very small fraction   \\ 
$\bar{p}$ and $\bar{n}$  & medium to hard annihilation 
& small fraction  \\ \hline 
\end{tabular}
\end{center}
\end{table}

We are limited by the absence of an independent 
antineutron sample that we can study to 
define our selection criteria. However, antiprotons 
also annihilate with nucleons in the calorimeter. 
Therefore, antiproton annihilation showers compose the 
shower sample in data and Monte Carlo which 
we use to define our antineutron selection criteria as 
well as to gauge how well the Monte Carlo models 
antineutron annihilation showers.  

\subsection{Shower Parameters}	

        We use the CCFC clustering package to select 
antineutron candidates. The CCFC shower package was not optimized 
to separate annihilation showers from other showers 
in the calorimeter. Nevertheless, we find that the parameters 
previously optimized to include photons, such as E9OE25, 
the list of nearby showers, and track-to-shower matching, 
are useful in excluding them as well.  

        The shower parameters used are:

\renewcommand{\theenumi}{\arabic{enumi}}
\begin{enumerate}
\item E9OE25 $<$ cut 1, a pre-determined 
value: E9OE25 has been defined in our 
discussion of $\pi^0$'s. 
It is very close to 1 for photons, and, we find, 
farthest from 1 for annihilation 
showers, overlapped {\pizero}'s-which are not merged, 
but have the two photon showers very near each other, 
and $K_{L}$'s. An energy dependent cut, cut 1, 
on E9OE25, is applied to reject 99$\%$ of isolated photons. 
\item NNESH: List of 
nearby showers (which does not include the shower to 
which the list belongs). The area encompassed increases with 
energy. It is in this list that we find showers near 
the antineutron that are most likely to be 
hadronic split-offs. We call the shower 
associated with the list the main shower, 
and sum the energy of it and all the others in NNESH, 
we call a group. In {\bdeca} CLEO II signal MC, 
showers in the NNESH list are tagged to the parent 
shower in excess of 93 $\%$ of the time.
\item $|(cos(\theta)|$: 
The calorimeter can be divided by polar angle $\theta$ into 
sections: good barrel, bad barrel, barrel/endcap overlap, 
bad endcap, good endcap, and near beampipe. We use the good 
barrel section throughout, which corresponds to 
$|(cos(\theta)| <$ 0.71.
\item Match level: The track to shower 
match levels in CCFC's NTRSH array are:
\renewcommand{\theenumi}{\arabic{enumi}}
\begin{enumerate}
\item Match level 1: 
shower center $<$ 8 cm from the track projection.
\item Match level 2: 
not level 1, but $\ge$ 1 member crystal
\item Match level 3: 
not level 1 nor level 2.
\end{enumerate}
We use match 3 showers for antineutron candidates and 
match 1 and 2 showers for antiprotons. 
\end{enumerate}
        
\subsection{Antiproton Showers in Data}

	Lacking a sample of antineutrons to study in 
data which is independent of the sample we will 
use to measure {\bdeca} and {\bdecb}, 
we use antiproton annihilation showers which have been 
matched to a track. 
Antiprotons, like antineutrons, annihilate with matter 
in the calorimeter. The annihilation group of showers that results from 
this process has a distinctive signature: the antibaryon interaction 
is associated with a main shower, which has most of the energy 
of the annihilation, and the hadronic splitoffs from 
the interaction are associated with satellite showers near 
the point of interaction, each making a small contribution to the 
group energy. 
        
\begin{figure}[ht]
   \centering \leavevmode
        \epsfysize=8cm
   \epsfbox[20 143 575 699]{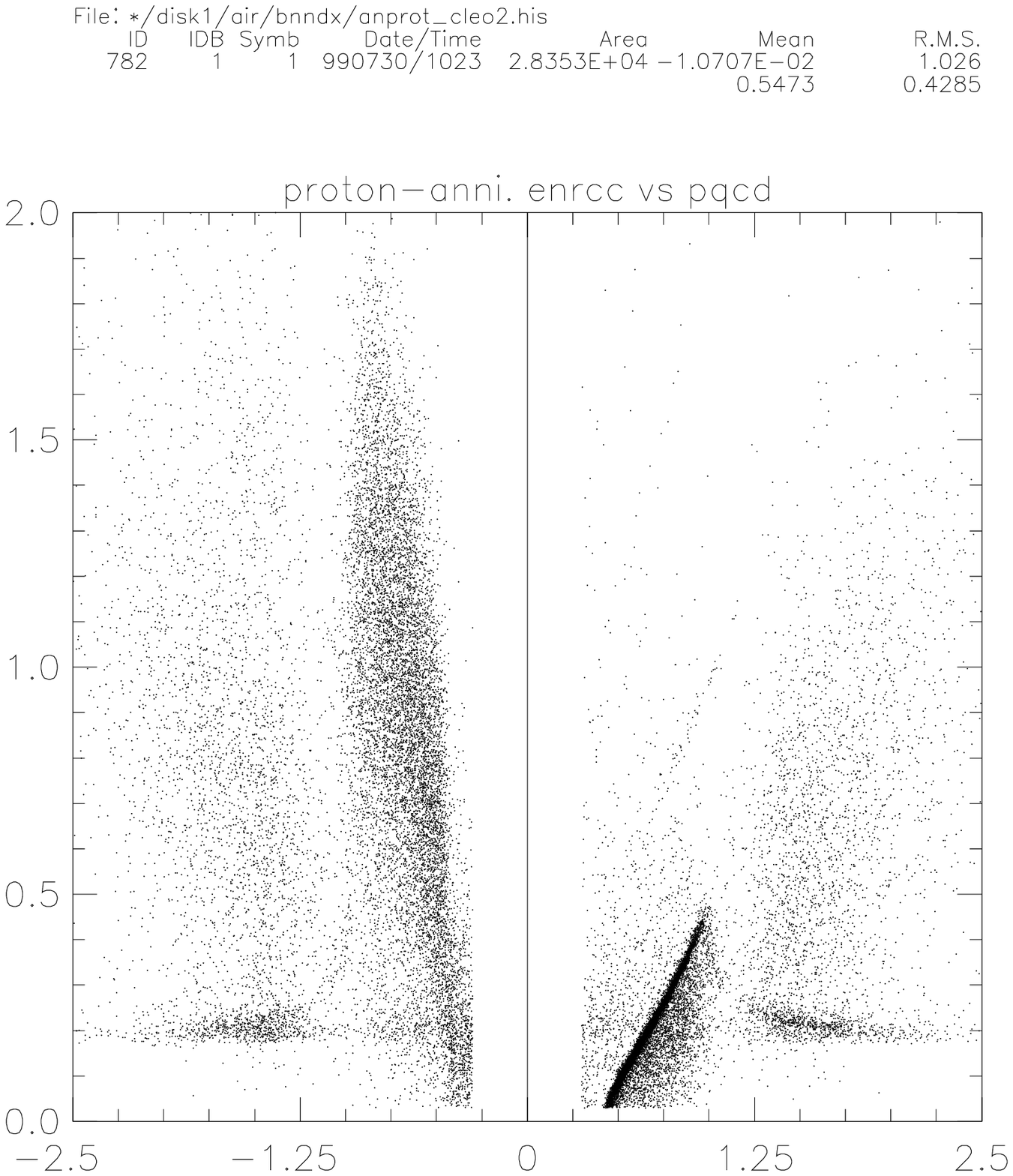}
\vskip 10 pt 
   \caption{${E}_{main}$ vs. PQCD for 
protons and antiprotons in CLEO II \label{scatter} }
\end{figure}

        In Figure \ref{scatter} we plot shower energy 
${E}_{main}$ vs. signed momentum (PQCD) for protons and antiprotons, 
allowing all values of E9OE25, 
in the CLEO II data from a {\dst} skim. Continuum and {\bbar} 
hadronic events are combined in this plot. The use of a {\dst} 
skim is a result of convenience. 

	The selection criteria used to generate 
Figure \ref{scatter} is as follows: 

\renewcommand{\theenumi}{\arabic{enumi}}
\begin{enumerate}
\item $|PQCD(track)|$ $>$ 300 MeV.
\item $L_{proton}$ $>$ 0.9.
\item Track-to-shower match level 1.
\item $|\sigma_{dE/dx,electron}|$ $>$ 2.2 to 
reject electron fakes. 
\item $|(cos(\theta)| <$ 0.71: good barrel.
\item All values of E9OE25 allowed.
\end{enumerate}

\vskip 50pt

\begin{figure}[ht]
   \centering \leavevmode
        \epsfysize=8cm
   \epsfbox[20 143 575 699]{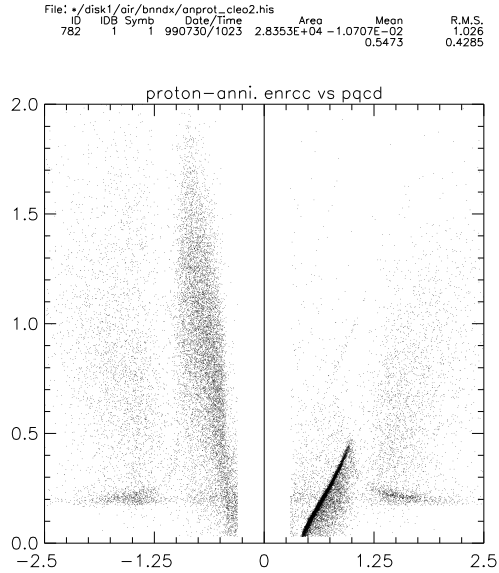}
\vskip 10 pt 
   \caption{${E}_{main}$ vs. PQCD for 
protons and antiprotons in CLEO II \label{scatter2} }
\end{figure}

	In Figure \ref{scatter2} we use the same 
selection criteria as was used for Figure \ref{scatter} 
with the exception of a rejection cut, E9OE25 $<$ cut 1, 
to suppress non-annihilation showers. 
The antiproton annihilation showers are typically 
those with ${E}_{main}$ $>$ 300 MeV and in the track 
momentum range 500 MeV to 900 MeV. Note that for these 
showers there is only a loose 
correlation between the reconstructed energy of the 
shower and the momentum of the antiproton that produced it. 
We can identify an annihilation shower, but have a 
very poor measurement of the momentum of the particle that 
produced it. 

        The horizontal line at shower energy 200 MeV, sloping 
upwards at PQCD 1 GeV, corresponds to minimum ionizing protons 
and antiprotons. The rest of this line in the lower momentum 
range corresponds to minimum ionizing pions. 
In the case of protons, before an E9OE25 cut is applied, the 
diagonal line corresponds to protons captured in the calorimeter. 
1.0 GeV momentum is the threshold for this capture. Neutrons, 
analogously to protons, do not annihilate, and, since we have 
neither momentum information nor distinguishable showers, the 
study of B decays with neutrons is beyond CLEO's capabilities. 

\subsection{Antineutron Selection Criteria}

	Using antiproton annihilation showers, we devise 
the antineutron selection criteria to be used when 
reconstructing {\bdeca}, as shown in Table \ref{showercuts}. 
Although this selection criteria allows a shower to 
have no nearby daughters, in which case 
${E}_{main} = {E}_{group}$, the majority of 
antineutron showers have at least one nearby daughter 
in Monte Carlo, as well as in our exclusively 
reconstructed $B^0$ signal events. 

\begin{table}[htb]
\caption{ Antineutron shower selection criteria.
\label{showercuts} }
\begin{center}
\begin{tabular}{|c|}
\hline
Track-to-shower match level 3 \\ 
E9OE25 $<$ cut 1\\ 
$|cos(\theta_{shower})|$ $<$ 0.71\\ 
${E}_{main}$ $>$ 500 MeV \\
${E}_{group}$ $>$ 800 MeV \\ \hline 
\end{tabular}
\end{center}
\end{table}

	We test this selection criteria in a generic 
$B\bar{B}$ Monte Carlo before and after the 
requirement that there be a proton 
with $|PQCD(track)|$ $>$ 300 MeV and $L_{proton}$ $>$ 0.9 in 
the event. The generic Monte Carlo sample is discussed in 
Section \ref{genmc}. In Table \ref{unmatched} we show the results before 
and after applying the antineutron selection criteria cuts 
for a sample before the proton requirement is applied.

\begin{table}[htb]
\caption{Shower population in a generic $B\bar{B}$ 
Monte Carlo sample without proton requirement
\label{unmatched}}
\begin{center}
\begin{tabular}{|c|c|c|}
\hline
Shower & Annihilation cuts & Annihilation cuts \\ 
type & without $E_{group}$ cut & with $E_{group}$ $>$ 800 MeV \\ \hline 
$\gamma$ from {\pizero} & 44.5 & 46.2\\ 
$K_{L}$                  & 33.0& 22.2\\ 
${\bar{p}}$              & 0.5 &  0.6\\ 
${\pi}^{\pm}$,${K}^{\pm}$& 3.3 &  1.9\\ 
other($\omega$,etc)      & 1.4 &  1.3\\ 
${\bar{n}}$              &17.3 & 27.8\\ \hline
TOTAL                    &100.0&100.0\\ \hline 
\end{tabular}
\end{center}
\end{table}
 
	According to our generic Monte Carlo simulation, 
even before a proton requirement is applied 27.8$\%$ 
of the annihilation-like showers in a $B\bar{B}$ event 
are antineutrons. We next apply the proton requirement to 
a generic $B\bar{B}$ Monte Carlo sample. The results 
are shown in Table \ref{unmatched2}. 

\begin{table}[htb]
\caption{Shower population in a generic $B\bar{B}$ 
Monte Carlo sample with proton requirement
\label{unmatched2} }
\begin{center}
\begin{tabular}{|c|c|c|}
\hline
Shower & Annihilation cuts & Annihilation cuts \\ 
type & without $E_{group}$ cut & with $E_{group}$ $>$ 800 MeV \\ \hline 
$\gamma$ from {\pizero}  & 16.3& 11.4\\ 
$K_{L}$                  &  9.5&  3.9\\ 
${\bar{p}}$              &  2.7&  1.8\\ 
${\pi}^{\pm}$,${K}^{\pm}$&  2.0&  0.9\\ 
other($\omega$,etc)      &  0.7&  0.5\\ 
${\bar{n}}$              & 68.8& 81.5\\ \hline
TOTAL                    &100.0&100.0\\ \hline
\end{tabular}
\end{center}
\end{table}

        The above results are not surprising: use of 
baryon number conservation by selecting a proton 
increases the probability of an event having an
antineutron from a few $\%$ to 50 $\%$, 
while leaving all other annihilation like backgrounds 
at near the same level. 
Use of ${E}_{group}$ cut reduces $K_{L}$ 
contribution by 1/3 in the Non-baryon sample and 
by 1/2 in the baryon sample.
        
\subsection{Antineutron Backgrounds}

\begin{figure}[ht]
   \centering \leavevmode
        \epsfysize=5cm
   \epsfbox[20 143 475 599]{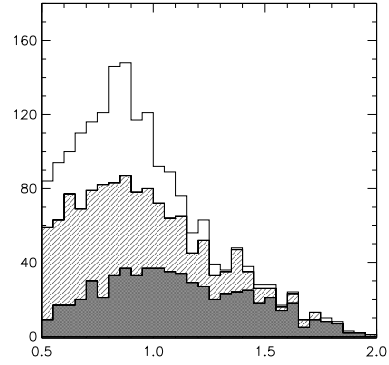}
   \caption{$E_{main}$ in GeV without proton 
requirement for $\bar{n}$'s, 
$\gamma$'s from {\pizero}'s, and $K_{L}$'s 
\label{bgshower}}
\end{figure}

\begin{figure}[ht]
   \centering \leavevmode
        \epsfysize=5cm
   \epsfbox[20 143 475 599]{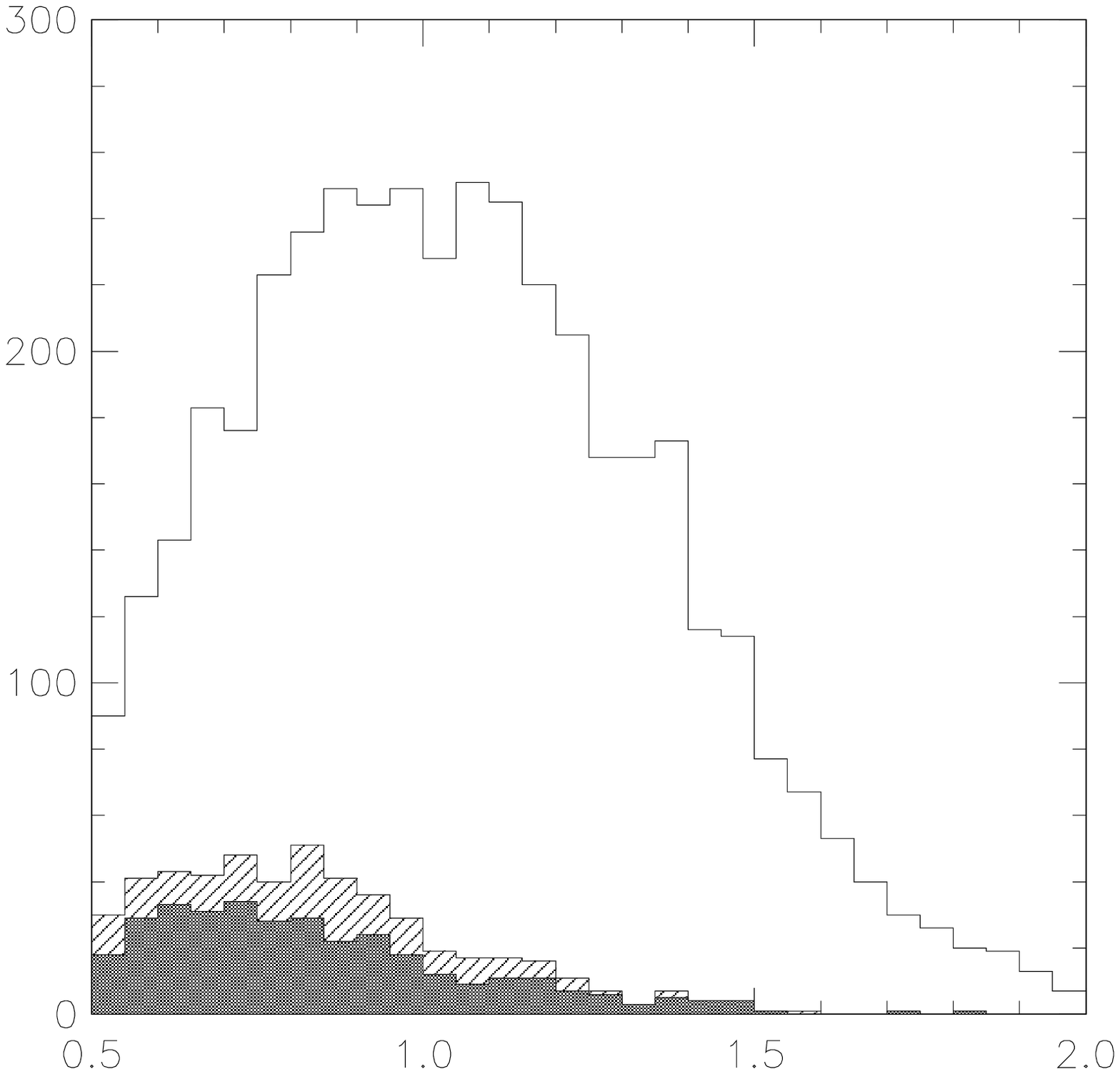}
   \caption{$E_{main}$ in GeV with proton 
requirement for $\bar{n}$'s, 
$\gamma$'s from {\pizero}'s, and $K_{L}$'s 
\label{bgshower2} }
\end{figure}

	In Figures \ref{bgshower} and \ref{bgshower2} 
we compare the shower energy spectrum of 
each of the two major backgrounds 
to antineutrons-$\gamma$ from {\pizero}'s, 
and $K_{L}$-to that of antineutrons. All antineutron 
selection criteria cuts, including $E_{group}$ $>$ 800 MeV, 
have been applied. Figure \ref{bgshower} is for 
case before proton requirement has been applied, and 
Figure \ref{bgshower2} is for the case after 
proton requirement has been applied. 
The sample is, again, generic $B\bar{B}$ Monte Carlo. 
The backgrounds that have not been included make 
up (4.3,3.0) $\%$ of the (before,after) proton requirement 
distributions. In Figure \ref{bgshower} the solid distribution 
is for $\bar{n}$'s, the slanted lines distribution 
is for $\gamma$'s from {\pizero}'s, and the 
white distribution is for $K_{L}$'s.

	In Figure \ref{bgshower2} the white distribution 
is for $\bar{n}$'s, the solid distribution 
is for $\gamma$'s from {\pizero}'s, and the 
slanted lines distribution is for $K_{L}$'s. 
The large contamination in the low range of the 
$E_{main}$ spectrum, as shown in Figure \ref{bgshower} 
and Figure \ref{bgshower2} does not affect {\bdeca} 
since for this $B$ meson mode the antineutron shower 
energy spectrum is dominantly in the 
range 1.0 $<$ $E_{main}$ $<$ 1.5 GeV.

\pagebreak

\vskip 100pt

\chapter{Measurement of $B^0 \rightarrow D^{*-}p\bar{p}\pi^+$ }

	Modes in {\bDnnx} have not been previously measured. 
The reconstruction efficiency multiplied by the 
number of $B$ mesons in our dataset 
for {\bdecb} and {\bdeca}, assuming branching fractions similar to 
the ones measured for {\blcppi} and {\blcppipi} \cite{jjo}, 
is at a level we can measure, which lends credibility to 
a search for both modes. 

\section{Monte Carlo Reliability}

	The reconstruction process at CLEO, as well 
as at all other high energy detectors, involves 
a reconstruction efficiency $\epsilon_{Data}$. For any given 
$B$ decay, only a fraction of the number present 
in the data can be partially or fully 
reconstructed. If $x$ number of B decays in our data 
are {\bbaryons} decays, only $\epsilon_{Data} \times x$ 
are measured. In the case of $B \rightarrow p X$, for example, 
there is a significant contamination due 
to protons from beam gas interactions at 
the low end of the momentum spectrum which limits 
$B \rightarrow p X$ to be carried out using
antiprotons only, and then multiplying the 
result by 2. 

	By tuning a wide range of parameters, 
the Monte Carlo simulation can be made to closely 
resemble the data, in which case the assumption 
$\epsilon_{Data}$ $\approx$ $\epsilon_{MC}$ is reasonable. 
$\epsilon_{MC}$ is reliable because several 
processes which are measured at CLEO allow considerable 
ease of simulation as well as large samples which 
can be reliable separated from backgrounds. Processes such as 
$e^+e^- \rightarrow e^+e^-$, $e^+e^- \rightarrow \mu^+\mu^-$, 
$e^+e^- \rightarrow \pi^+\pi^-$, 
$e^+e^- \rightarrow \gamma\gamma$, and 
$e^+e^- \rightarrow \gamma\gamma\gamma$ allow very 
accurate Monte Carlo modeling of individual particles, 
which in turn can be combined to model many B decays well. 

	On this level of accuracy a second group of decays 
is studied to widen the scope of our simulation: 
$K_{S} \rightarrow \pi^+\pi^-$, $\phi \rightarrow K^+K^-$, 
{\dsdec} with {\dzdecc}, and $\Lambda \rightarrow p\pi$ 
all have very low backgrounds. Many particle separation 
studies are based on these decays. An example is 
the use of a low background sample of $\bar{\Lambda}$'s to 
study antiproton annihilation showers, 
as will be discussed later in this chapter.

\section{Reconstruction Procedure}
	
	$B$ mesons are produced with $\approx$ 300 MeV 
momentum at CLEO, which implies that 
$E_{beam}$ is very close to $E_{B}$. This 
momentum is small when compared 
with the $B$ meson mass of 5.28 GeV. 
Since the beam energy at CLEO is measured with a 
2 MeV resolution, we can constraint the mass of the $B$ 
meson candidate to be equal to the beam 
energy $E_{beam}$. We therefore use the beam 
contrained mass when reconstructing {\bdecb}:
\vskip 10pt

{\centerline{ ${M}_{BC}$ = $\sqrt{ {E_{beam}}^{2} - 
\sum_{i}^{3} {{p}_{i}}^{2}  }$ }}       

\vskip 10pt
where $E_{beam}$ is the beam energy, on average 5.29 GeV, and 
$\sum_{i}^{3} {{p}_{i}}^{2}$ is the sum of daughter momenta squared. 

        The energy difference between the beam energy and 
the energy of the reconstructed B candidate, defined by: 
\vskip 10pt
{\centerline{ ${\Delta}_{E}$ = $E_{beam}$ - $E_{reconstructed B}$}}  
\vskip 10pt
is centered at 0 and has a Gaussian width similar to the 
${M}_{BC}$ Gaussian width, whereas background events are 
much more likely to form a random distribution in ${\Delta}_{E}$. 
Selecting $B$ candidates using ${\Delta}_{E}$ cuts based 
on the Monte Carlo is useful in separating signal from background. 
As modelled by our Monte Carlo ${\Delta}_{E}$ for {\bdecb} is 
centered near zero and has a Gaussian width $\sigma <$ 100 MeV, 
which allows us to separate the signal from backgrounds 
that differ by a miss-measured extra pion. The methodology 
we use in measuring {\bdecb} is analogous to that used 
in previous exclusive reconstructions at CLEO \cite{dpi,jorge}.

\section{Monte Carlo Study}

	In order to find the detector efficiency for the 
reconstruction of the modes we are searching for, we 
generate Monte Carlo in which one of the $B$ mesons 
in the event is forced to decay to the mode we are 
reconstructing. We refer to this sample as signal 
Monte Carlo. In this signal Monte Carlo 
$B$ mesons decay according to phase space. 

We also use a sample of generic Monte Carlo $B\bar{B}$ 
events on which we run the {\bdecb} analysis code. 
The generic sample is composed of the $B$ decays 
that have previously been measured as well as 
randomnly generated events assuming inclusive 
momentum distributions. There are no {\bDnnx} 
modes in the generic Monte Carlo sample. We 
use the generic Monte Carlo sample to model 
our backgrounds due to modes of the $D^0$ we 
are not reconstructing.  

Applying the particle selection criteria outlined 
in Chapter 3 for the {\bdecb} decay daughters, 
we fit the $\Delta E$ distribution 
to a single Gaussian for signal and a 1st order 
Chebyshev polynomial for background. We show in 
Table \ref{dele} and Table \ref{dele2} the result 
of these fits. 

\begin{table}[htb]
\caption{$\Delta E$ fit results for {\bdecb} signal MC in CLEO II
\label{dele} }
\begin{center}
\begin{tabular}{|c|c|c|c|}
\hline
Decay mode & Central value (MeV) 
& $\sigma$ (MeV) & $\Delta E$ cut (MeV) \\ \hline 
{\dzdecc}  & 0.6 & 11.8 & $\pm$ 35 \\ 
{\dzdecb}  & -2.3 & 16.8 & $\pm$ 50 \\
{\dzdeca}  & 1.1 & 9.8 & $\pm$ 29 \\ \hline
\end{tabular}
\end{center}
\end{table}

\begin{table}[htb]
\caption{$\Delta E$ fit results for {\bdecb} signal MC in CLEO II.5
\label{dele2} }
\begin{center}
\begin{tabular}{|c|c|c|c|}
\hline
Decay mode & Central value (MeV) 
& $\sigma$ (MeV) & $\Delta E$ cut (MeV) \\ \hline
{\dzdecc}  & -1.0 & 9.8 & $\pm$ 29 \\ 
{\dzdecb}  & -3.1 & 13.4 & $\pm$ 40 \\ 
{\dzdeca}  & -1.6 & 6.1 & $\pm$ 18 \\ \hline
\end{tabular}
\end{center}
\end{table}

        We define mode/dataset specific $\Delta E$ cuts 
which are applied to ${M}_{BC}$. 
The ${M}_{BC}$ distribution is then fitted to a single Gaussian 
to determine each ${\epsilon}_{MC}$, as shown in 
Table \ref{delcut} and Table \ref{delcut2}.

\begin{table}[htb]
\caption{ ${M}_{BC}$ distribution fit results for {\bdecb} 
signal MC in CLEO II \label{delcut} }
\begin{center}
\begin{tabular}{|c|c|c|}
\hline
Decay mode & $\epsilon_{MC}$ ($\%$) & $\sigma$ (MeV) \\ \hline 
{\dzdecc}& 8.69 $\pm$ 0.21 & 2.63 \\ 
{\dzdecb}& 4.48 $\pm$ 0.15 & 2.86 \\ 
{\dzdeca}& 3.89 $\pm$ 0.14 & 2.65 \\ \hline
\end{tabular}
\end{center}
\end{table}

\begin{table}[htb]
\caption{ ${M}_{BC}$ distribution fit results for {\bdecb} 
signal MC in CLEO II.5 \label{delcut2} }
\begin{center}
\begin{tabular}{|c|c|c|}
\hline
Decay mode & $\epsilon_{MC}$ ($\%$) & $\sigma$ (MeV) \\ \hline
{\dzdecc}& 4.32 $\pm$ 0.22 & 2.55  \\
{\dzdecb}& 1.88 $\pm$ 0.14 & 2.60  \\ 
{\dzdeca}& 1.05 $\pm$ 0.08 & 2.50  \\ \hline 
\end{tabular}
\end{center}
\end{table}

	The $\pi_{soft}$ in {\dsdec} for the 
CLEO II.5 dataset is the most sensitive particle to 
errors in our Monte Carlo simulation. The replacement 
of the PTL detector by the SVX detector decreased 
by approximately a factor of 2 the reconstruction 
efficiency for the $\pi_{soft}$'s used here, and 
may have introduced a systematic uncertainty.

\section{Results in Data}

\begin{figure}[ht]
   \centering \leavevmode
        \epsfysize=10cm
   \epsfbox[20 143 575 699]{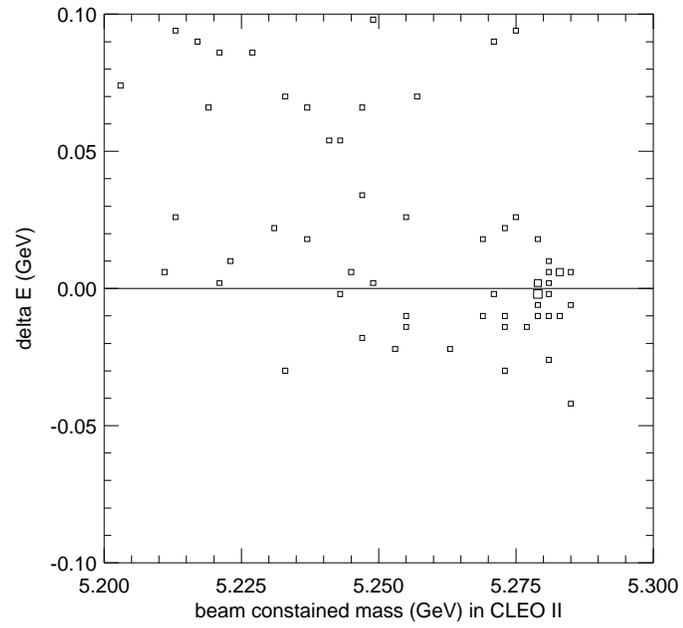}
\vskip 10pt
   \caption{$\Delta E$ vs $M_{BC}$ distribution 
for {\bdecb} in CLEO II ON resonance data. 
\label{cleo20scatter} }
\end{figure}

\begin{figure}[ht]
   \centering \leavevmode
        \epsfysize=10cm
   \epsfbox[20 143 575 699]{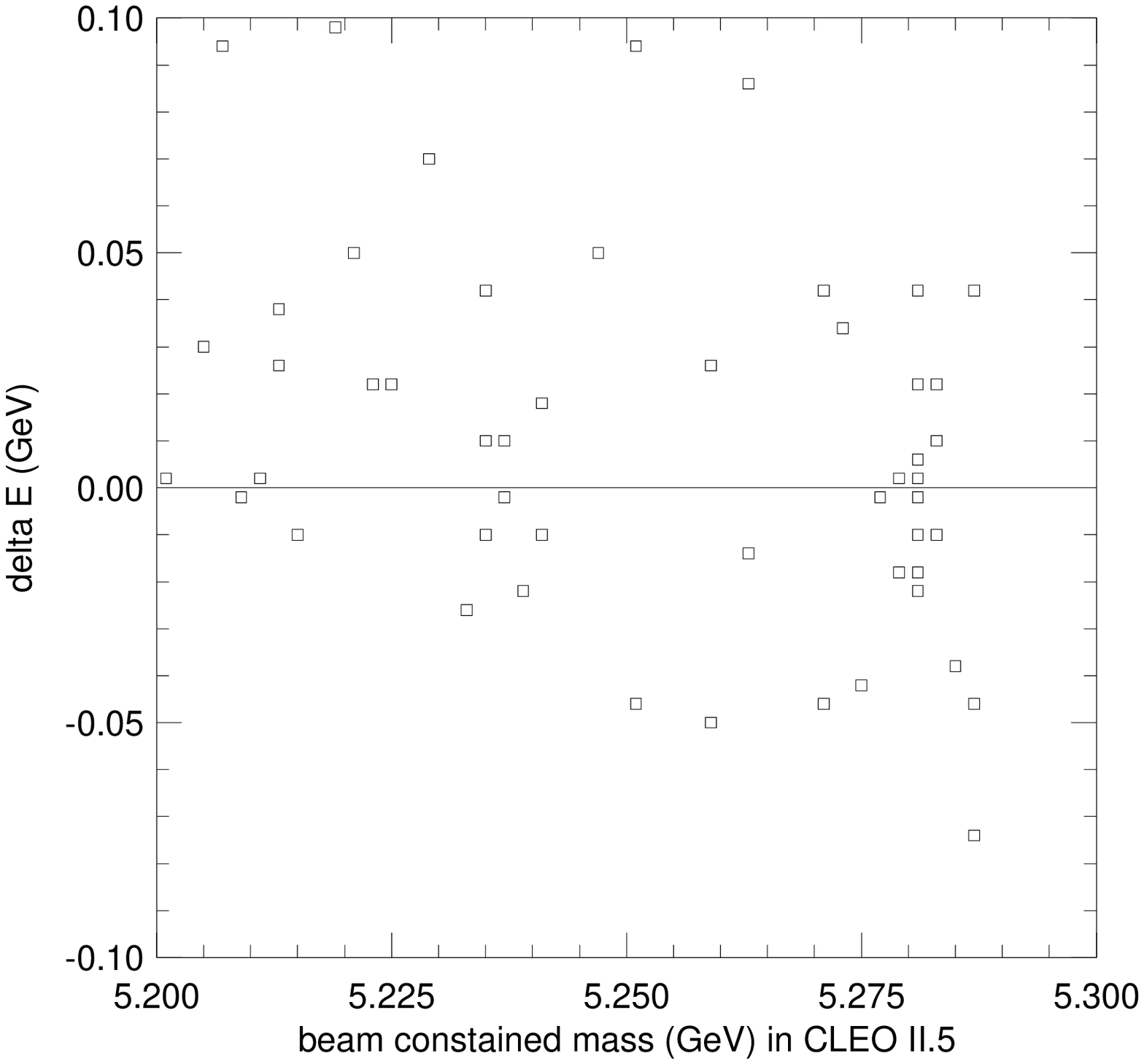}
\vskip 10pt
   \caption{$\Delta E$ vs $M_{BC}$ distribution 
for {\bdecb} in CLEO II.5 ON resonance data 
\label{cleo25scatter} }
\end{figure}

\begin{figure}[ht]
   \centering \leavevmode
        \epsfysize=14cm
   \epsfbox[20 143 575 699]{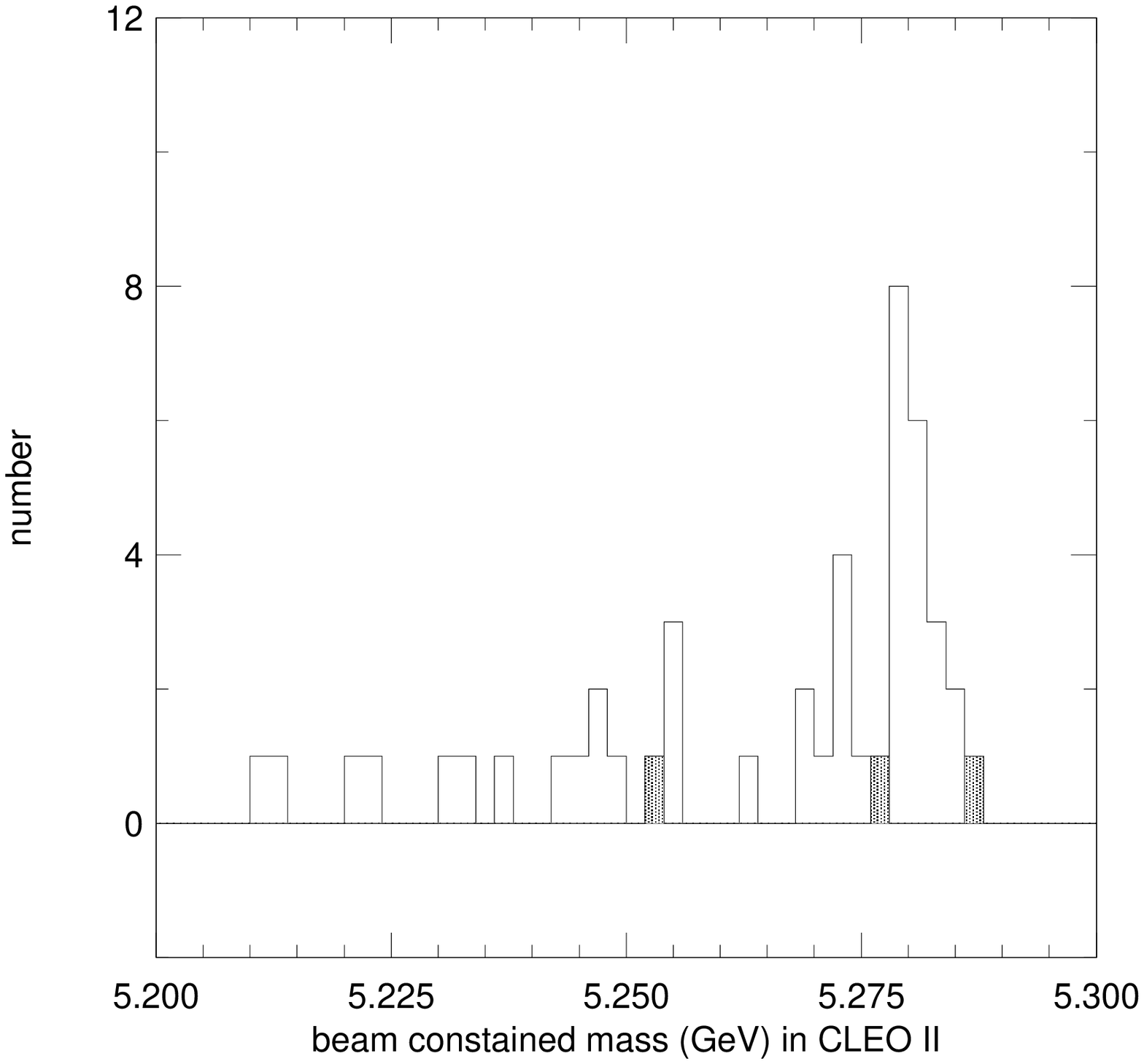}
\vskip 10pt
   \caption{$M_{BC}$ (in GeV) 
for {\bdecb} in CLEO II
\label{pppids20} }
\end{figure}

\begin{figure}[ht]
   \centering \leavevmode
        \epsfysize=14cm
   \epsfbox[20 143 575 699]{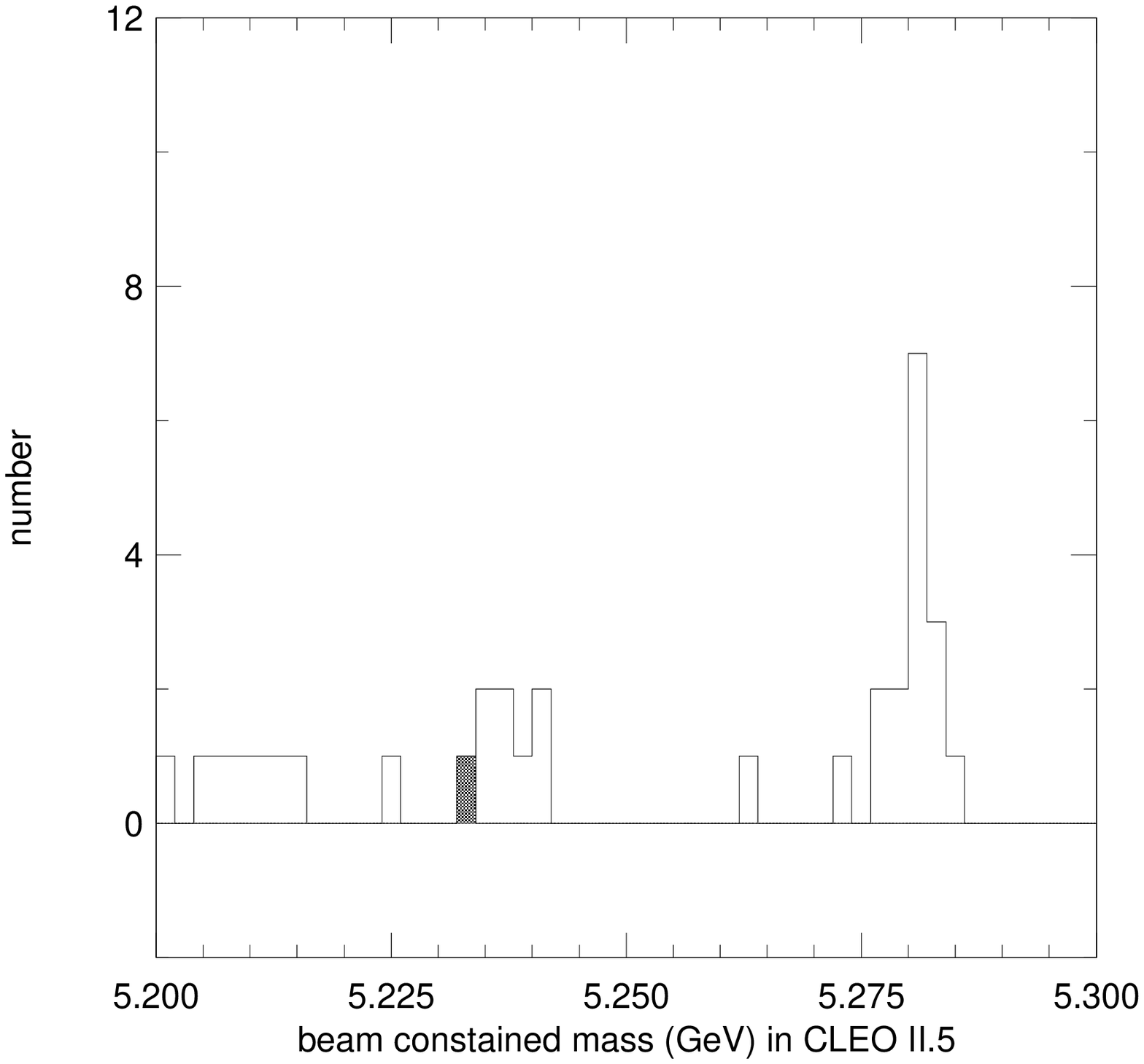}
\vskip 10pt
   \caption{ $M_{BC}$ (in GeV)
for {\bdecb} in CLEO II.5. 
\label{pppids25} }
\end{figure}

	In Figure \ref{cleo20scatter} and Figure 
\ref{cleo25scatter} we show the 
$\Delta E$ vs $M_{BC}$ distribution for {\bdecb} 
in, respectively, CLEO II and CLEO II.5. 
In Figure \ref{pppids20} and Figure 
\ref{pppids25} we show the 
$M_{BC}$ distribution for {\bdecb} in, 
respectively, CLEO II and CLEO II.5, after the 
mode dependent $\Delta E$ cuts have been applied. 
The mode-by-mode $M_{BC}$ distributions are 
combined for each of Figure \ref{pppids20} and Figure 
\ref{pppids25}, in which (dark) events are from 
continuum. The continuum background to {\bdecb} is 
statistically insignificant. If this had not 
been the case, continuum subtraction would 
have been neccessary to subtract this background 
from the  $M_{BC}$ distribution.

\clearpage

        In Table \ref{pppifits} we attempt various fits to the 
combined CLEO II and CLEO II.5 $M_{BC}$ distribution for {\bdecb}:

\renewcommand{\theenumi}{\arabic{enumi}}
\begin{enumerate}
\item Allowing $M_{BC}$ and $\sigma_{M_{BC}}$ to float.
\item Fixing $M_{BC}$ = 5.28 GeV and allowing 
$\sigma_{M_{BC}}$ to float.
\item Allowing $M_{BC}$ to float and fixing 
$\sigma_{M_{BC}}$ according to expectations from signal MC.
\item Fixing both $M_{BC}$ and $\sigma_{M_{BC}}$ 
as prescribed in 2. and 3. above.
\end{enumerate}

        The result of these fits are shown in 
Table \ref{pppifits}. The variation in yield is 
small.

\begin{table}[htb]
\caption{ Results in the CLEO II/II.5 data of various 
fits to $M_{BC}$ for {\bdecb} \label{pppifits} }
\begin{center}
\begin{tabular}{|c|c|}
\hline
fitting options: & fit results  \\ \hline 
floating values &   \\ \hline
$M_{BC}$ (in MeV) & 
${5,280.7}^{+0.43}_{-0.45}$\\
$\sigma$  (in MeV) & 
${2.05}^{+0.43}_{-0.33}$ \\ 
Fit yield & 
${30.52}^{+6.55}_{-5.92}$ \\ \hline 
fixed $M_{BC}$ &   \\ \hline
$\sigma$  (in MeV) & 
${2.40}^{+0.69}_{-0.53}$ \\ 
Fit yield & 
${31.88}^{+7.34}_{-6.54}$ \\ \hline  
fixed $\sigma$ &   \\ \hline
$M_{BC}$ (in MeV) & 
${5,280.5}^{+0.55}_{-0.56}$ \\ 
$\sigma$  (in MeV) & 2.65 \\
Fit yield & 
${32.44}^{+6.68}_{-6.02}$ \\ \hline 
fixed $M_{BC}$ and $\sigma$ & \\ \hline 
Fit yield & 
${32.98}^{+6.75}_{-6.02}$ \\ 
${\cal B}$({\bdecb}) $\times$ $10^{-4}$ & 
${6.6}^{+1.3}_{-1.2}$ \\  \hline 
\end{tabular}
\end{center}
\end{table}

        ${\cal B}$({\bdecb}) is calculated using:
\vskip 20pt

{\centerline{ ${\cal B}$({\bdecb}) = 
{\large{ $\frac{ Fitted Yield }{ 
{ (2 \times B^0\bar{B^0} \times \epsilon_{MC}) }_{CLEO II} + 
{ (2 \times B^0\bar{B^0} \times \epsilon_{MC}) }_{CLEO II.5} }$
}}}}
\vskip 20pt

        The product ${ (B\bar{B} \times \epsilon_{MC}) }_{CLEO II}$ 
is $\approx$ the product 
${ (B\bar{B} \times \epsilon_{MC}) }_{CLEO II.5}$. The number 
of events for $M_{BC}$ $>$ 5.275 GeV is 21 for CLEO II, and 
15 for CLEO II.5. Fitting each dataset separately would 
yield a measurable ${\cal B}$({\bdecb}).

        In table \ref{pppievents} we quote the number 
of background events in the range 
5.2 GeV $<$ ${M}_{BC}$ $<$ 5.275 GeV, and the number 
of signal events for ${M}_{BC}$ $>$ 5.275 GeV. 
The $D^0$ mode with the largest background 
is {\dzdecb}. Signal region contains $\approx$ 10 $\%$ 
background events.

\begin{table}[htb]
\caption{{\bdecb}: number of events found per mode
\label{pppievents} }
\begin{center}
\begin{tabular}{|c|c|c|c|c|}
\hline
Mode & Background & Signal & Background & Signal \\    
 & region & region & region & region \\ 
 & CLEO II & CLEO II & CLEO II.5  & CLEO II.5 \\ \hline  
{\dzdecc}&  7 &  6 &  0 & 6  \\ 
{\dzdecb}& 14 & 10 & 15 & 6  \\ 
{\dzdeca}&  4 &  5 &  3 & 3  \\ \hline 
\end{tabular}
\end{center}
\end{table}

\clearpage

\begin{figure}[ht]
   \centering \leavevmode
        \epsfysize=12cm
   \epsfbox[20 143 575 699]{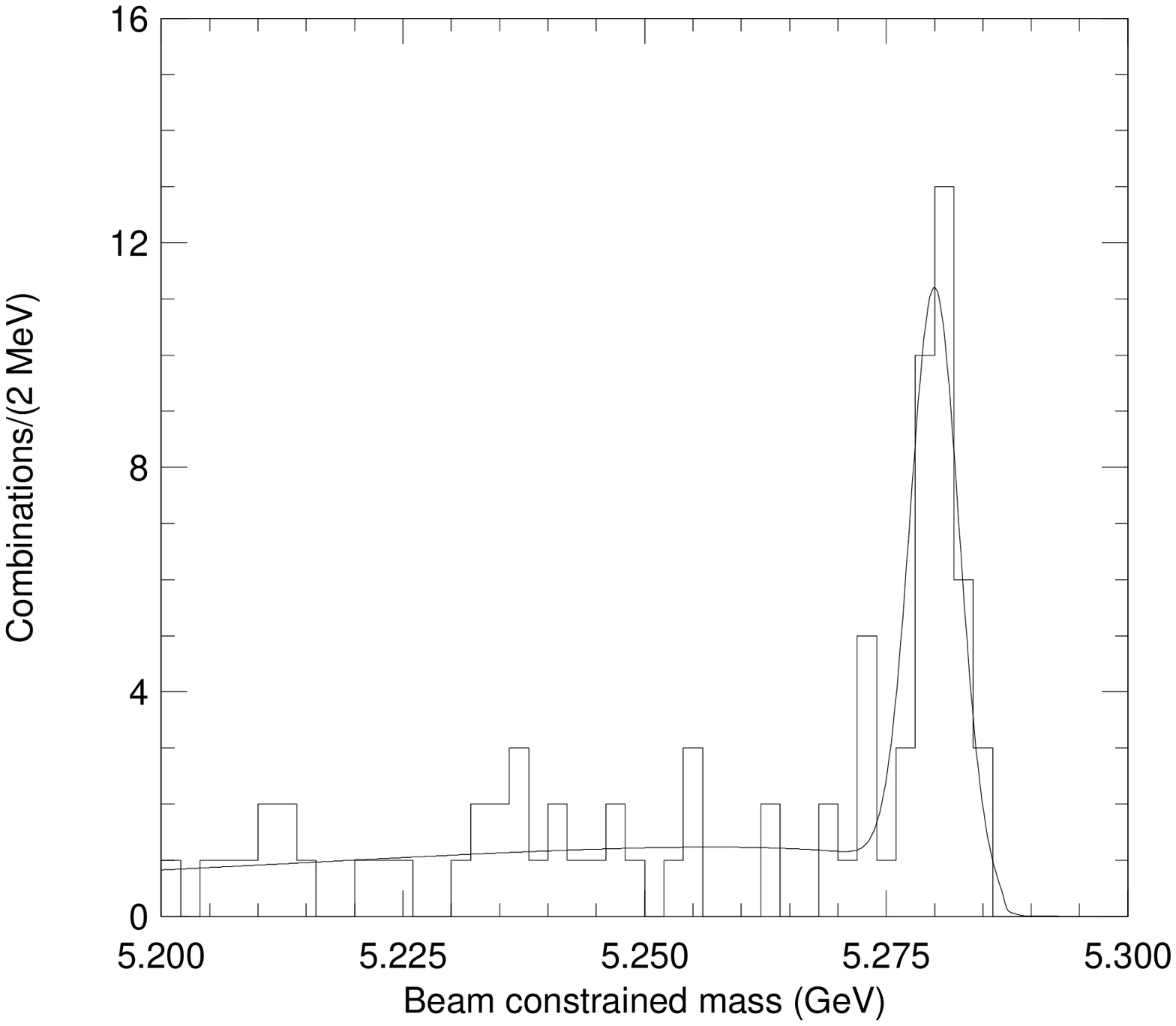}
\vskip 30pt
   \caption{ $M_{BC}$ for {\bdecb} in data }
\end{figure}

\clearpage

\begin{figure}[ht]
   \centering \leavevmode
        \epsfysize=7cm
   \epsfbox[20 143 575 699]{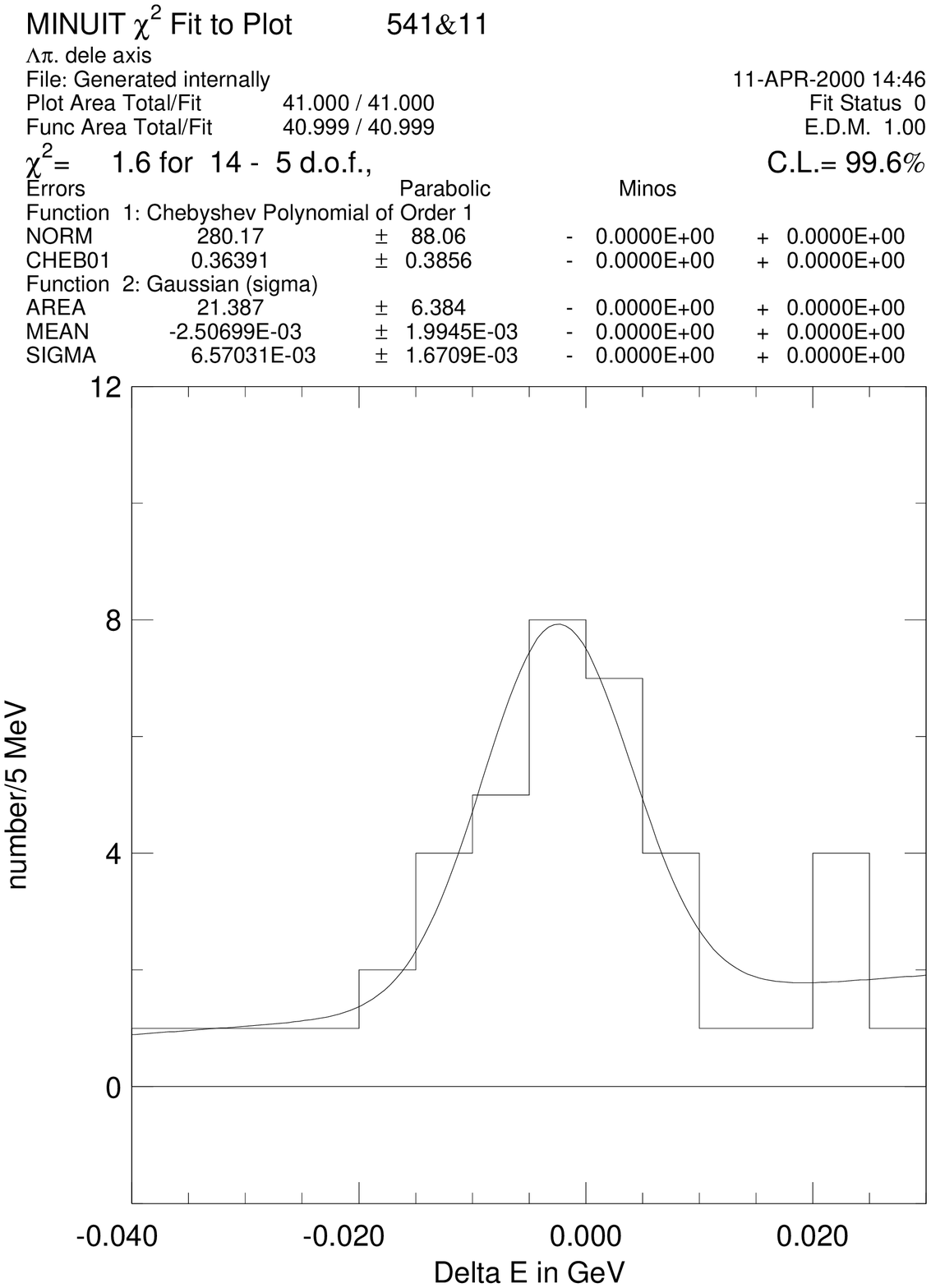}
\vskip 40pt
   \caption{$\Delta E$ (in GeV) for {\bdecb} in 
CLEO II/II.5 data 
\label{bpppidsdele} }
\end{figure}

\begin{figure}[ht]
   \centering \leavevmode
        \epsfysize=7cm
   \epsfbox[20 143 475 599]{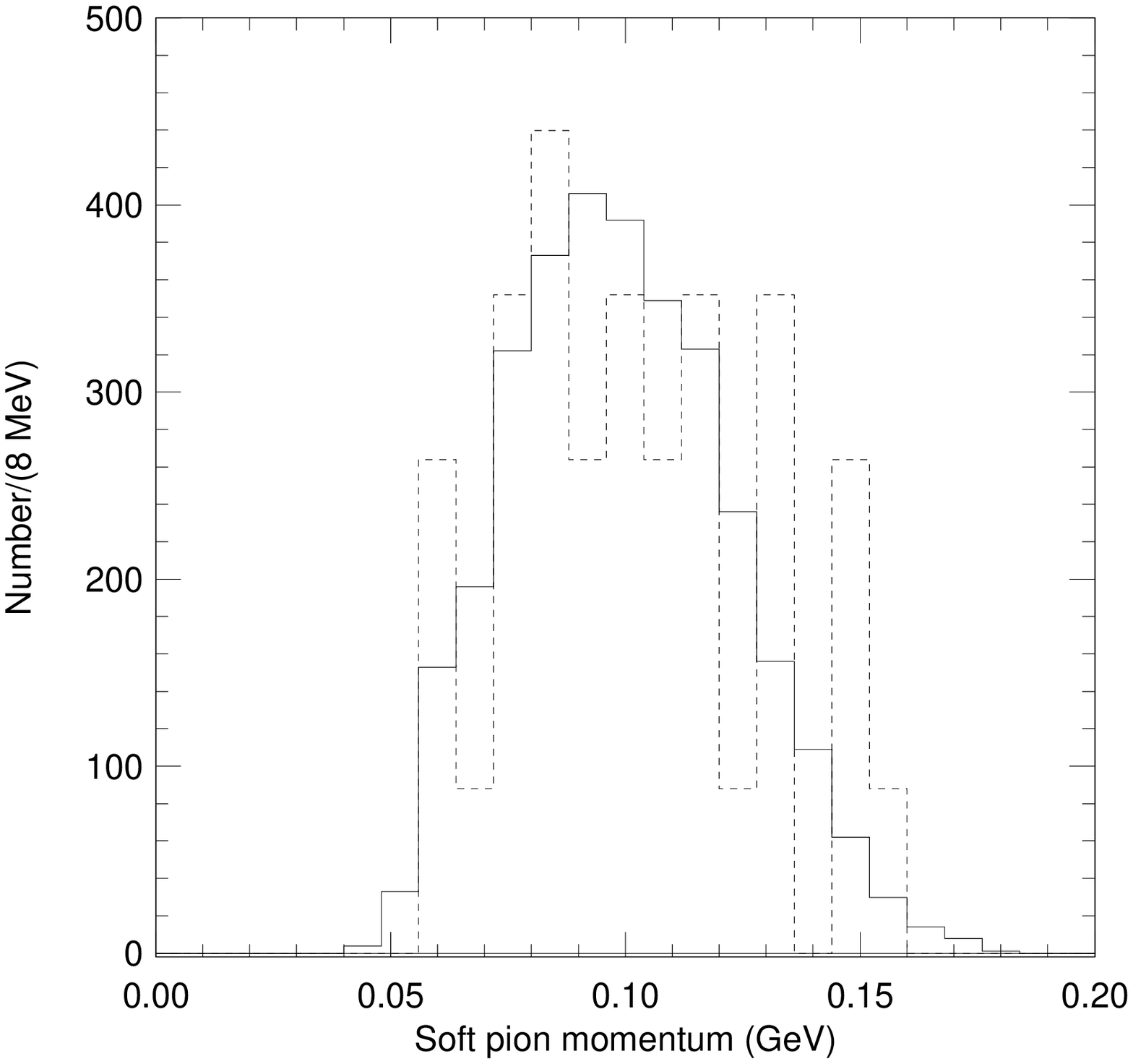}
\vskip 40pt
   \caption{{\pisoft} from {\dst} momentum 
for {\bdecb} in data and Monte Carlo 
\label{bpppidspisoft} }
\end{figure}

        We find the $\Delta$ E distribution 
for {\bdecb}, as shown in Figure \ref{bpppidsdele} 
to be Gaussian, as expected from signal MC. 
We also find the momentum spectrum for {\pisoft} 
from the {\dst} decay, as shown in 
Figure \ref{bpppidspisoft}, to agree with expectations 
from signal MC and to be very soft. In Figure 
\ref{bpppidspisoft} the solid distribution is 
CLEO II Monte Carlo and the dashed 
distribution is CLEO II/II.5 data. Only events 
with $M_{BC}$ $>$ 5.27 are plotted for both 
figures.

\vskip 60pt

\clearpage

\section{Resonant Substructure}

In order to determine whether we are 
using the correct $\epsilon_{MC}$ to  
measure ${\cal B}$({\bdecb}), we need to 
know if we are measuring {\bdecb} or a mode 
with intermediate particles decaying 
to the same combination of decay daughters. 
We therefore search for possible contributions 
to the resonant substructure of {\bdecb}. 
Any measurable resonant substructure 
would have to be subtracted from the 
inclusive measurement of ${\cal B}$({\bdecb}), 
and the $\epsilon_{MC}$ would likewise 
have to be adjusted. 

\subsection{Two-body Decay and Possible Strong Resonances} 

        We search for two types of resonances: 

        1.      A heavy charmed baryon decaying 
strongly to $\bar{p}$ + {\ds}. No significant 
peaking is observed. 

        2.      A resonance of the virtual W 
decaying to $p \bar{p} \pi^+$, {\bdecb} being a two body decay. 
No significant peaking is observed.

        We study the $M_{p + \bar{p} + \pi^+}$ spectrum 
in data to check if it is consistent with phase 
space or two-body decay. For the two-body sample 
we generate MC with a fictitious 
heavy particle that decays to $p \bar{p} \pi^+$ 
for {\bdecb}. It has a mass 
of 2.6 GeV and a width of 200 MeV. 

        No conclusive evidence is found for a measurable 
contribution to ${\cal B}$({\bdecb}) from these 
possible contributions. The $\epsilon_{MC}$ for 
{\bdecb} varies slightly depending on the resonance substructure. 
We allow a 5$\%$ systematic uncertainty to account 
for these variations.

\subsection{$\Delta$ Baryon Contributions in the 
Form of {\bdelpp} and {\bdelz} \label{del1} }

        Only the non-resonant {\bdecb} $\epsilon_{MC}$'s are shown 
in table \ref{del1}.    The $\epsilon_{MC}$ does not vary 
by more than 5 $\%$ for any of the three possible 
assumptions: {\bdecb}, {\bdelpp}, and {\bdelz}. 
Any contribution from the latter two would imply a 
lumping of events in the 1.0 to 1.3 GeV range 
for $M_{p {\pi}^+}$ or $M_{\bar{p} {\pi}^+}$. 
The $\%$ makeup of $M_{p {\pi}^+}$ and $M_{\bar{p} {\pi}^+}$ 
is nearly identical in Monte Carlo, and we quote only the former. 
There are 41 events in data in signal region.

\begin{table}[htb]
\caption{  $\%$ composition of generated (phase 
space MC) and events reconstructed (data) of 
$M_{p {\pi}^+}$ and $M_{\bar{p} {\pi}^+}$
 \label{deltas}}
\begin{center}
\begin{tabular}{|c|c|c|c|c|c|}
\hline
Mass & Signal  & Data &  & Data &  \\
in GeV & MC ($\%$) & $M_{p {\pi}^+}$ & $\%$ &
$M_{\bar{p} {\pi}^+}$ & $\%$ \\ \hline 
1.0 - 1.1 & 7.1 $\pm$ 2.1   & 0  & 0.0  & 2 & 4.9 \\ 
1.1 - 1.2 & 11.2 $\pm$ 0.8  & 3  & 7.3  & 4 & 9.8 \\ 
1.2 - 1.3 & 11.9 $\pm$ 0.7  & 8  & 19.5 & 7 & 17.1 \\ 
1.3 - 1.4 & 11.4 $\pm$ 0.6  & 8  & 19.5 & 9 & 22.0 \\ 
1.4 - 1.5 & 10.2 $\pm$ 0.6  & 11 & 26.8 & 7 & 17.1 \\ 
1.5 - 1.6 & 9.2 $\pm$ 0.6   & 3  & 7.3  & 6 & 14.6 \\ 
1.6 - 1.7 & 11.3 $\pm$ 0.6  & 3  & 7.3  & 3 & 7.3 \\ 
1.7 - 1.8 & 6.4 $\pm$ 0.6   & 3  & 7.3  & 3 & 7.3 \\ \hline  
\end{tabular}
\end{center}
\end{table}

        We place an upper limit of 
$<$ 2 events for $M_{p {\pi}^+}$, and $<$ 3 events for 
$M_{\bar{p} {\pi}^+}$ as contributions from 
{\bdelpp}, and {\bdelz} respectively. These upper limits 
are educated guesses, and {\bfseries{do not}} have any 
effect on the systematic uncertainty of our measurement 
of ${\cal B}$({\bdecb}), since our quoted value includes 
the resonant substructure. 
Our quoted systematic uncertainty allows 
for the possibility of $\Delta$ baryon contributions to the 
resonant substructure of {\bdecb}.

\section{Backgrounds}

 	We use generic Monte Carlo, which is discussed in 
Section \ref{genmc}, to study possible backgrounds to 
{\bdecb}. Several characteristics of {\bdecb} result in 
expected low backgrounds as well as an accurate Monte Carlo 
simulation:

\renewcommand{\theenumi}{\arabic{enumi}}
\begin{enumerate}
\item {\bdecb} has a proton and an antiproton as 
decay daughters, the baryon-antibaryon constraint is 
strictly applied. 
\item {\dst}'s have low $B\bar{B}$ and continuum 
backgrounds. 
\item The high combined mass of the decay daughters 
is an added suppressant of continuum backgrounds. 
\end{enumerate}

        The $\Delta$E cut suppresses modes of type 
{\bdecb}X, where X is any additional number of neutral and 
charged pions, but it does not suppress their 
combinatoric backgrounds. Since, as 
previously mentioned, {\bDnnx} decays are not accounted 
for in the generic MC sample we study, it is 
reasonable to assume that if 
75 $\%$ of the background in the range 
5.2 GeV $<$ $M_{BC}$ $<$ 5.275 GeV is not 
accounted for by the generic MC sample, 
the most likely source of this background is 
combinatoric background to {\bDnnx} decays. 
The small amount of non-{\bDnnx} background 
is shown as a solid filled region in 
Figure \ref{bpppidsgenon}, in which filled 
overlay on data is combinatoric non-{\bDnnx} background 
from generic MC.

\begin{figure}[ht]
   \centering \leavevmode
        \epsfysize=7cm
   \epsfbox[20 143 575 699]{rubierafit4.8.ps}
\vskip 10pt
   \caption{$M_{BC}$ in data and generic Monte Carlo 
for {\bdecb} (in GeV) 
 \label{bpppidsgenon} }
\end{figure}

\clearpage

\section{Systematic Uncertainties} 

	In Table \ref{bpppidssyst} we show the 
systematic uncertainties we consider to contribute 
significant errors to our measurement of ${\cal B}$({\bdecb}). 
{\bdecb} has an average of 6.6 tracks. 

\begin{table}[htb]
\caption{Estimate of systematic uncertainties (in $\%$) 
for {\bdecb}
\label{bpppidssyst}}
\begin{center}
\begin{tabular}{|c|c|}
\hline
Source & Uncertainty (in $\%$)  \\ \hline 
{\dzz} branching fractions & 0.6 \\
{\dst} branching fraction & 1.4  \\
{\dst} reconstruction & 5.0 \\ 
Monte Carlo statistics  & 5.0  \\ 
$\#$ of {\bbar}'s & 2.0  \\ 
tracking (1$\%$/track)  & 6.6  \\ 
PRLEV proton ID  & 8.0  \\ 
$\Delta$ signal contribution & 5.0 \\ 
$\Delta$ background contribution & 5.0  \\ 
Phase space versus two body & 5.0  \\  \hline 
TOTAL  & 15.5 $\%$ \\ \hline 
\end{tabular}
\end{center}
\end{table}

\chapter{Measurement of $B^0 \rightarrow D^{*-}p\bar{n}$}

	As mentioned in Chapter 4, we have selected the 
{\bDnnx} modes which are expected to have the lowest 
backgrounds as well as the highest reconstruction 
efficiency. When reconstructing {\bdeca}, however, 
we encounter a stumbling block not encountered 
in the reconstruction of {\bdecb} by virtue 
of the former mode having an antineutron-or a neutron in 
the case of its charge conjugate-as one of 
its decay daughters. Our inability to separate 
neutron showers from their backgrounds reduces 
the reconstruction efficiency for {\bdeca} by 50$\%$. 
The antineutron reconstruction efficiency, however, 
is sufficiently large (in the range of 30 to 40$\%$). 
We do not have an accurate  measurement of the antineutron 
energy, yet we do have a well measured direction 
of its shower, which we use to reconstruct 
{\bdeca}. The reconstruction method we use is 
analogous to that used in an unpublished reconstruction 
of $B \to J/\Psi K_L$ by CLEO \cite{andy2}. 
In Table \ref{showerc} we outline our 
antineutron selection criteria which we derived 
in Chapter 3.

\begin{table}[htb]
\caption{Antineutron shower selection criteria.\label{showerc} }
\begin{center}
\begin{tabular}{|c|}
\hline
Track-to-shower match level type 3 \\ 
E9OE25 $<$ cut 1\\ 
 $|(cos(\theta)|$ $<$ 0.71, or good barrel \\ 
${E}_{main}$ $>$ 500 MeV \\ 
${E}_{group}$ $>$ 800 MeV \\ \hline 
\end{tabular}
\end{center}
\end{table}

\section{Reconstruction Procedure}

	Instead of ${M}_{BC}$, we define $m_{B^0}$, in 
which we set $\Delta$E = 0 and assign the missing energy to 
the antineutron using the directional cosines of its shower. 
Since the measured shower energy for an antineutron, 
even after summing the energy in the list of nearby showers, 
fails to match the total energy of the antineutron, 
we only use the electromagnetic shower 
energy as part of our selection criteria. 

The reconstruction steps for this mode are:

\renewcommand{\theenumi}{\arabic{enumi}}
\begin{enumerate}

\item $E_{\bar{n}}$=$E_{beam}$-$E_{D^*+p}$ is assigned to the 
antineutron. This energy difference is 
the equivalent of setting $\Delta$E = 0, or 
$E_{beam}$ = $E_{reconstructed B}$.

\item 3-momentum magnitude of antineutron candidate 
$p_{\bar{n}}$=$\sqrt{E_{\bar{n}}^2-m_{\bar{n}}^2}$, 
with $m_{\bar{n}}$ = 0.9395 GeV.  

\item 3-momentum magnitude of antineutron \textbf{$p_{\bar{n}}$} 
times x,y,z directional cosines of shower 
energy are assigned to, respectively, 
x,y,z components of \textbf{$p_{\bar{n}}$}.

\item 4-momentum of B candidate $p_{B^0}$ = $p_{\bar{n}}$ + $p_{D^*}$ 
+ $p_{proton}$.

\item Mass of B candidate $m_{B^0}$ = $\sqrt{E_{beam}^2-p_{B^0}^2}$.

\end{enumerate}

        As in the {\bdecb} reconstruction, 
we use Monte Carlo samples in which the B meson decays 
according to phase space. The resonant substructure of 
{\dzdecb} is same as in {\bdecb}. 

\section{{\dsspn} in Monte Carlo}

        In reconstructing {\bdeca} we can also be 
reconstructing {\bdsspn} with {\dsspn}. The {\bdsspn} 
with {\dsspn} contribution is part of our signal, 
yet it is not a mode within the resonant substructure 
of {\bdeca}. It is a separate mode of the $B$ meson with 
${\cal B}$({\bdsspn}) = $9.6 \pm 3.4 \times 10^{-3}$ 
\cite{pdg99}. {\dsspn}, a Feynman diagram 
of which is shown in Figure \ref{dspnfeyn}. 
Yet another contribution which can be in the 
$M_{p + \bar{n}}$ distribution is 
{\bdsstpn} with {\dsspn}, which is a background. 
${\cal B}$({\bdsstpn}) = 
$2.0 \pm 0.7 \times 10^{-2}$ \cite{pdg99}. 
The effect of either on {\bdeca} 
cannot be assesed accurately since we do not 
know ${\cal B}$({{\dsspn}). If we assume 
${\cal B}$({{\dsspn}) is $\approx$ 1$\%$, 
the effect on {\bdeca} can be significant enough 
to affect our measurement of ${\cal B}$({\bdeca}).

\begin{figure}[ht]
   \centering \leavevmode
        \epsfysize=4cm
   \epsfbox[170 525 575 699]{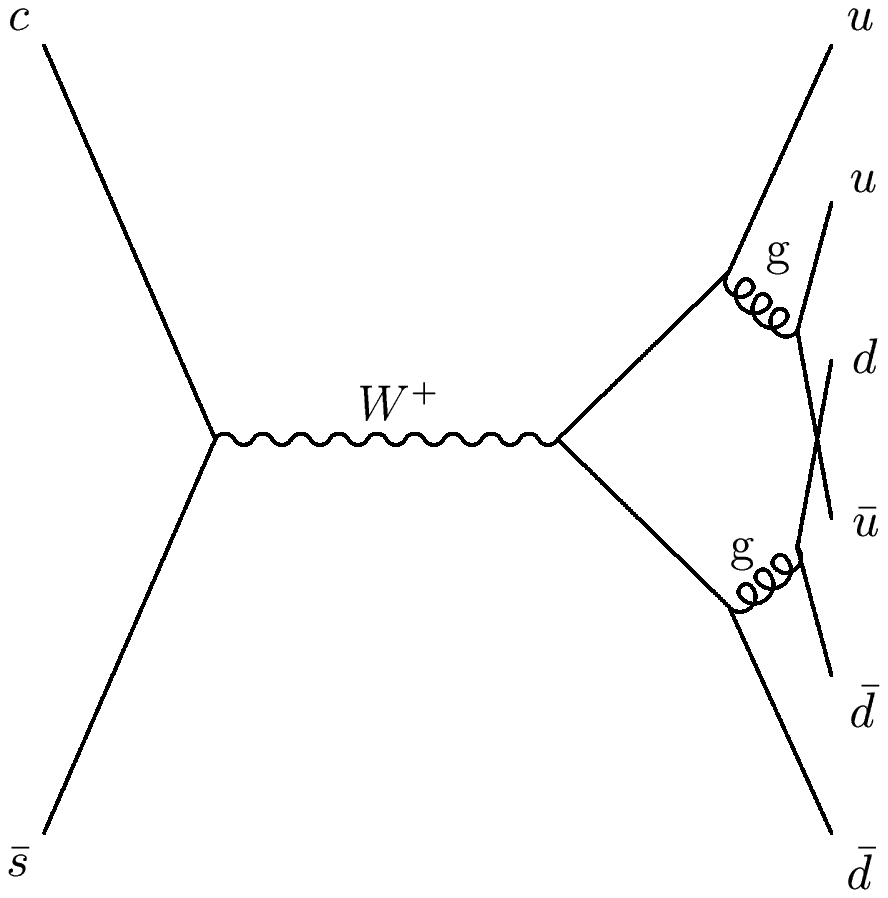}
\vskip 90pt
   \caption{A Feynman diagram for {\dsspn} } 
\label{dspnfeyn}
\end{figure}

The $M_{p + \bar{n}}$ distribution should 
not peak appreciably in {\bdeca}, and a narrow 
Gaussian peak with 5 MeV $\sigma$ in {\bdsspn} with {\dsspn}. 
The {\dsspn} Monte Carlo peak for {\bdsspn} is shown 
in Figure \ref{dspn1}.

\begin{figure}[ht]
   \centering \leavevmode
        \epsfysize=8cm
   \epsfbox[20 143 575 699]{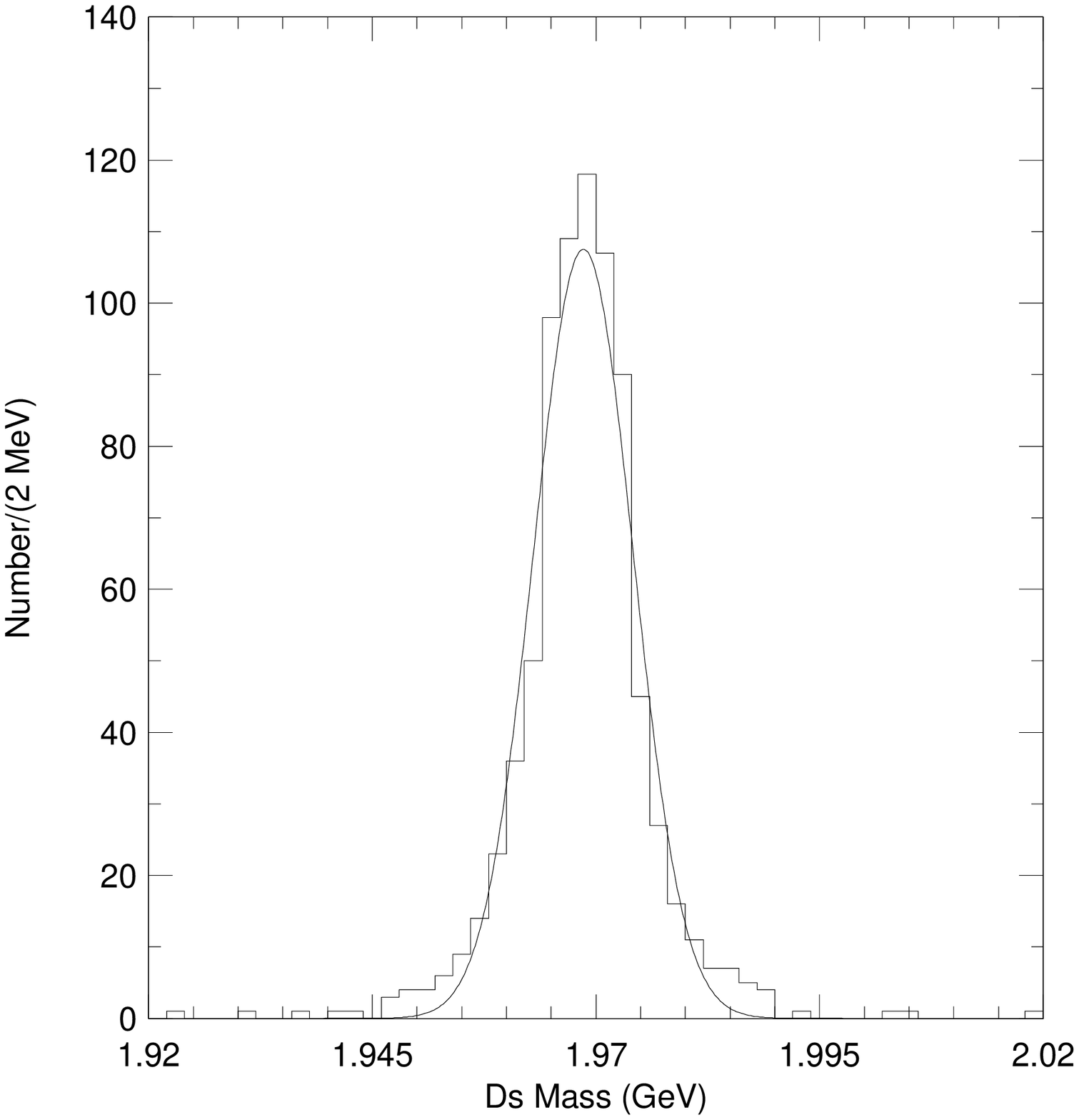}
\vskip 10pt
   \caption{$M_{p + \bar{n}}$ in GeV. 
{\bdsspn} with {\dsspn} 
\label{dspn1} }
\end{figure}

In Figure \ref{dspn2} we show the smeared $M_{p + \bar{n}}$ 
distribution resulting from reconstructing 
{\bdeca} in a signal Monte Carlo {\bdsstpn} with {\dsspn} 
sample. ${D^{*+}_{s}} \rightarrow {D^{+}_{s}} \gamma$ = 1. 
The missing soft photon will cause this background 
sample to be considerably broader than 
the {\bdsspn} with {\dsspn} signal contribution. 

\begin{figure}[ht]
   \centering \leavevmode
        \epsfysize=8cm
   \epsfbox[20 143 575 699]{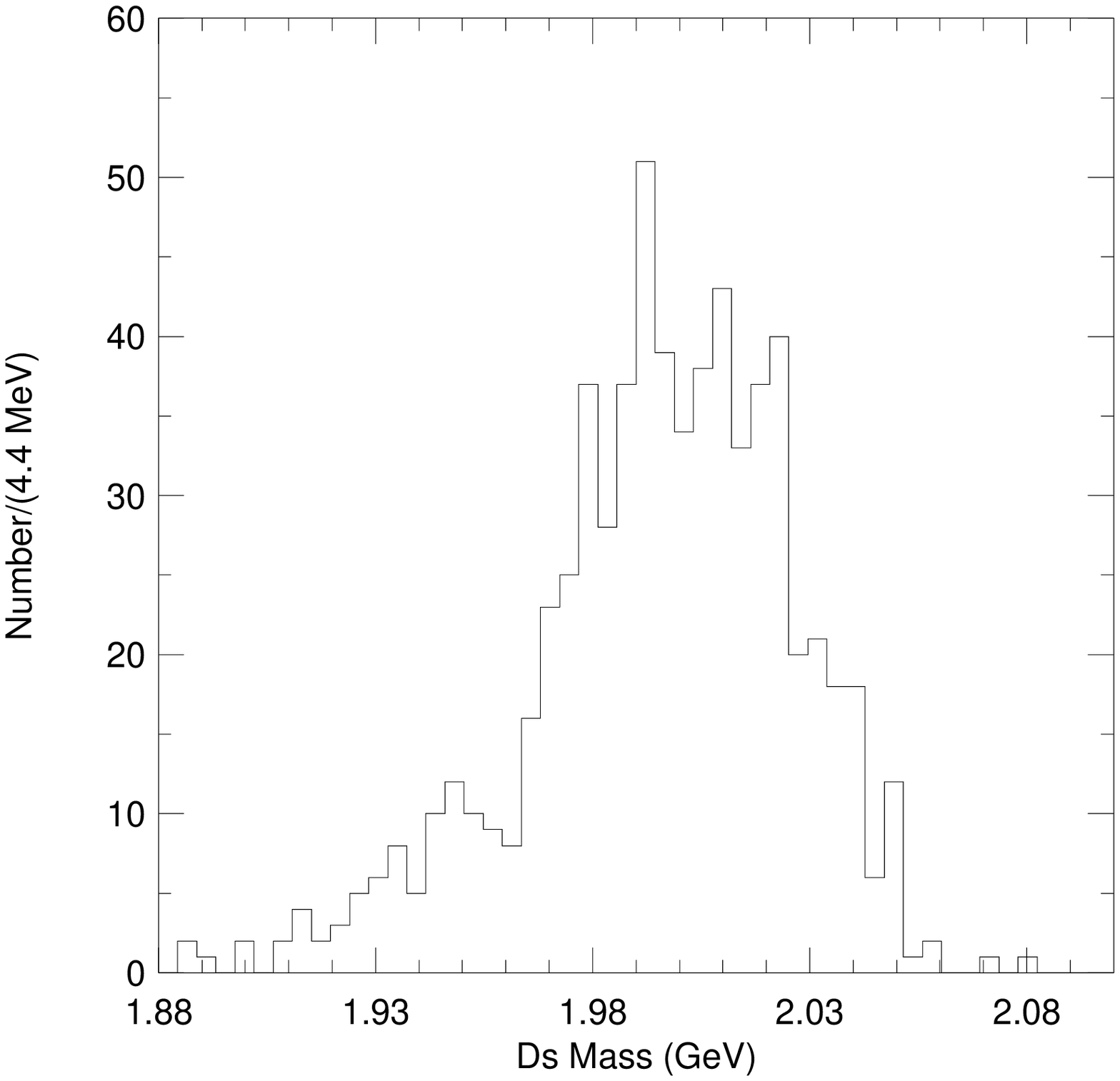}
\vskip 10pt
   \caption{$M_{p + \bar{n}}$ (in GeV) from 
a reconstruction of {\bdeca} in a signal Monte Carlo 
{\bdsstpn} with {\dsspn} \label{dspn2} }
\end{figure}

        The {\bdsstpn} contribution is a background to {\bdeca} 
as well as {\bdsspn} with {\dsspn}. 
Our choice regarding how to deal with these 
contributions is to exclude both possible contributions 
({\bdsspn} with {\dsspn} and {\bdsstpn} with {\dsspn}) 
in quoting our final value for ${\cal B}$({\bdeca}).

\begin{figure}[ht]
   \centering \leavevmode
        \epsfysize=8cm
   \epsfbox[20 143 575 699]{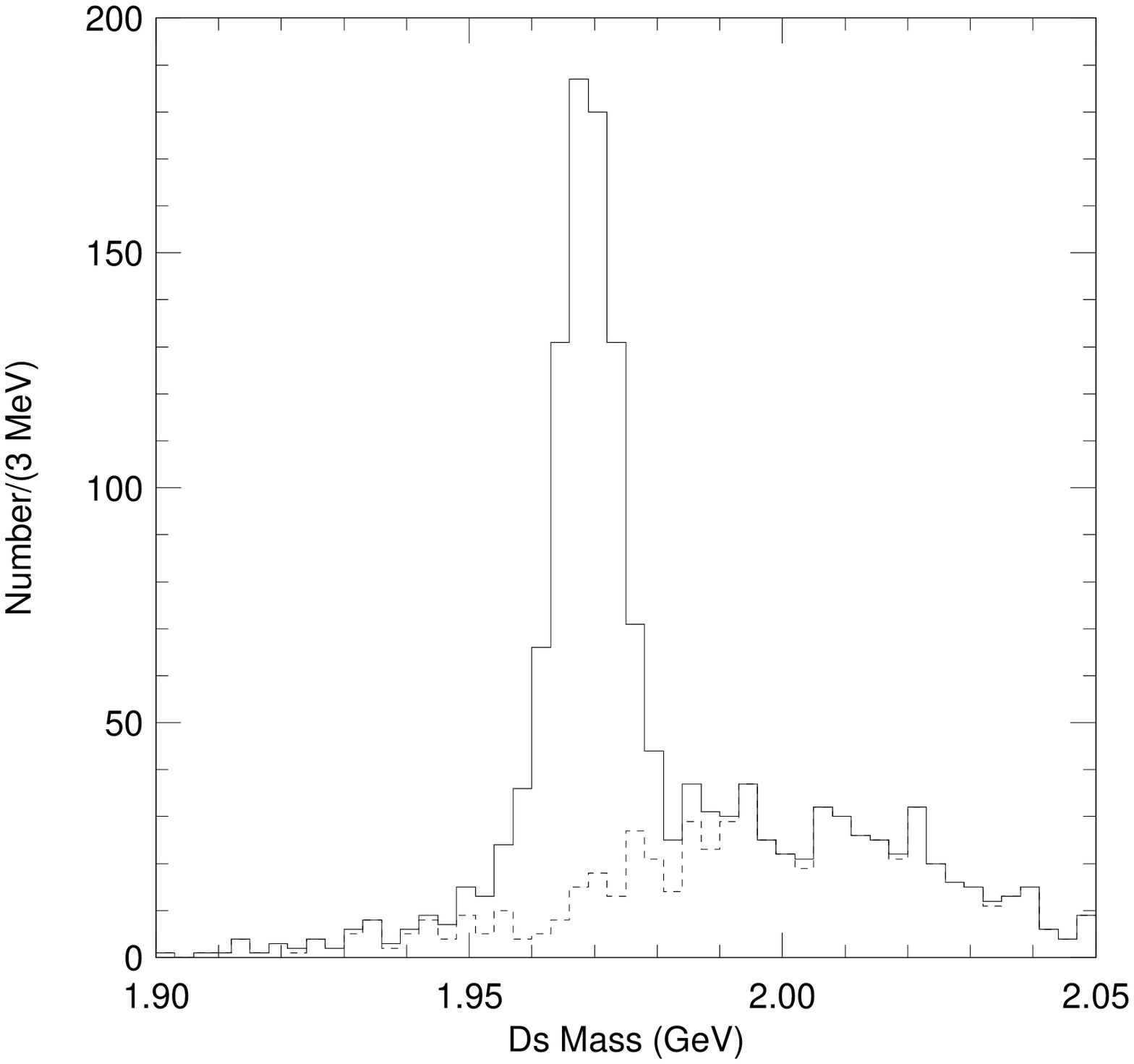}
\vskip 10pt
   \caption{
(white) is {\bdsspn} with {\dsspn}. 
(dashed) is {\bdsstpn} with {\dsspn}. 
$M_{p + \bar{n}}$ (in GeV)
\label{overl} }
\end{figure}

\begin{figure}[ht]
   \centering \leavevmode
        \epsfysize=8cm
   \epsfbox[20 143 575 699]{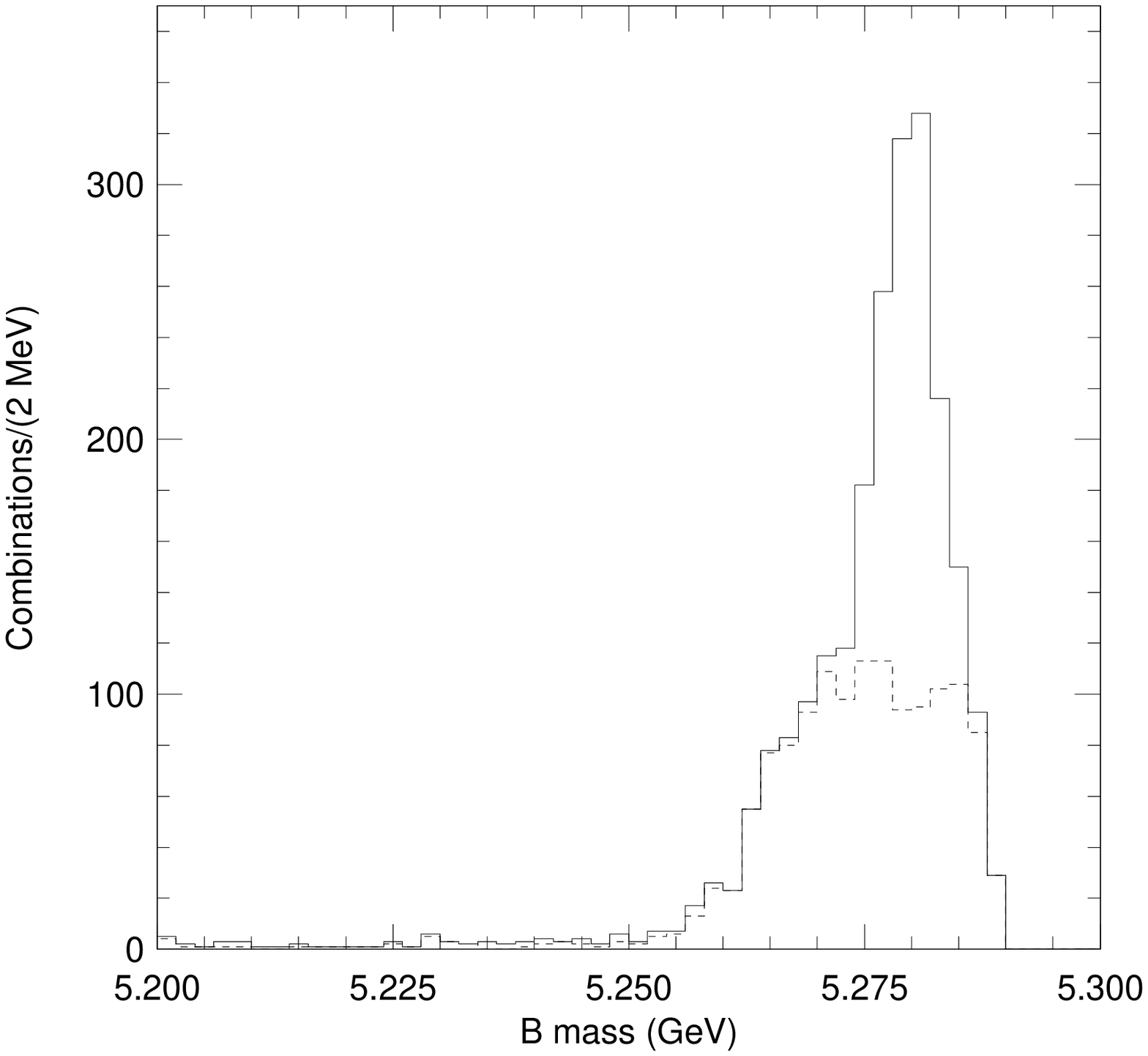}
\vskip 10pt
   \caption{
(white) is {\bdsspn} with {\dsspn}. 
(dashed) is {\bdsstpn} with {\dsspn}. 
$m_{B^0}$ (in GeV) \label{overl2} }
\end{figure}

       Both {\dsspn} contributions can be excluded 
with 1.91 GeV $<$ $M_{p + \bar{n}}$ $<$ 2.04 GeV. In 
Tables \ref{variations} and \ref{variations2} we 
outline the ${\epsilon}_{MC}$'s for {\bdeca}. 
These ${\epsilon}_{MC}$'s take into consideration 
that we are not reconstructing the conjugate 
mode-with a neutron. 

\begin{table}[htb]
\caption{ {\bdeca} signal MC ${\epsilon}_{MC}$'s and widths 
in CLEO II. 
\label{variations} }
\begin{center}
\begin{tabular}{|c|c|c|}
\hline
Mode     &  $\%$   & MeV    \\  \hline 
both {\dsspn} contributions excluded. & &  \\ \hline
{\dzdecc}& 7.47 $\pm$ 0.28 & 3.10   \\
{\dzdecb}& 2.88 $\pm$ 0.17 & 3.40   \\ 
{\dzdeca}& 3.33 $\pm$ 0.19 & 3.11   \\ \hline 
both {\dsspn} contributions included, & &  \\
without any background correction. & &  \\ \hline
{\dzdecc}& 8.12 $\pm$ 0.32 & 3.14   \\
{\dzdecb}& 3.09 $\pm$ 0.19 & 3.00   \\
{\dzdeca}& 3.52 $\pm$ 0.20 & 3.05   \\ \hline
\end{tabular}
\end{center}
\end{table}

\begin{table}[htb]
\caption{{\bdeca} signal MC ${\epsilon}_{MC}$'s and widths 
in CLEO II.5  
\label{variations2} }
\begin{center}
\begin{tabular}{|c|c|c|}
\hline
Mode     &  $\%$   & MeV     \\  \hline
Both {\dsspn} contributions excluded. & &  \\ \hline
{\dzdecc}& 4.51 $\pm$ 0.17 & 2.95  \\ 
{\dzdecb}& 1.60 $\pm$ 0.13 & 3.23  \\ 
{\dzdeca}& 1.68 $\pm$ 0.14 & 2.79 \\ \hline 
Both {\dsspn} contributions included, & &  \\
without any background correction. & &  \\ \hline
{\dzdecc}& 4.99 $\pm$ 0.18 & 2.92   \\ 
{\dzdecb}& 1.82 $\pm$ 0.14 & 3.35 \\ 
{\dzdeca}& 1.83 $\pm$ 0.15 & 2.78  \\ \hline
\end{tabular}
\end{center}
\end{table}

\clearpage

\section{Results in Data}

        We look for {\bdeca} with or without the mass 
region 1.91 GeV $<$ $M_{p + \bar{n}}$ $<$ 2.04 GeV. We 
compare in Figure \ref{bpndsoverl} these two cases in CLEO II/II.5. 

\begin{figure}[ht]
   \centering \leavevmode
        \epsfysize=8cm
   \epsfbox[20 143 575 699]{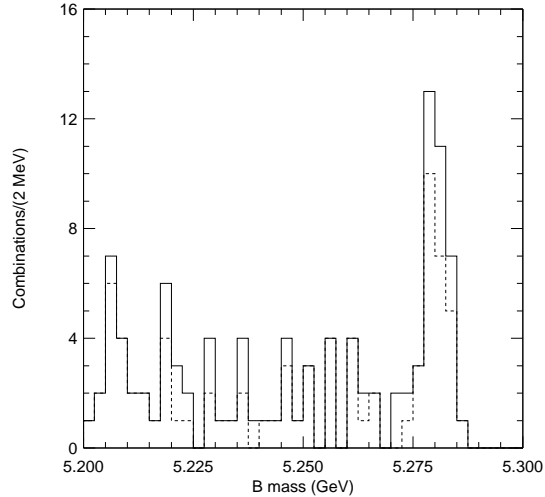}
\vskip 10pt
   \caption{(white) is inclusive of all contributions. 
(dashed) is after exclusion of both {\dsspn} contributions 
\label{bpndsoverl} }
\end{figure}

        The {\bdeca} $m_{B^0}$ distribution is 
less statistically significant without the {\dsspn} 
contributions. However, the cost of lost events for the 
purposes of our measurement is warranted for the 
following reasons:

        1.      The signal events we are left with have an 
insignificant probability of being something other 
than {\bdeca}. 

        2.      We properly account for the lost events 
in our ${\epsilon}_{MC}$'s.

        In Figure \ref{bpndsoverl2} we show $m_{B^0}$ 
for ON resonance and OFF resonance for the case we use 
to quote ${\cal B}$({\bdeca}): without both contributions. 

\begin{figure}[ht]
   \centering \leavevmode
        \epsfysize=8cm
   \epsfbox[20 143 575 699]{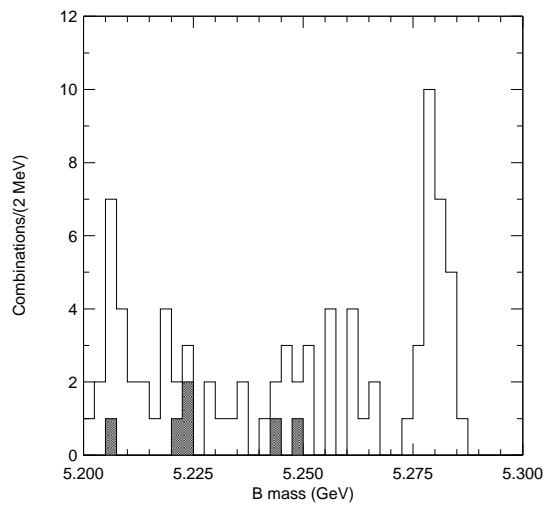}
\vskip 10pt
   \caption{(white) is ON resonance, (solid) is OFF resonance
\label{bpndsoverl2} }
\end{figure}

        The number of OFF resonance events is not 
significant enough to affect our measurement. 
We choose not to subtract these from the 
ON resonance distribution.

        In Figure \ref{bpnds20} and 
Figure \ref{bpnds25} we show $m_{B^0}$ 
distributions by dataset.

\begin{figure}[ht]
   \centering \leavevmode
        \epsfysize=8cm
   \epsfbox[20 143 575 699]{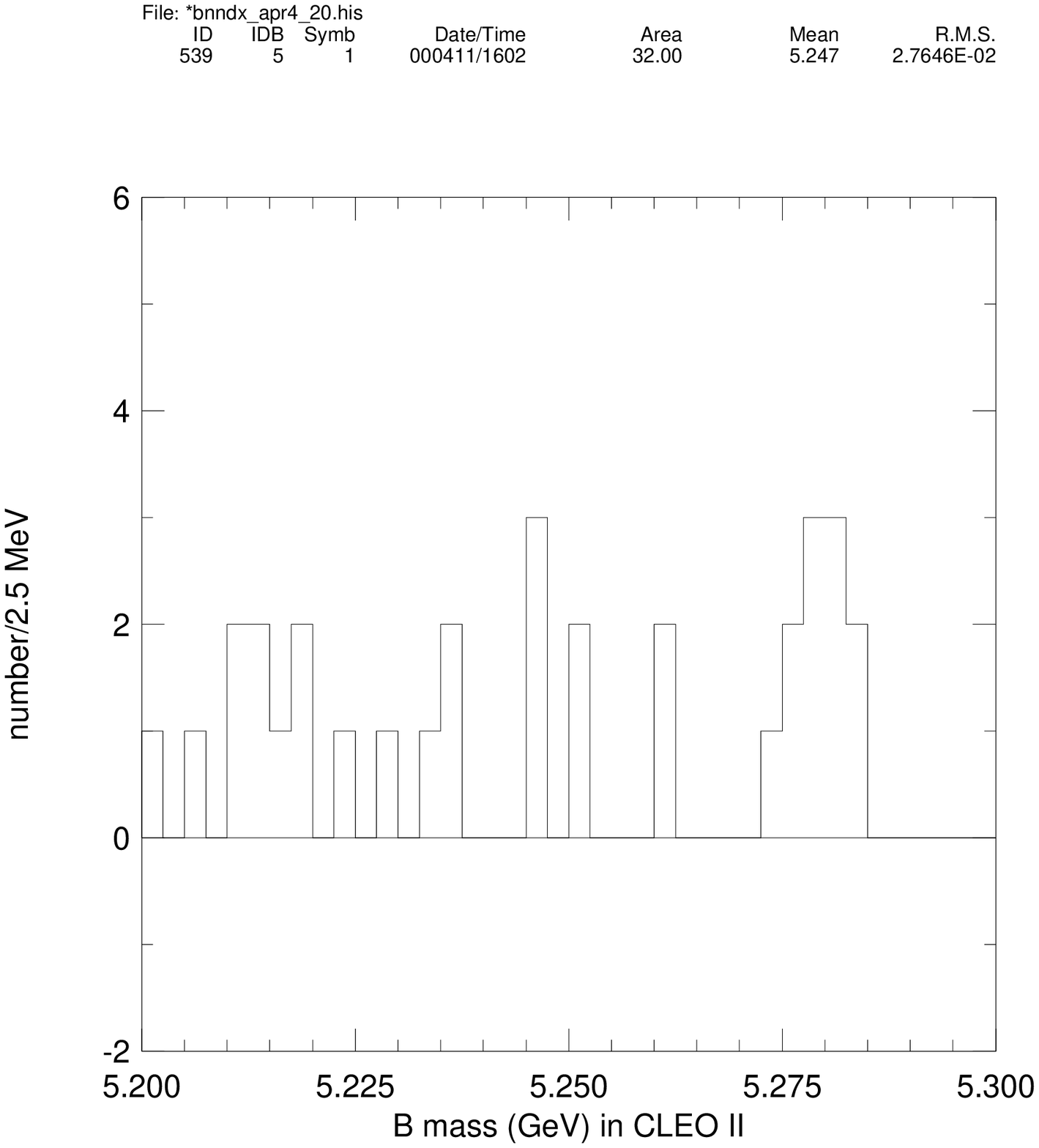}
\vskip 10pt
   \caption{$M_{B^0}$ for {\bdeca} in 
CLEO II ON resonance data
\label{bpnds20} } 
\end{figure}

\begin{figure}[ht]
   \centering \leavevmode
        \epsfysize=8cm
   \epsfbox[20 143 575 699]{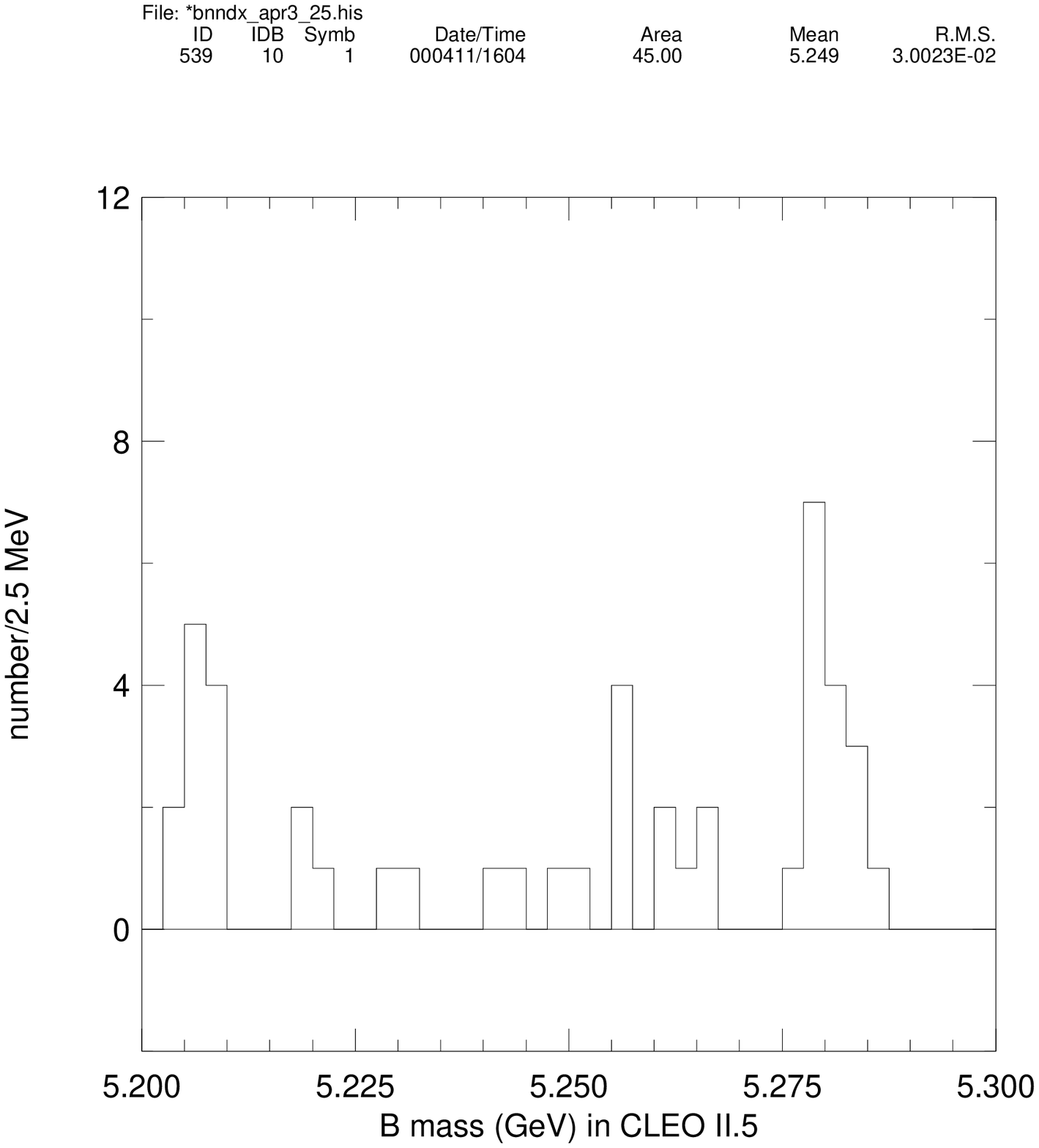}
\vskip 10pt
   \caption{$M_{B^0}$ for {\bdeca} in 
CLEO II.5 ON resonance data
\label{bpnds25} } 
\end{figure}

\clearpage

\section{{\dsspn} in Data}

        In Figure \ref{dspnmass} we plot $M_{p + \bar{n}}$ 
for the region where we expect to see {\dsspn} as 
a small width signal for {\bdsspn}, and as a shallow 
background for {\bdsstpn}. The horizontal lines 
demarcate the mass region 
1.91 GeV $<$ $M_{p + \bar{n}}$ $<$ 2.04 GeV, 
which we are excluding. 
Events with $m_{B^0}$ $>$ 5.275 GeV region 
are shown in Figure \ref{dspnmass}. 
The vertical lines demarcate the mass 
region 1.91 GeV $<$ $M_{p + \bar{n}}$ $<$ 2.04 GeV, 
which we are excluding. Not shown are events with 
$M_{p + \bar{n}}$ $>$ 2.27 GeV. We expect to 
find $\approx$ 10$\%$ of {\bdeca} signal events 
in this region, which translates to 3 events.

\begin{figure}[ht]
   \centering \leavevmode
        \epsfysize=8cm
   \epsfbox[20 143 575 699]{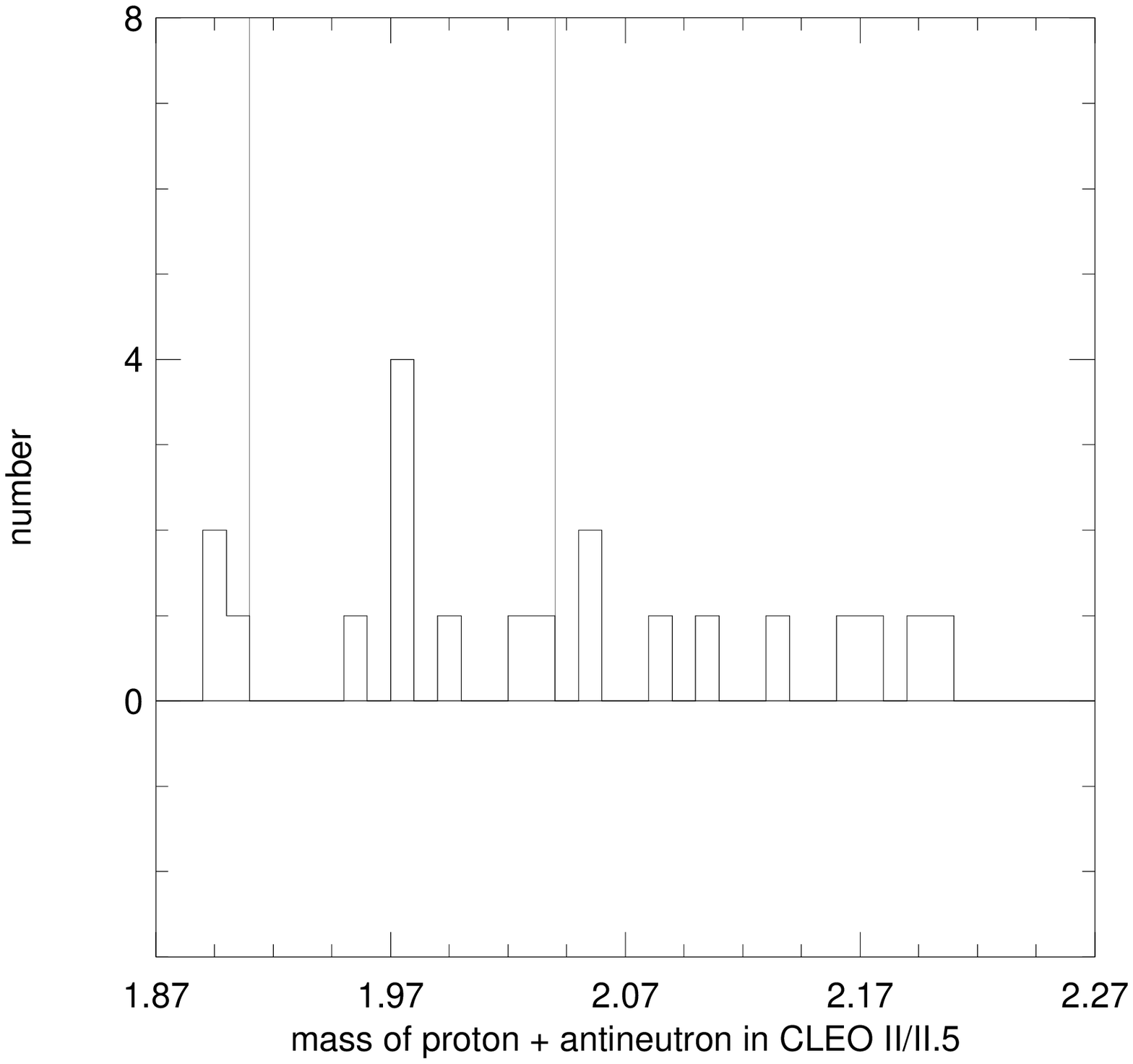}
\vskip 10pt
   \caption{ $M_{p + \bar{n}}$ for {\bdeca} (in GeV) 
in CLEO II/II.5
 \label{dspnmass} }
\end{figure}

        We find 8 events where a {\dsspn} 
signal is expected. This sample is not sufficiently 
significant to label these events as signal. 
The scatter plot in Figure \ref{dspnscatter} 
contains more information than Figure \ref{dspnmass}: 
there are other regions in this plot that are 
equally statistically significant to the region 
where we expect to find {\dsspn}. If we are to claim 
a signal in one, we should be able to claim one 
in any of these others. 

\begin{figure}[ht]
   \centering \leavevmode
        \epsfysize=10cm
   \epsfbox[20 143 575 699]{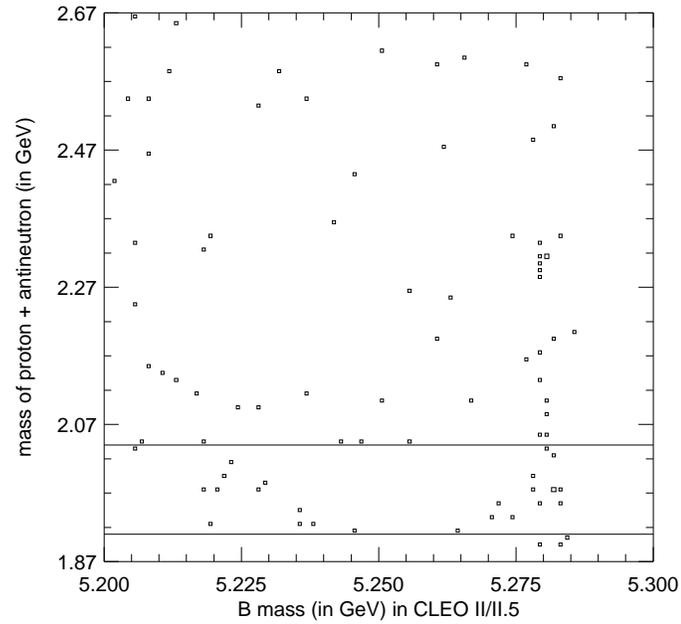}
\vskip 10pt
   \caption{ $M_{p + \bar{n}}$ 
vs. $m_{B^0}$ (both in GeV) \label{dspnscatter}} 
\end{figure}

        We can, however, estimate an upper limit 
for {\dsspn}. The most conservative estimate is to 
call all 8 events in the expected region signal, 
which leads to a value of 11.5 for a 90$\%$ confidence 
level measurement. 11.5 signal events correspond 
to ${\cal B}$({\dsspn}) $<$ 5.4 $\%$ using 
the PDG value for ${\cal B}$({{\bdsspn}).

\clearpage

\section{${\cal B}$({\bdeca}) Measurements.}

        We use the same equation to calculate 
${\cal B}$({\bdeca}) as we did for {\bdecb}:
\vskip 20pt

{\centerline{ ${\cal B}$({\bdecb}) = 
{\large{ $\frac{ Fitted Yield }{ 
{ (2 \times B^0\bar{B^0} \times \epsilon_{MC}) }_{CLEO II} + 
{ (2 \times B^0\bar{B^0} \times \epsilon_{MC}) }_{CLEO II.5} }$
}}}}
\vskip 20pt

        As we mentioned when calculating ${\cal B}$({\bdecb}), 
the product ${(B\bar{B}\times\epsilon_{MC})}$ 
is almost the same for CLEO II and CLEO II.5. 
The number of events for this mode for 
$M_{BC}$ $>$ 5.275 GeV is 10 for CLEO II, and 16 for CLEO II.5. 

        The combined $m_{B^0}$ distribution can be 
fitted to yield results that are consistent with 
Monte Carlo expectations. 

\begin{figure}[ht]
   \centering \leavevmode
        \epsfysize=8cm
   \epsfbox[20 143 575 699]{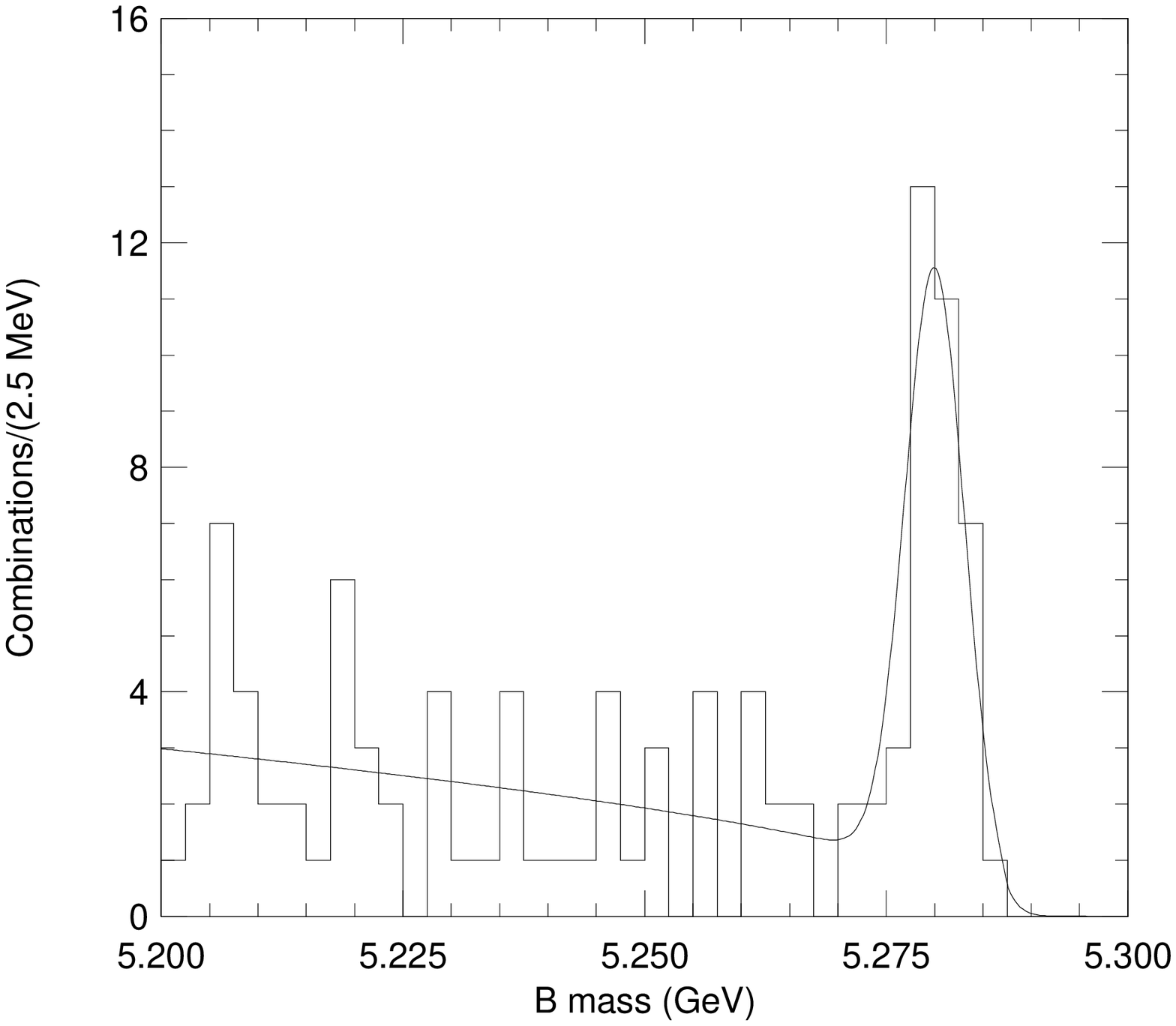}
\vskip 20pt
   \caption{Inclusive $m_{B^0}$ in data for {\bdeca} (in GeV) 
}
\end{figure}

\begin{figure}[ht]
   \centering \leavevmode
        \epsfysize=8cm
   \epsfbox[20 143 575 699]{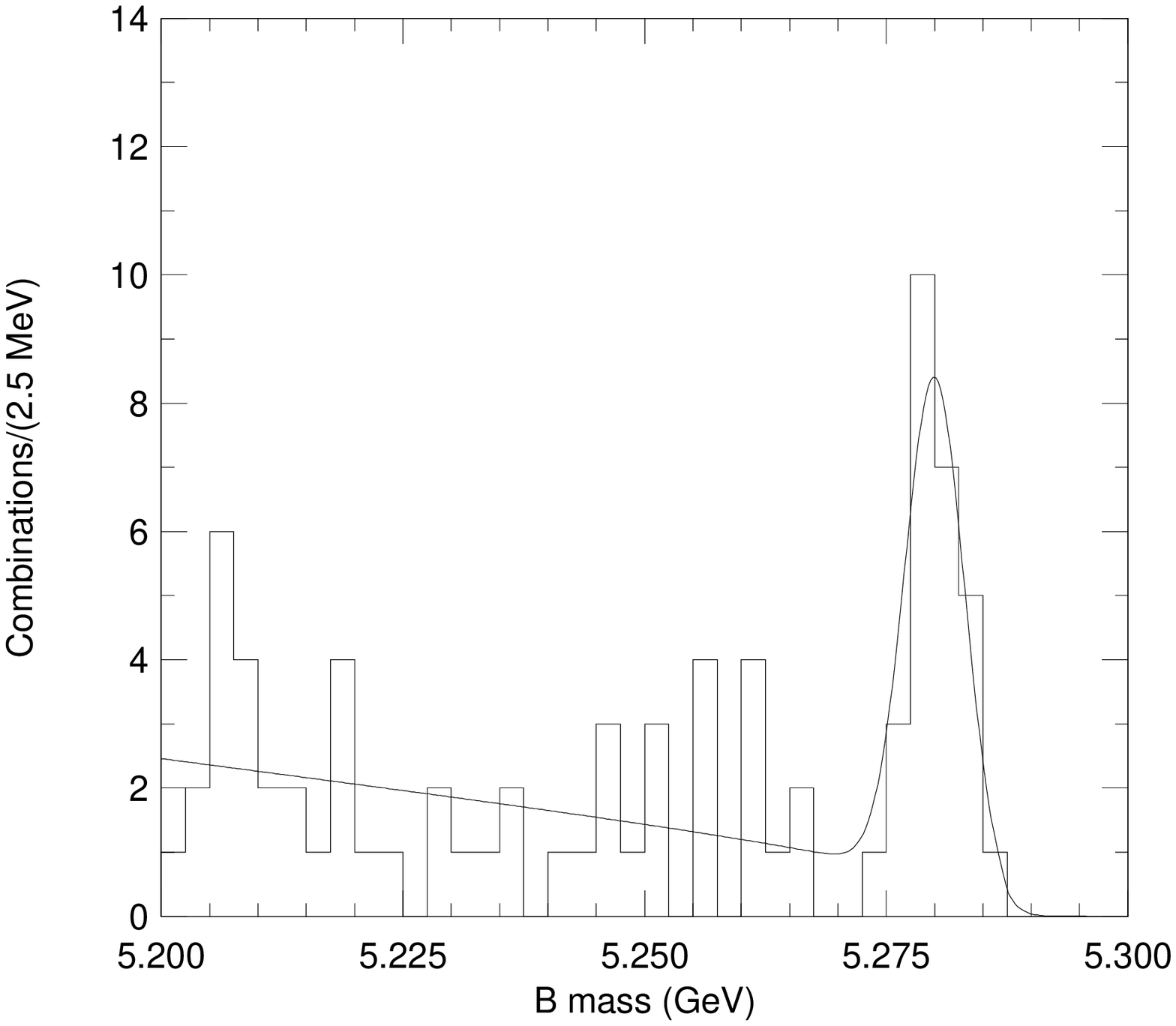}
\vskip 20pt
   \caption{ $m_{B^0}$ in data for {\bdeca} in GeV 
excluding both 
{\dsspn} contributions }
\end{figure}

\begin{table}[htb]
\caption{ Measurements of {\bdeca} branching fractions
\label{brs} }
\begin{center}
\begin{tabular}{|c|c|c|}
\hline
Mode & Yield & Branching Fraction \\ \hline
{\bdeca} inclusive & ${32.94}^{+6.56}_{-5.92}$ & 
${14.9}^{+3.0}_{-2.7}$ $\times$ $10^{-4}$ \\ 
{\bdeca} without both contributions & ${24.02}^{+5.60}_{-4.96}$ &  
${11.9}^{+2.8}_{-2.5} \times 10^{-4}$ \\ \hline 
\end{tabular}
\end{center}
\end{table}

        The difference in branching fractions does not imply 
that what has been excluded is only {\dsspn}. We do not 
attempt to derive an upper limit by this method. 
As in the case of our measurement of {\bdecb}, 
we find the {\bdeca} signal events are consistent with 
phase space decay. We look for a resonance of type 
${(cud)+(d\bar{d})}_{strong} \rightarrow \bar{n} + D^{*-}$, 
and fail to find it.

        The fits are quoted in Table \ref{pnfits}. 
$m_{B^0}$ is fitted to a single Gaussian 
for signal, and an Argus function for background. The offset 
is fixed at 0.0025, and the $E_{beam}$ is fixed at 5.29 GeV 
in all the fits. 
As in the case of {\bdecb}, in table \ref{pnfits}, 
leaving some parameters fixed, but not others 
allows us to test the validity of our results. 
The excluded region is 
1.91 GeV $<$ $M_{p + \bar{n}}$ $<$ 2.04 GeV.

\begin{table}[htb]
\caption{Results of various fits to $m_{B^0}$ for 
{\bdeca} with or without the {\dsspn} contributions
\label{pnfits} }
\begin{center}
\begin{tabular}{|c|c|c|}
\hline
Fitting options & No {\dsspn} & Inclusive \\ \hline
floating values & & \\ \hline
$m_{B^0}$ (MeV) & 
${5,280.3}^{+0.60}_{-0.61}$ &
${5,280.5}^{+0.50}_{-0.52}$ \\
$\sigma$  (MeV) & 
${2.59}^{+0.49}_{-0.41}$ & 
${2.43}^{+0.46}_{-0.37}$ \\
Fit Yield & 
${23.57}^{+5.61}_{-4.97}$ & 
${31.48}^{+6.52}_{-5.86}$ \\ \hline
fixed $m_{B^0}$ = 5.28 GeV & & \\ \hline 
$\sigma$  (MeV) & 
${2.62}^{+0.49}_{-0.42}$ & 
${2.55}^{+0.47}_{-0.40}$ \\
Fit Yield & 
${23.67}^{+5.62}_{-4.99}$ & 
${31.97}^{+6.57}_{-5.93}$ \\ \hline 
fixed $\sigma$ = 3.07 MeV & &\\ \hline 
$m_{B^0}$ (in MeV) & 
${5,280.3}^{+0.68}_{-0.67}$ & 
${5,280.3}^{+0.60}_{-0.60}$ \\ 
Fit Yield & 
${23.97}^{+5.60}_{-4.96}$ & 
${32.75}^{+6.56}_{-5.91}$ \\ \hline 
fixed $m_{B^0}$ and $\sigma$ & &  \\ \hline 
Fit Yield & 
${24.02}^{+5.60}_{-4.96}$ & 
${32.94}^{+6.56}_{-5.92}$ \\ 
${\cal B}$({\bdeca}) $\times$ $10^{-4}$: & & \\  
From raw yield &
${11.9}^{+2.8}_{-2.5}$ & 
${14.9}^{+3.0}_{-2.7}$ \\
With MC Correction Factor &
${14.8}^{+3.5}_{-3.1}$ & 
${18.3}^{+3.7}_{-3.3}$ \\  \hline 
\end{tabular}
\end{center}
\end{table}

        In the following section we discuss the 
correction factor we apply to the raw ${\cal B}$({\bdeca}) 
due to a discrepancy between data and Monte Carlo 
in the $\epsilon_{\bar{p}}$ for 
antiproton annihilation showers. The Correction Factor 
we use increases the raw ${\cal B}$({\bdeca}) quoted 
in Table \ref{pnfits} by a factor of 1.21 with an 
added systematic uncertainty of 4$\%$.

\section{Correction Factor}

	The Monte Carlo we use has not been optimized 
to model nucleon-antinucleon annihilation. We find a 
discrepancy between the reconstruction efficiency 
for data and Monte Carlo for antiprotons. 
We consider the discrepancy 
credible and change the antineutron efficiency 
using a Correction Factor. We assume that the 
Monte Carlo fails to model antineutrons by 
the same amount as it does for antiprotons, even 
though the quark content of antiprotons and 
antineutrons is different. 

	We define ${\epsilon}_{annihilation}$ as the 
efficiency for an antibaryon shower to pass the 
annihilation shower cuts outlined in Table \ref{showercuts}. 
This quantity will be different for antiprotons 
and antineutrons. In the former case we choose 
a sample which is inevitably biased by the momentum 
range in which we can separate antiprotons from other 
charged tracks. In the latter case we 
use a sample which has shower backgrounds.

        We define a correction factor to account for 
the discrepancy between data and Monte Carlo for 
annihilation showers as: 
\vskip 20pt

{\centerline{ Correction Factor = 
{\large{ $\frac{ {\epsilon}_{annihilation,MC} }
{ {\epsilon}_{annihilation,Data} }$ }}
}}

\vskip 20 pt 

	This Correction Factor is found for antiproton 
momentum bins in a range consistent with the 
momentum range of antineutrons in {\bdeca} as 
found from the energy assigned to the antineutron 
candidate shower after assuming the antineutron 
mass. The number of antineutrons in each of these 
momentum bins is used to weigh the contribution of 
each bin to the Correction Factor, which is defined 
for the entire range.

\section{Use of a $\bar{\Lambda}$ Sample}

	The selection of a clean sample of antiprotons 
is achieved by reconstructing $\bar{\Lambda}$'s. In the 
decay $\bar{\Lambda} \to \bar{p}\pi^+$, due to the 
large difference in mass between the $\bar{p}$ and 
the $\pi^+$, the kinematics of the decay allow for 
a clear separation of the $\bar{p}$ tracks from 
the $\pi^+$ tracks. Furthermore, the single Gaussian width 
of the $m_{\bar{\Lambda}}$ distribution is 1 MeV 
and is affected by backgrounds insignificantly. 

	Antiprotons are selected with the same criteria 
used in the reconstruction of {\bdecb} except for 
the use of $L_{proton}$ $>$ 0.1. 
$\Lambda$'s are from the KNVF package \cite{cpp}. 
The $\bar{\Lambda}$ selection criteria is as shown in 
Table \ref{lambdas}.

\begin{table}[htb]
\caption{$\Lambda$ selection criteria\label{lambdas} }
\begin{center}
\begin{tabular}{|c|}
\hline
Both tracks pass TRKMNG flag tng(track) $\ge$ 0 \\ 
${\chi}^{2}$ of two tracks to form a vertex $\le$ 30\\
${\chi}^{2}$ for $\Lambda$ to point back to 
the interaction region $\le$ 30 \\
Significance of the two dimensional 
flight distance $<$ 3$\sigma$ \\
Flight distance $\ge$ 0.005 meter \\ \hline    
\end{tabular}
\end{center}
\end{table}

        We count the number of $\Lambda$'s in 
the signal region after applying a double Gaussian 
fit with fixed parameters as shown in Table \ref{lambdafit}.

\begin{table}[htb]
\caption{$\Lambda$ double Gaussian fixed parameters
\label{lambdafit} }
\begin{center}
\begin{tabular}{|c|c|}
\hline
Mass &  1.1158 GeV \\ 
${\sigma}^{1}$ & 0.00162 \\ 
Area2/area1 & 0.46 \\  
${\Delta}_{MEAN}$  & 0 \\ 
${\sigma}^{2}$/${\sigma}^{1}$ & 0.425 \\ \hline  
\end{tabular}
\end{center}
\end{table}

        In Table \ref{anprot1} we find the $\%$ of 
antiproton showers in a $\Lambda$ sample which pass 
the annihilation shower cuts. 
In Table \ref{anprot1} we find ${\epsilon}_{annihilation}$ 
for MC and Data $\bar{\Lambda}$ samples. The samples are 
a combination of ON and OFF resonance. The ON and 
OFF resonance ${\epsilon}_{annihilation}$'s were found to 
yield the same results.

\begin{table}[htb]
\caption{${\epsilon}_{annihilation}$: ($\%$) 
of match 1 or 2 antiprotons in 
$\bar{\Lambda}$'s passing 
annihilation shower selection cuts for six
momentum spectra
\label{anprot1} }
\begin{center}
\begin{tabular}{|c|c|c|c|c|}
\hline
& CLEO II  & CLEO II & CLEO II.5 & CLEO II.5  \\  
Momentum range & MC & Data & MC & Data \\ 
& $\%$ & $\%$ & $\%$ & $\%$ \\  \hline
300-500 MeV 
& 28.3 $\pm$ 1.3 & 13.2 $\pm$ 0.3 & 26.5 $\pm$ 1.0 
& 13.3 $\pm$ 0.2 \\ 
500-700 MeV 
& 65.1 $\pm$ 1.7 & 52.7 $\pm$ 0.9 & 65.0 $\pm$ 2.6 
& 52.2 $\pm$ 0.6 \\
700-900 MeV 
& 70.8 $\pm$ 1.8 & 60.6 $\pm$ 1.0 & 71.5 $\pm$ 2.8 
& 60.9 $\pm$ 0.7 \\
900-1100 MeV 
& 65.3 $\pm$ 1.7 & 52.9 $\pm$ 0.9 & 63.4 $\pm$ 2.5 
& 50.7 $\pm$ 0.6 \\
1100-1300 MeV 
& 50.6 $\pm$ 2.0 & 41.9 $\pm$ 1.0 & 52.2 $\pm$ 2.1 
& 41.4 $\pm$ 0.7 \\
1300-1700 MeV 
& 47.4 $\pm$ 1.9 & 44.0 $\pm$ 1.0 & 46.3 $\pm$ 1.8 
& 41.0 $\pm$ 0.7 \\
1700-2100 MeV 
& 56.0 $\pm$ 2.0 & 54.0 $\pm$ 1.0 & 55.1 $\pm$ 1.9 
& 53.2 $\pm$ 0.7 \\ \hline
\end{tabular}
\end{center}
\end{table}

        The Correction Factor is weighed according to 
the fraction of antineutrons in a momentum range as 
produced in Monte Carlo. The solid distribution 
in Figure \ref{ancorrection} is data and the dashed 
distribution is signal Monte Carlo.

\clearpage

\begin{figure}[ht]
   \centering \leavevmode
        \epsfysize=8cm
   \epsfbox[20 143 575 699]{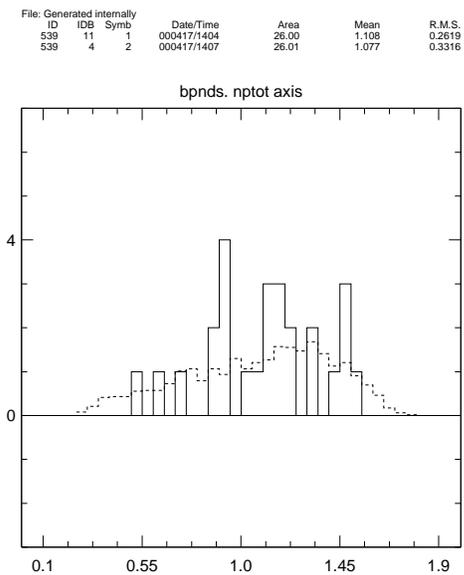}
   \caption{ \label{ancorrection} Antineutron momentum 
distribution (in GeV) }
\end{figure}

\clearpage

        In Table \ref{anprot2} we calculate the Correction 
Factor as previously defined, as well as the fraction of 
antineutrons found in signal Monte Carlo for each 
momentum range. 
 
\begin{table}[htb]
\caption{Correction Factor (C.F.) as a function of $p_{proton}$ 
and $\%$ of antineutrons found in signal Monte Carlo for each 
momentum range in CLEO II/II.5
\label{anprot2} }
\begin{center}
\begin{tabular}{|c|c|c|}
\hline
Momentum range & $\%$ of $\bar{n}$ in MC & C.F.\\  \hline
300-500 MeV 
& 9.1 & 2.07 $\pm$ 0.11 \\ 
500-700 MeV 
& 10.0 & 1.25 $\pm$ 0.04 \\
700-900 MeV 
& 14.7 & 1.17 $\pm$ 0.04 \\
900-1100 MeV 
& 15.2 & 1.24 $\pm$ 0.04 \\
1100-1300 MeV 
& 16.3 & 1.23 $\pm$ 0.06 \\
1300-1700 MeV 
& 20.7 & 1.10 $\pm$ 0.05  \\
1700-2100 MeV 
& 15.7 & 1.04 $\pm$ 0.05  \\ 
300-1700 MeV 
& 100.0 & 1.21 $\pm$ 0.05 \\ \hline
\end{tabular}
\end{center}
\end{table}

        We will use 1.21 as the Correction Factor by which to 
increase ${\cal B}$({\bdeca}), which is an equally weighed 
average of CLEO II and CLEO II.5, and add to our list of 
systematic errors a 4$\%$ contribution due to this correction.

\clearpage

\subsection{Backgrounds in Signal and Generic Monte Carlo} 

	To check that our antineutron selection criteria 
does not contaminate our sample with showers from decays 
similar to {\bdeca}, we run our {\bdeca}, with {\dzdecc} 
reconstruction code on signal Monte Carlo for 
{\bdecc} with equal numbers of $D_{1}{(2420)}^{0}$ 
and ${D_{2}}^{*}{(2460)}^{0}$, {\bpnpzds}, 
and {\bdelz} with ${\Delta}^0 \rightarrow n {\pi}^0$. 
The result for $m_{B^0}$ are overlayed in solid on our data 
results, in white, for {\bdeca} in Figure \ref{signalbkg}. 

\begin{figure}[ht]
   \centering \leavevmode
        \epsfysize=8cm
   \epsfbox[20 143 575 699]{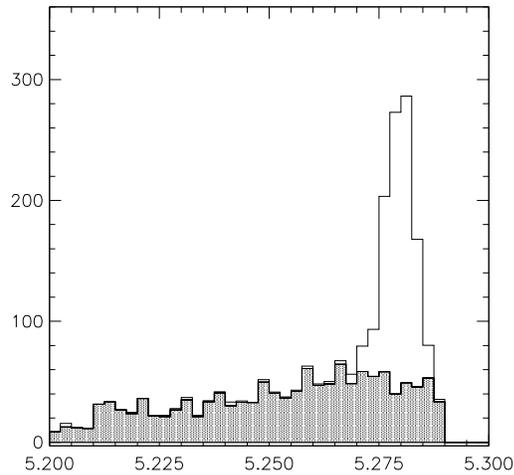}
   \caption{$m_{B^0}$ in GeV for {\bdeca} selection 
criteria applied to selected signal Monte Carlo background 
modes
\label{signalbkg}}
\end{figure}

	We also run the {\bdeca} code for all three 
{\dzz} modes on our generic Monte Carlo sample, as 
shown in Figure \ref{genericbkg}. The generic Monte Carlo 
sample is discussed in Section \ref{genmc}. 
Scaling the generic Monte Carlo sample 
to the size of our dataset, we find it to account 
for 40$\%$ of the background. 

        Within the generic MC sample $\approx$ 25$\%$ is 
combinatoric background events from each of {\dzdeca} 
and {\dzdecc}. The remaining events, 50$\%$, are from {\dzdecb}. 
Since {\bdnnx} decays are not included in 
the generic MC, these events are from random combinations 
of a {\bbaryons} decay and a $B \rightarrow$ {\dst} $X$ decay.  

\begin{figure}[ht]
   \centering \leavevmode
        \epsfysize=8cm
   \epsfbox[20 143 575 699]{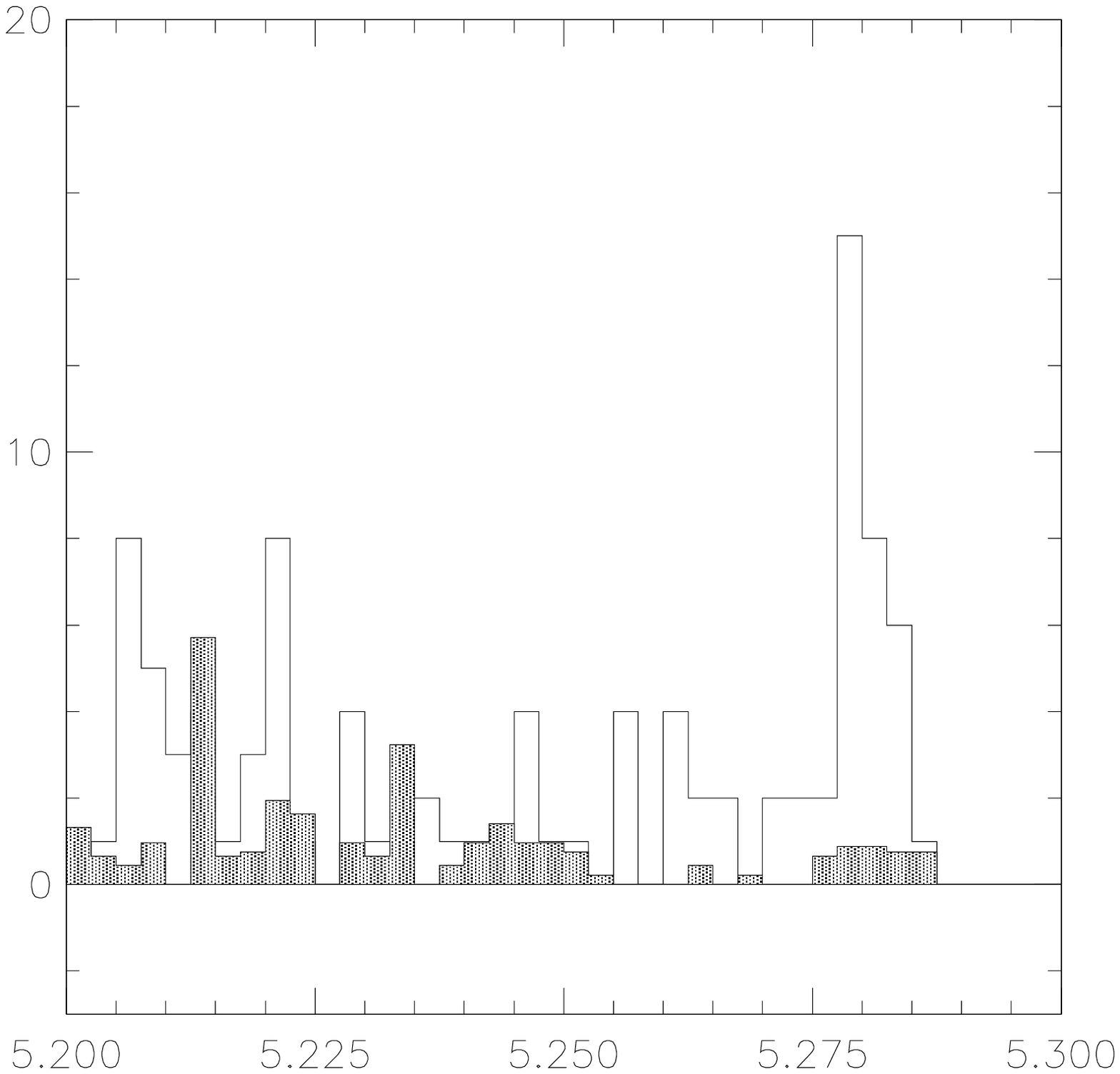}
   \caption{$m_{B^0}$ for {\bdeca} in GeV. (white) 
is data, (solid) is generic MC 
\label{genericbkg}}
\end{figure}

        We are unable to account for 60$\%$ of the background 
found in data in the $m_{B^0}$ distribution for {\bdeca}. 
However, this background is small, and the 
$m_{B^0}$ distribution we find is consistent with 
a signal at the nominal B mass.

\section{Antineutron Directional Cosine Resolution} 

        We compare the generated and reconstructed 
directional cosines for the antineutron candidates 
in signal Monte Carlo. 
We find the median error to be 37.8 milliradians. 
Since a single 5-cm calorimeter cell corresponds to 
$\approx$ 50 milliradians, the systematic error 
in our measurement of the directional cosines 
of the antineutron candidate shower is insignificant. 

\section{$\bar{B}^0 \rightarrow D^{*+} {\bar{p}} n$. }

        Neutrons do not annihilate in the calorimeter. 
Since our selection criteria for antineutron showers rejects 
the vast majority of neutrons, we derive a new set of cuts specific 
to neutrons. The selection criteria we 
used is shown in Figure \ref{neutroncuts}. The most 
important difference with antineutrons 
is the selection of low energy photon-like showers.

\begin{table}[htb]
\caption{Neutron shower selection criteria\label{neutroncuts} }
\begin{center}
\begin{tabular}{|c|}
\hline
Track-to-shower match level type 3 \\ 
E9OE25 $>$ cut 1\\ 
 $|(cos(\theta)|$ $<$ 0.71, or good barrel\\ 
${E}_{main}$ $<$ 500 MeV \\ \hline 
\end{tabular}
\end{center}
\end{table}

	We are unable to define selection criteria 
that allow us to separate neutrons from 
soft photons. This background overwhelms the signal, making unatainable 
the reconstruction of $\bar{B}^0 \rightarrow D^{*+} {\bar{p}} n$. 
We plot $E_{main}$ for neutrons (solid distribution) and antineutrons 
(dashed distribution) in Figure \ref{neutronbmass}. 

\begin{figure}[ht]
   \centering \leavevmode
        \epsfysize=7cm
   \epsfbox[20 143 575 699]{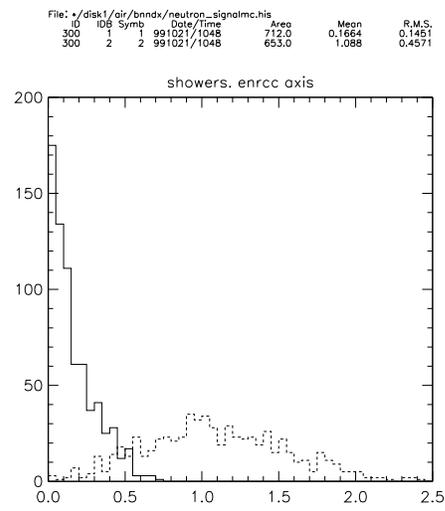}
\vskip 10 pt 
   \caption{$E_{main}$ (in GeV) for neutrons and antineutrons 
in {\bdeca} signal Monte Carlo \label{neutronbmass} }
\end{figure}

\clearpage

\section{Systematic Uncertainties} 

        In Table \ref{bpndssyst} we show the 
systematic uncertainties we consider to contribute 
significant errors to our measurement of ${\cal B}$({\bdeca}). 
{\bdeca} has an average of 4.6 tracks. 

\begin{table}[htb]
\caption{ Estimate of systematic uncertainties (in $\%$) 
for {\bdeca}
\label{bpndssyst}}
\begin{center}
\begin{tabular}{|c|c|}
\hline
Source & Uncertainty (in $\%$) \\ \hline 
{\dzz} branching fractions & 0.6 \\
{\dst} branching fraction & 1.4 \\ 
{\dst} reconstruction  & 5.0 \\
Monte Carlo statistics  & 5.0 \\ 
$\#$ of {\bbar}'s & 2.0 \\ 
tracking (1$\%$/track)   & 4.6 \\ 
PRLEV proton ID   & 4.0 \\
$\Delta$ background contribution & 5.0 \\
Phase space versus two body  & 3.0 \\
antineutron miss-ID &  15.0 \\ 
TOTAL  &  18.8 $\%$ \\ \hline 
\end{tabular}
\end{center}
\end{table}

\chapter{Conclusion}
	This work comprises the second successful 
exclusive reconstruction of {\bbaryons} modes, after the 
exclusive reconstructions of {\blcppi} and 
{\blcppipi} carried out by CLEO in 1997 \cite{jjo}. 
In Table \ref{allbb} we summarize all the exclusive 
{\bbaryons} measurements to date. We rank the modes 
from largest to smallest branching fraction central value. 
The first error is statistical and the second is systematic. 
The $\Lambda_c$ modes have a third uncertainty, which is 
systematic and is due to the error in the determination 
of ${\cal B}(\Lambda_c^+\to pK^-\pi^+)$. 

\begin{table}[htb]
\caption{Exclusive measurements in {\bbaryons} to date
\label{allbb}}
\begin{center}
\begin{tabular}{|c|c|}
\hline
Mode & Branching Fraction ($\times 10^{-4}$) \\ \hline
{\bdeca} & ${14.5}^{+3.4}_{-3.0} \pm 2.7$ \\ 
{\blcppipi} & ${13.3}^{+4.6}_{-4.2} \pm 3.1 \pm 2.1$ \\ 
{\bdecb} & ${6.6}^{+1.3}_{-1.2} \pm 1.0$ \\
{\blcppi} & ${6.2}^{+2.3}_{-2.0} \pm 1.1 \pm 1.0$ \\ \hline
\end{tabular}
\end{center}
\end{table}

	As shown in Table \ref{allbb}, 
the size of the branching fractions 
we are measuring in this work are of the same order 
of magnitude as those previously measured in 
decay modes including a $\Lambda_c$. We 
are unable to extrapolate the exclusive results to 
compare the relative magnitude of the inclusive 
modes with or without a $\Lambda_c$. For instance, 
the available phase space for the production of more 
light mesons is larger in the case of modes with 
a $\Lambda_c$ than in the case without, which 
may cause the $B \to \Lambda_c X$ inclusive rate 
to be larger than the {\bDnnx} inclusive rate. However, we 
think it reasonable to assume that the measurements 
we have at our disposal to date are representative 
of the inclusive modes. We infer that {\bDnnx} modes 
contribute significantly to the total 
{\bbaryons} decay rate.

\section{{\bbaryons} phenomenology}

	The {\bbaryons} theoretical models 
attempted to date \cite{jarfi,deshpande} 
rely on the assumption that the $B$ meson 
decays to two decay daughters, which is 
the two-body assumption. 
Unlike two-body $B$ meson modes, 
most {\bbaryons} modes, including all four that 
have been measured to date, have at least three decay 
daughters. Most {\bbaryons} modes also have a 
varied resonant substructure. The assumption 
that the hadronization process takes place 
at a late stage in the decay, which allows 
for a substantial simplification of the 
equations due to the suppressed dependence 
on the exchange of gluons and light quarks, 
is justified in hard two-body decays. However, in 
many-body decays there can be re-scattering 
of the hadronizing quarks and multiple 
exchanges of soft gluons and virtual 
quarks. Notwithstanding the calculational 
difficulties of many-body {\bbaryons} decays 
such as the ones we measure here, progress 
continues to be made using HQET to 
explain $B$ meson decay. The focus 
has been on decays for which HQET can be used to 
extract the most information, and which can 
be used to search for CP violation \cite{mathias,wise}. 
Future attempts to theoretically explain {\bbaryons} 
are needed.

\section{Possible Future {\bbaryons} Modes at CLEO}

	In CLEO II $D^0$ reconstruction from $B$ 
decays is plagued by high backgrounds due to poor 
separation between kaons and pions. 
The substantial improvements in charged particle 
separation (pion-kaon-proton) in the CLEO III data 
should allow for the successful reconstruction of 
$B^+ \rightarrow {\bar{D}}^0 p \bar{n}$ and 
$B^+ \rightarrow {\bar{D}}^0 p \bar{p} \pi^+$. 

	The ability to reconstruct antineutron modes 
should allow for the reconstruction of modes 
of type  $B \rightarrow (\Lambda_{c}, \Sigma_{c}) \bar{n} X$. 
Modes of type 
$B \rightarrow ({\Lambda_{c}}^*, {\Sigma_{c}}^*) \bar{p} X$, 
and $B \rightarrow \Xi_{c} \bar{\Lambda_{c}}$, 
which have low reconstruction efficiencies in CLEO II/II.5, 
are also worth pursuing. 

	{\bbaryons} for $b \rightarrow sg$, and 
$b \rightarrow u$ modes will also be of interest and 
likely within range. If previous measurements of the 
meson modes are used as a guide, the modes 
$B^+ \rightarrow \bar{\Lambda} p$, 
$\bar{B}^0 \rightarrow p \bar{n} \pi^-$, 
and $\bar{B}^0 \rightarrow p \bar{p} \pi^+ \pi^-$ are 
worth pursuing.  

	Semileptonic $b \rightarrow c$ {\bbaryons} 
decays, such as $B^+ \rightarrow {\Lambda_{c}}^- p l^+ \nu_{l}$, 
with $l=e,\mu$, due to the low reconstruction efficiency 
at CLEO for low momentum leptons, have not been 
measured. The substantially larger dataset expected 
in CLEO III may compensate for the low reconstruction 
efficiency. A mode such as 
$B^+ \rightarrow \bar{p} p l^+ \nu_{l}$ should be feasible. 
This latter mode can be used as an auxiliary in the 
measurement of $|V_{ub}|$. 

\section{Significance of Results}

We have found the first
evidence of decay modes of the $B^0$ of the type 
$B \to DN\bar{N}\pi$. We measure the branching fractions ${\cal B}$
({$B^0$$\rightarrow$ $D^{*-}$ $p$ ${\bar{p}}$ ${\pi}^+$}) = 
(${6.5}^{+1.3}_{-1.2}$ $\pm$ 1.0) $\times$ $10^{-4}$, and 
${\cal B}$({
$B^0$$\rightarrow$ $D^{*-}$ $p$ ${\bar{n}}$
}) = (${14.8}^{+3.5}_{-3.1}$ $\pm$ 3.0) 
$\times$ $10^{-4}$. 
These measurements indicate the fraction of baryonic decays of 
$B$ mesons that do not proceed via $\Lambda_c^+$ may be of 
approximately the same magnitude as those that do.

      \appendix		
   \backmatter		

\chapter{Biographical Sketch}

Antonio Rubiera was born in Havana, Cuba on January 21, 
1967. He came to the U.S. by the Mariel boatlift 
in 1980. He attended Shenandoah Jr. High and 
Miami Sr. High in Miami, Florida. From 1986 to 1991 
he attended Cornell University, obtaining a 
B.S. and an M.Eng., both in Electrical Engineering. 
From 1991 to 1994 he resided in Miami, where he 
worked for the Latin America branch of 
Ingersoll-Rand Co. He pursued a PhD in physics 
at the University of Florida from 1995 to 2000.

	
\end{document}